\documentclass[a4paper,11pt]{article}
\usepackage{jcappub} % for details on the use of the package, please
\usepackage[T1]{fontenc} % if needed
\usepackage{graphicx}% Include figure files
\usepackage{dcolumn}% Align table columns on decimal point
\usepackage{bm}% bold math
\usepackage{amsmath}
\usepackage{amsfonts} % for kokuban-shotai
\usepackage{latexsym}
\usepackage{bbm}
\usepackage{color}
\usepackage{amssymb}
\usepackage{amsthm}
\usepackage{epsf}
\usepackage{epsfig}
\usepackage{caption3}
\usepackage{appendix}
\usepackage{cases}
\usepackage{subfigure}
\usepackage{multirow}
\usepackage{float}
\usepackage{array}

\title{Phase transition dynamics and gravitational wave spectra of strong first-order phase transition in supercooled universe}

\author[a,b]{Xiao Wang,}
\author[c]{Fa Peng Huang,}
\author[a,b]{Xinmin Zhang}

\affiliation[a]{Theoretical Physics Division, Institute of High Energy Physics, Chinese Academy of Sciences, 19B Yuquan Road, Shijingshan District, Beijing 100049, China}
\affiliation[b]{School of Physics, University of Chinese Academy of Sciences, Beijing 100049, China}
\affiliation[c]{Department of Physics and McDonnell Center for the Space Sciences, Washington University, St.
	Louis, MO 63130, USA}

\emailAdd{wangxiao2016@ihep.ac.cn}
\emailAdd{fapeng.huang@wustl.edu}
\emailAdd{xmzhang@ihep.ac.cn}
\abstract{Phase transition dynamics may play important roles in the evolution history of the
	early universe, such as its possible roles in electroweak baryogenesis and dark matter.
	We systematically discuss and clarify the important details of the
	phase transition dynamics during a strong first-order phase transition (SFOPT).
	We classify the SFOPT into four types: slight supercooling,
	mild supercooling, strong supercooling, and ultra supercooling.
	Using different characteristic temperatures, length scales and
	bubble wall velocities, the corresponding gravitational wave (GW) spectra are investigated in details.
	We emphasize the essential importance of using the correct characteristic temperature and length scale when the phase transition dynamics and GW spectra are calculated.
	Especially, for strong supercooling and ultra supercooling cases, there are obvious differences of the phase transition strength and GW spectra between the results calculated at the nucleation temperature and those derived at the percolation temperature.
	For ultra supercooling case, we propose a criterion to quantify whether the phase transition can terminate.
	Besides the model-independent discussions, we also study three representative models as concrete examples to clearly show the subtle points therein.}


\arxivnumber{2003.08892}

\begin{document}
	\maketitle
	\flushbottom
\section{Introduction}
Motivated by the current theoretical and experimental status of particle cosmology,
the precise study of phase transition dynamics becomes more and more important.
From the aspect of theory, a strong first-order phase transition (SFOPT) is
essential to electroweak baryogenesis~\cite{Trodden:1998ym}, dark matter~\cite{Huang:2017kzu,Huang:2017rzf,Baker:2019ndr,Chway:2019kft},\footnote{For example, the dynamics of phase transition can essentially determine the property of dark matter as in refs.~\cite{Huang:2017kzu,Baker:2019ndr,Chway:2019kft}.
Cosmological first-order phase transition in the early universe is an alternative formation mechanism of primordial black hole \cite{Hawking:1982ga,Kodama:1982sf,Konoplich:1999qq,Khlopov:2000js}, which can provide some fractions of the observed dark matter density.} stochastic phase transition gravitational wave (GW) signal, and the true shape of Higgs potential, which are the central issues in the frontier research fields of cosmology and particle physics.
From the aspect of experiments, the discovery of Higgs boson~\cite{Aad:2012tfa,Chatrchyan:2012xdj} at LHC and GW at aLIGO~\cite{Abbott:2016blz} open a new window to study the above problems and new physics beyond standard model (SM) by electroweak phase transition GW.
Especially, the approved Laser Interferometer Space Antenna (LISA) (launch in 2034 or even earlier)~\cite{Audley:2017drz} project makes it extremely important to precisely study the electroweak phase transition dynamics and their corresponding GW signals.
Besides LISA~\cite{Audley:2017drz,Caprini:2015zlo,Caprini:2019egz,LISA:documents}, there are more and more proposals of GW experiments, such as
Taiji~\cite{Hu:2017mde,Guo:2018npi}, TianQin~\cite{Luo:2015ght,Hu:2018yqb},
Deci-hertz Interferometer Gravitational wave Observatory (DECIGO)~\cite{Seto:2001qf,Kawamura:2011zz}, Ultimate-DECIGO (U-DECIGO)~\cite{Kudoh:2005as}, and Big Bang Observer (BBO)~\cite{Corbin:2005ny}.

To completely and correctly quantify the phase transition dynamics, there are several important characteristic parameters,
including the characteristic temperature and the characteristic length scale.
We clarify the importance of using the correct characteristic parameters to obtain more precise GW spectra.
There are several characteristic temperatures, such as the critical temperature $T_c$, the nucleation temperature $T_n$, the percolation temperature $T_p$ and so on.
Each temperature usually represents different physical process.
For different physical process, we should use the most appropriate one.

To conveniently describe and study the different types of SFOPT, we classify the SFOPT into four classes:
\begin{enumerate}
	\item Slight supercooling for   $\alpha \leq 0.1$
	\item Mild supercooling~\cite{Linde:2005ht} for   $0.1 \leq \alpha \leq 0.5$
    \item Strong supercooling for  $0.5 \leq \alpha \leq 1$
	\item Ultra supercooling~\cite{Jinno:2019jhi} for $\alpha \geq 1$
\end{enumerate}

For slight supercooling case, the nucleation and percolation temperatures are close.
However, for strong supercooling and ultra supercooling, there may exist obvious hierarchies between the two temperatures.
We clarify the essential differences of using different temperatures to determine the
phase transition dynamics and the corresponding GW signals. Especially for the strong supercooling and ultra supercooling cases,
the differences between the results obtained from the inappropriate temperature and the correct temperature are obvious.
We explicitly study different cases by comparing the
hierarchies between these characteristic temperatures.
For example, in the ultra supercooling case, the percolation temperature $T_p$ may be obviously smaller than the nucleation temperature $T_n$. To calculate more reliable GW spectra, the percolation temperature $T_p$ should be used in lieu of the nucleation temperature $T_n$ during the calculations of  the phase transition dynamics and GW spectra. The criterion of the phase transition termination also should be carefully considered for the ultra supercooling case.

After the general discussions, we take three concrete and representative models as examples.
Since we focus on the phase transition dynamics and relevant GWs, the collider constraints and predictions for the models are beyond our consideration in this work.
Firstly, we study the benchmark effective scenario where the Higgs sextic term contributes
to a SFOPT~\cite{Zhang:1992fs,Grojean:2004xa,Huang:2015izx,Huang:2016odd,Cao:2017oez}.
With this dimension-six effective operator, the new Higgs potential is provided, and a SFOPT can be naturally triggered with strong strength.
From the perspective of SM effective field theory, this effective scenario can represent general properties of many models, like singlet, doublet, triplet extended Higgs model and composite Higgs model~\cite{Cao:2017oez}.
Secondly, we study a renormalizable toy model where the thermal barrier is formed by the cubic term.
This toy model can be treated as approximations of many renormalizable models with light new particles beyond SM.
Lastly, we study the logarithm model where the effective thermal barrier is formed by the logarithm term.
From the typical models, we can clearly see that it is crucial to use the correct temperature to obtain more reliable GW spectra in the calculations.
We study the general properties for the strong supercooling and ultra supercooling cases.
And for the strong supercooling and ultra supercooling cases, better definition of the phase transition duration is needed.
In most viable parameter spaces of these three models, the long-lasting sound wave is not favored, and hence their GW signals are suppressed.

This work is organized as follows. In section~\ref{PTdy}, we clarify the important quantities of the phase transition dynamics, which are the characteristic temperature, characteristic length scale, energy budget, and the bubble expansion mode.
Then the three mechanisms to generate the GWs during phase transition are discussed in section~\ref{GW}, and the formulae of the GW spectrum for each source are given.
In section~\ref{models}, we study three representative benchmark models with the detailed phase transition dynamics and the precise GW spectra.
The alternative criterion for the completion of ultra supercooling is proposed, and the possible effects to the GW generation are discussed in section~\ref{Discu}.
Section~\ref{Conclu} gives our conclusion.

\section{Phase transition dynamics}
\label{PTdy}

\begin{figure}[t]
	\centering
	
	\subfigure{
		\begin{minipage}[t]{0.5\linewidth}
			\centering
			\includegraphics[scale=0.5]{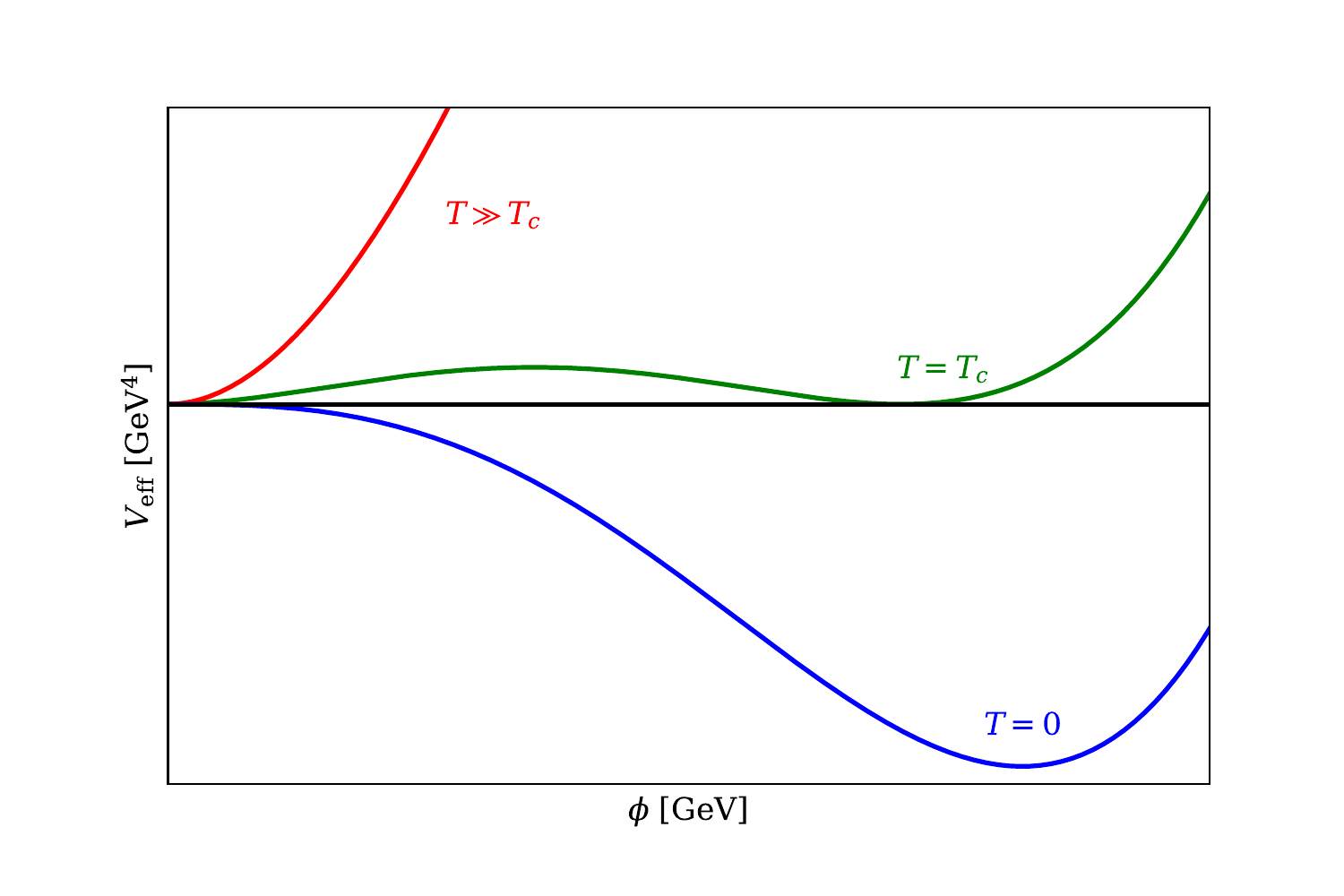}
		\end{minipage}%
	}%
	\subfigure{
		\begin{minipage}[t]{0.5\linewidth}
			\centering
			\includegraphics[scale=0.5]{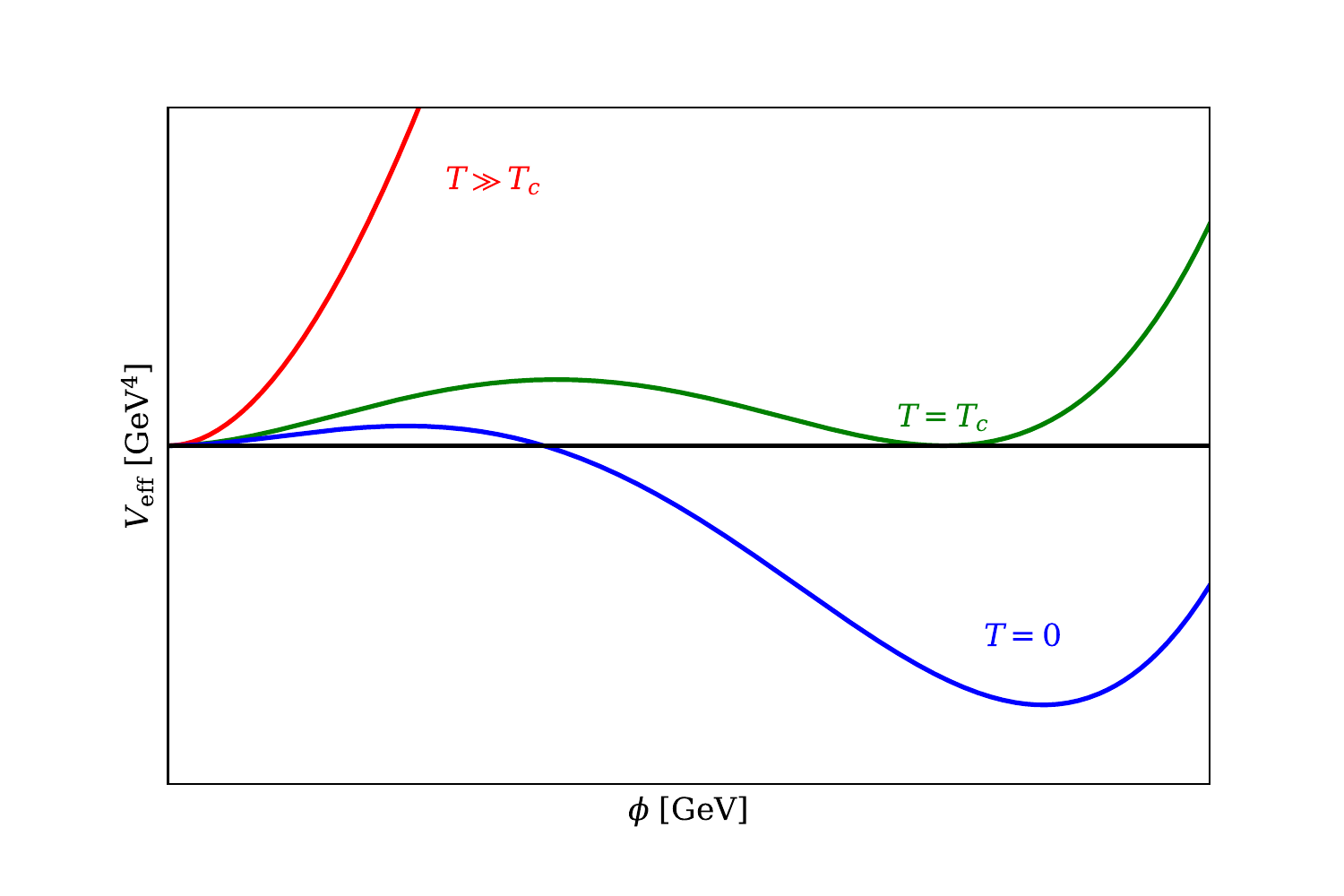}
			%\caption{fig2}
		\end{minipage}%
	}%
	\quad
	\subfigure{
		\begin{minipage}[t]{0.5\linewidth}
			\centering
			\includegraphics[scale=0.5]{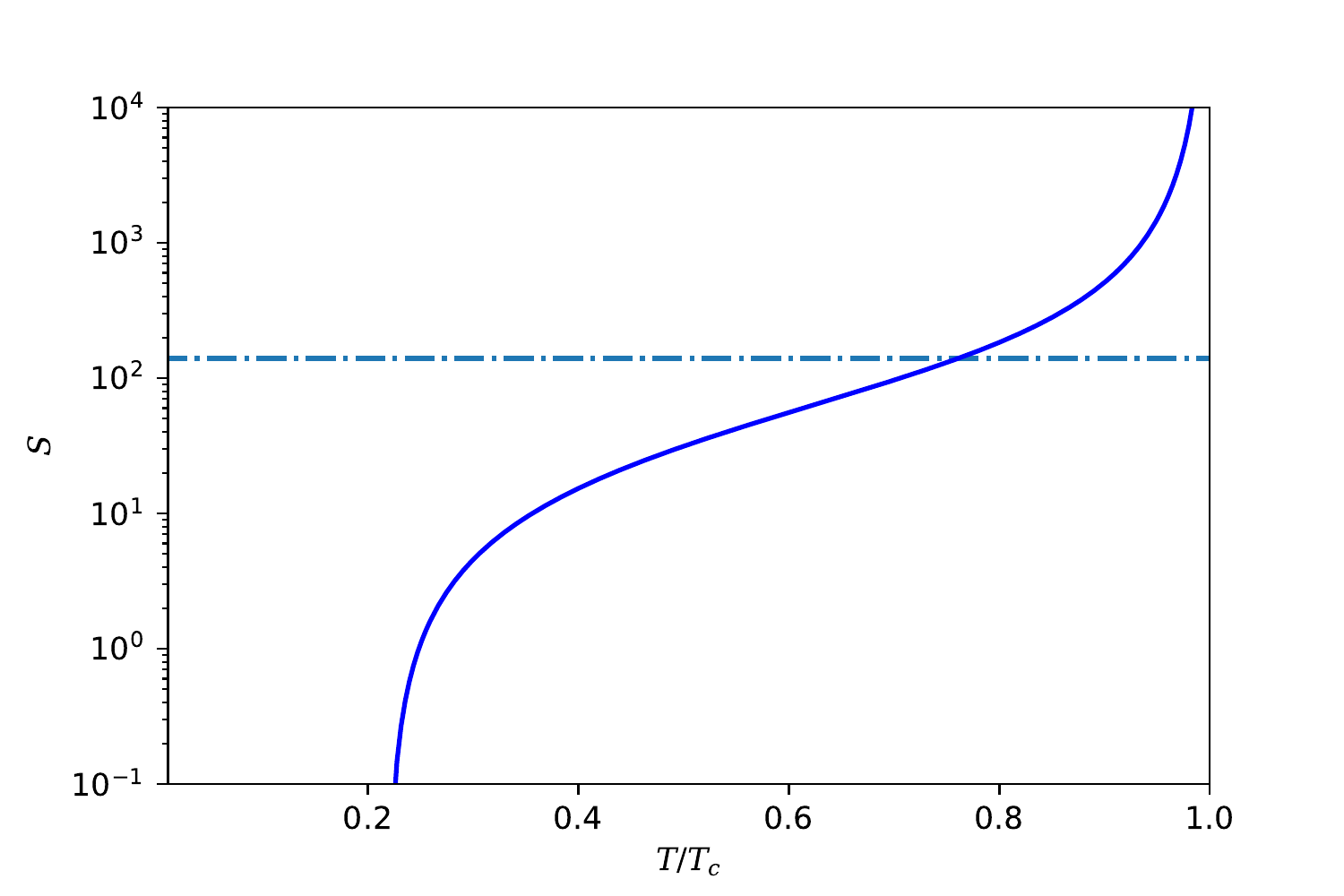}
		\end{minipage}
	}%
	\subfigure{
		\begin{minipage}[t]{0.5\linewidth}
			\centering
			\includegraphics[scale=0.5]{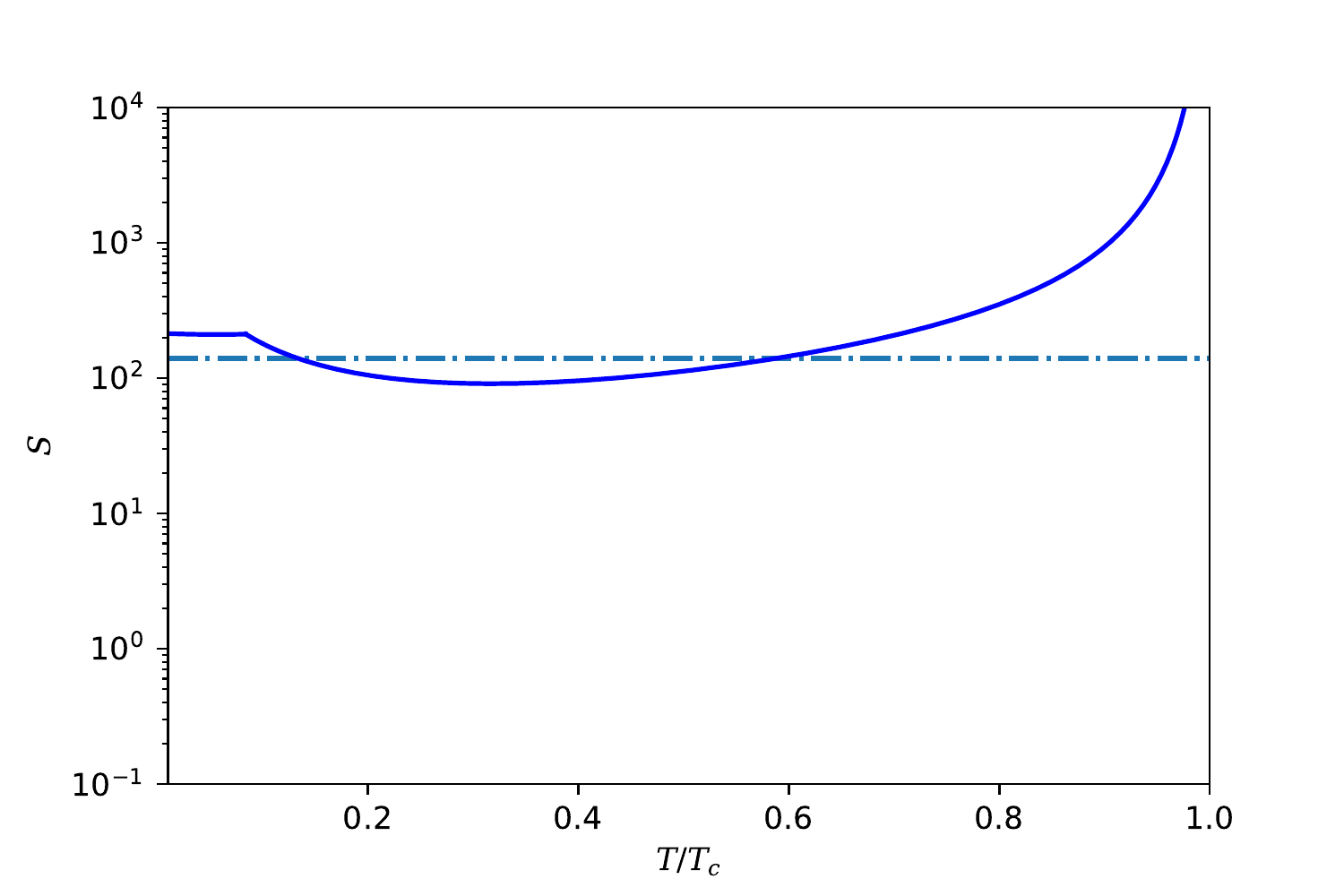}
		\end{minipage}
	}%
	
	\centering
	\caption{The evolution of the bounce action and the finite-temperature effective potential for two cases. Left: this represents an effective potential without zero-temperature potential barrier and its evolution of bounce action. Right: it depicts an effective potential with zero-temperature potential barrier and its evolution of bounce action.}\label{fg1}
\end{figure}
The essential quantity of phase transition dynamics is the bubble nucleation rate (the decay rate of the false vacuum) per unit time per unit volume
\begin{equation}
\Gamma \approx \Gamma_0e^{-S},
\end{equation}
where $S$ is the bounce action, and the temperature dependent factor $\Gamma_0$ will be given later.
The most general form of the bounce action should be \cite{Salvio:2016mvj}
\begin{equation}
S(T) = 4\pi\int_{1/T}^{0}d\tau\int_{0}^{\infty}drr^2\left[\frac{1}{2}\left(\frac{\partial\phi}{\partial\tau}\right)^2 + \frac{1}{2}\left(\frac{\partial\phi}{\partial r}\right)^2 + V_{\rm eff}(\phi, T)\right],
\end{equation}
where $\tau = it$ is the Euclidean time.
$V_{\rm eff}$ is the effective potential derived by concrete models with the thermal effective field theory \cite{Quiros:1999jp,Laine:2016hma}.
Generally the effective finite-temperature potential is composed of the following part
\begin{equation}
V_{\rm eff}(\phi, T) \equiv V_{\rm tree}(\phi) + V_{CW}(\phi) + V_{CT}(\phi) + V_T(\phi, T)\,\,\,, \label{eq.effpot}
\end{equation}
where $V_{\rm tree}$ is the tree-level potential derived by replacing the fields by classical background field $\phi$.
$V_{CW}$ is the zero-temperature one-loop corrections to the tree-level potential and conventionally called Coleman-Weinberg potential.
$V_{CT}$ is the counter-term potential to cancel the ultraviolet divergence in the one-loop corrections at zero temperature.
And the thermal correction including the daisy resummation at finite temperature is accommodated into $V_T$.
To perform daisy resummation, there exist two schemes, the Arnold-Espinosa scheme \cite{Arnold:1992rz} and Parwani scheme \cite{Parwani:1991gq}.
The effective potential and the corresponding results depend on the resummation scheme.
More works are needed to solve this problem.

From the finite-temperature effective potential, we can derive the bounce action by solving the following
equation-of-motion (EOM)
\begin{equation}
\frac{\partial^2\phi}{\partial\tau^2} + \frac{\partial^2\phi}{\partial r^2} + \frac{2}{r}\frac{\partial\phi}{\partial r} = \frac{\partial V_{\rm eff}}{\partial\phi}
\end{equation}
with the boundary conditions
\begin{equation}
\frac{\partial\phi}{\partial\tau}\Big|_{\tau = 0, \pm\frac{1}{2T}}=0,\quad\frac{\partial\phi}{\partial r}\Big|_{r = 0}=0, \quad \lim\limits_{r\rightarrow\infty}\phi(r) = \phi_{\rm false}\,\,.
\end{equation}
This EOM is a partial derivative equation which requires time-consuming numerical calculations.
Conventionally, we use an approximated bounce action to perform the calculations. The bounce action is approximated as \cite{Linde:1980tt,Linde:1981zj}
\begin{equation}
S(T) \approx \min[S_4(T), S_3(T)/T]\,\,,
\end{equation}
where $S_4(T)$ is bounce action for the $O(4)$-symmetric bounce solution
\begin{equation}
S_4 = 2\pi^2\int_{0}^{\infty}d\tilde{r}\tilde{r}^3\left[\frac{1}{2}\left(\frac{d\phi}{d\tilde{r}}\right)^2 + V_{\rm eff}\right],
\end{equation}
where $\tilde{r} = \sqrt{\tau^2 + r^2}$. The EOM and boundary conditions are
\begin{equation}
\frac{d^2\phi}{d\tilde{r}^2}+\frac{3}{\tilde{r}}\frac{d\phi}{d\tilde{r}} = \frac{\partial V_{\rm eff}}{\partial\phi}, \quad \phi^\prime(0) = 0, \quad \phi(\infty) = \phi_{\rm false}\,\,.
\end{equation}
$S_3(T)$ is bounce action for the $O(3)$-symmetric bounce solution
\begin{equation}
S_3 = 4\pi\int_{0}^{\infty}drr^2\left[\frac{1}{2}\left(\frac{d\phi}{dr}\right)^2 + V_{\rm eff}\right]\,\,.
\end{equation}
The corresponding EOM and boundary conditions are
\begin{equation}
\frac{d^2\phi}{dr^2}+\frac{2}{r}\frac{d\phi}{dr} = \frac{\partial V_{\rm eff}}{\partial\phi}, \quad \phi^\prime(0) = 0, \quad \phi(\infty) = \phi_{\rm false}\,\,.
\end{equation}
For single-field bounce equation, the overshooting-undershooting method~\cite{Coleman:1977py,Callan:1977pt} is
widely used to numerically solve it.
For the multi-field case, the path deformation method~\cite{Wainwright:2011kj} is the most extensively used in various literatures.
Then, the decay rate of the false vacuum can be obtained as
\begin{equation}
\Gamma(T)\approx\begin{cases}
T^4\left(\frac{S_3}{2\pi T}\right)^{\frac{3}{2}}\exp\left(-\frac{S_3}{T}\right),& T > T_{\rm div}\\
T_{\rm div}^4\left(\frac{S_4}{2\pi}\right)^2\exp(-S_4),& T < T_{\rm div}
\end{cases},
\end{equation}
where $T_{\rm div}$ is the dividing temperature to distinguish $O(3)$ regime and $O(4)$ regime.
The decay rate is dominated by thermal tunneling process if the temperature is higher than $T_{\rm div}$.
When temperature become lower than the dividing temperature, it is dominated by vacuum tunneling process.
Figure~\ref{fg1} shows two typical behavior of the bounce action that corresponds to the evolution of the finite-temperature effective potential.
The left panel shows that for an effective potential without a zero-temperature potential barrier, the bounce action monotonously decreases with the decreasing of the temperature and should eventually become zero at some specific temperature.
On the contrary, for the right panel, the bounce action firstly decreases to a minimum value and then increases with decreasing of the temperature.
And after the temperature drops down to $T_{\rm div}$, the vacuum tunneling process dominates.
Then the bounce action is nearly a constant.
This process is induced by an effective potential with zero-temperature potential barrier, and usually this type of effective potential can generate the ultra supercooling.
For this process, the first derivative of bounce action with respect to temperature, which is related to the approximated time duration of phase transition, can become negative in relatively low temperature region.

\subsection{Characteristic temperature}
\begin{figure}[t]
	\centering
	\subfigure{
		\begin{minipage}[t]{1\linewidth}
			\centering
			\includegraphics[scale=0.6]{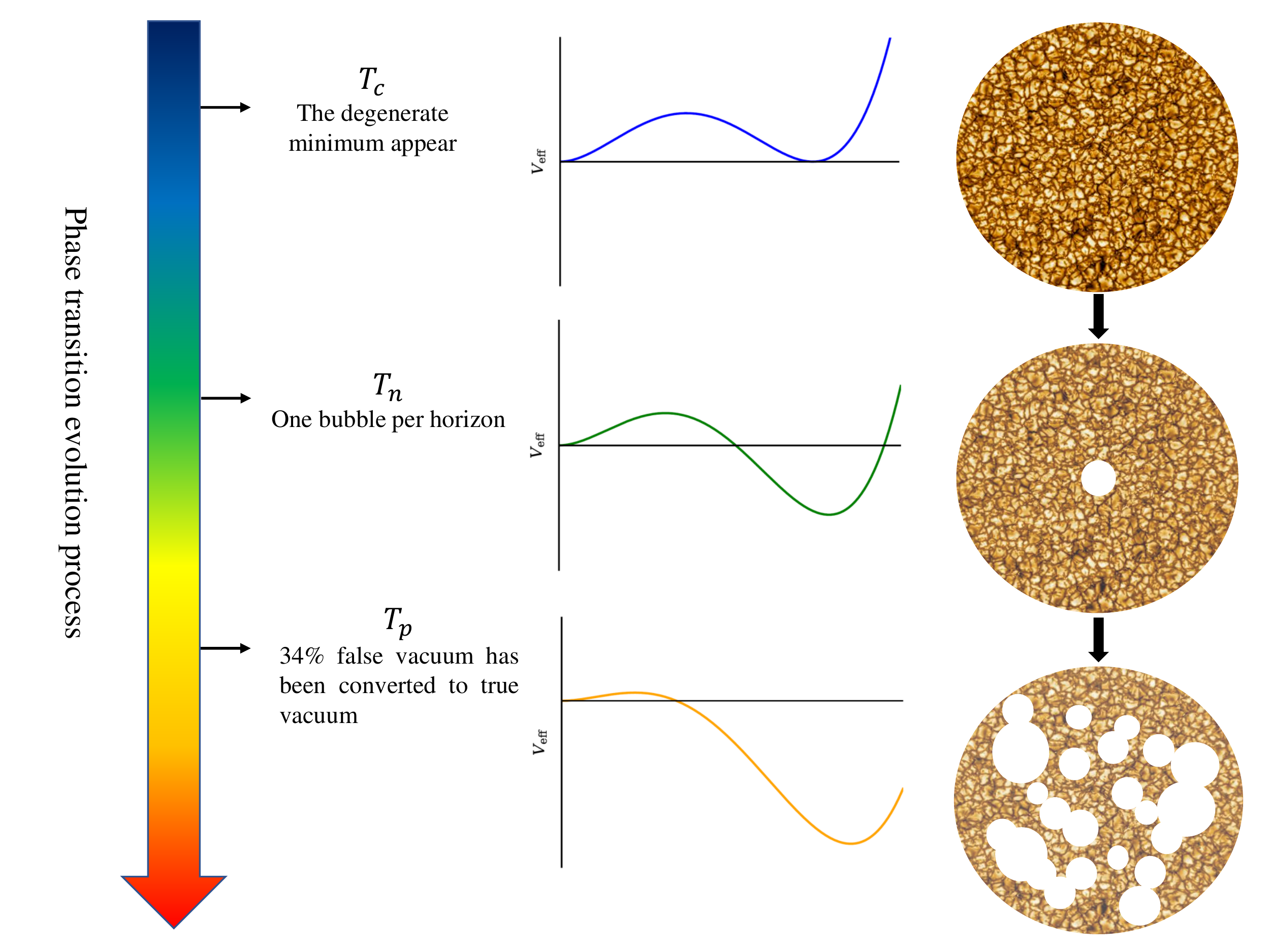}
			%\caption{fig2}
		\end{minipage}%
	}%
	\centering
	\caption{
The illustration of phase transition evolution process.
The middle panel shows the evolution of the effective potential at different temperatures.
The right panel represents the vacuum bubble evolution with the decreasing of the temperature during a SFOPT.
At the critical temperature, no bubbles have been generated, the whole Universe is in the false vacuum phase.
As temperature decreases to the nucleation temperature, for one Hubble volume, the accumulated number of bubbles reach order of unity.
With further decreasing of the temperature, bubbles expand and collide each other, and form a cluster-like structure at the percolation temperature, where $34\%$ false vacuum has been converted to the true vacuum.
	}\label{PTevo}
\end{figure}

To accurately study the phase transition dynamics, we need to clarify the following characteristic temperatures, which represent distinctive physical processes in a SFOPT, to illustrate the phase transition evolution.
\begin{enumerate}
	\item \emph{Critical temperature}. The temperature $T_c$ at which the effective potential has two degenerate minimums.
	\item \emph{Nucleation temperature}. The temperature $T_n$ at which one bubble is nucleated in one casual Hubble volume.
	\item \emph{Percolation temperature}. The temperature $T_p$ at which the probability of finding a point still in the false vacuum is 0.7 or 34\% false vacuum has been converted to the true vacuum.
	\item \emph{Minimum temperature}. The temperature $T_{\rm min}$ at which the temperature that minimizes the bounce action.
	\item \emph{Maximum temperature}. The temperature $T_{\max}$ at which the physical false vacuum volume begins to decrease.
\end{enumerate}
The nucleation temperature $T_n$ at which one bubble is nucleated per horizon on average, can be computed as
\begin{equation}
N(T_n) = \int_{t_c}^{t_n}dt\frac{\Gamma(t)}{H(t)^3} = \int_{T_n}^{T_c}\frac{dT}{T}\frac{\Gamma}{H^4}=1\,\,, \label{nucl}
\end{equation}
here we use the adiabatic time-temperature relation $dt=-dT/(TH(T))$.
There is another definition of nucleation tempeature which is treated as the moment when the nucleation rate first catches up with the Hubble rate,
\begin{equation}
\frac{\Gamma(T_n)}{H(T_n)^4} = 1\,\,.
\end{equation}
For the electroweak scale phase transition, this roughly corresponds to $S(T_n)\approx140$.
In this work, we use the first definition of the nucleation temperature.

The percolation temperature is defined as the moment at least 34\% of the false vacuum has been converted to the true vacuum, or at the time the probability of finding a point still in the false vacuum is 0.7.
To calculate the percolation temperature, we need to compute the probability of finding a point still in the false vacuum, given by \cite{Turner:1992tz,Megevand:2016lpr,Kobakhidze:2017mru,Cai:2017tmh,Ellis:2018mja}
\begin{equation}
P(t) = e^{-I(t)}, \quad I(t) = \frac{4\pi}{3}\int_{t_c}^{t}dt^\prime\Gamma(t^\prime)a(t^\prime)^3r(t,t^\prime)^3\,\,,
\end{equation}
where the scale factor $a(t^\prime)$ accounts for the expansion of the Universe, and the exponential corresponds to the bubble overlapping and yields the amount of the true vacuum volume per unit comoving volume.
$r(t,t^\prime)$ is the comoving distance of a bubble expansion from the earlier time $t^\prime$ to some later time $t$:
\begin{equation}
r(t, t^\prime) = \int_{t^\prime}^{t}\frac{v_b(\tilde{t})\tilde{t}}{a(\tilde{t})} \,\,, \label{corad}
\end{equation}
where $v_b(t)$ is the bubble wall velocity.
In our work, we assume the bubble wall achieves the terminal velocity very fast, so we can set $v_b$ as a constant, which can be a good approximation.
In the following calculations, we set $v_b = 1$ and $v_b=0.3$ as two default values to compare the effects of bubble wall velocity to percolation temperature and other important parameters in calculating the phase transition dynamics and GW signals.

For further discussions, we need to calculate the scale factor or the Hubble rate.
Based on radiation dominant scenario (the vacuum energy released by a SFOPT is negligible), the Hubble rate is computed in terms of radiation energy for most studies.
However, for a SFOPT with strong or ultra supercooling, the released vacuum energy during the phase transition process is possible to become important and even dominant at relatively low temperature. Then we can derive the Friedmann equation in terms of the vacuum and radiation energy densities $\rho_R$ and $\rho_V$,
\begin{equation}
H^2 = \frac{1}{3M_{\rm pl}^2}\left(\rho_R + \rho_V\right) \,\,,\label{hubble}
\end{equation}
where $M_{\rm pl} = 2.435\times 10^{18}$ GeV is the reduced Planck mass.
The strength parameter of phase transition $\alpha$ is defined as the ratio of vacuum to radiation energy density.
Hence, different definitions of the released vacuum energy during the phase transition has some effects on the Hubble rate, strength parameter and other related quantities.
We will discuss this later in detail.

\begin{figure}[t]
	\centering
	\subfigure{
		\begin{minipage}[t]{1\linewidth}
			\centering
			\includegraphics[scale=0.8]{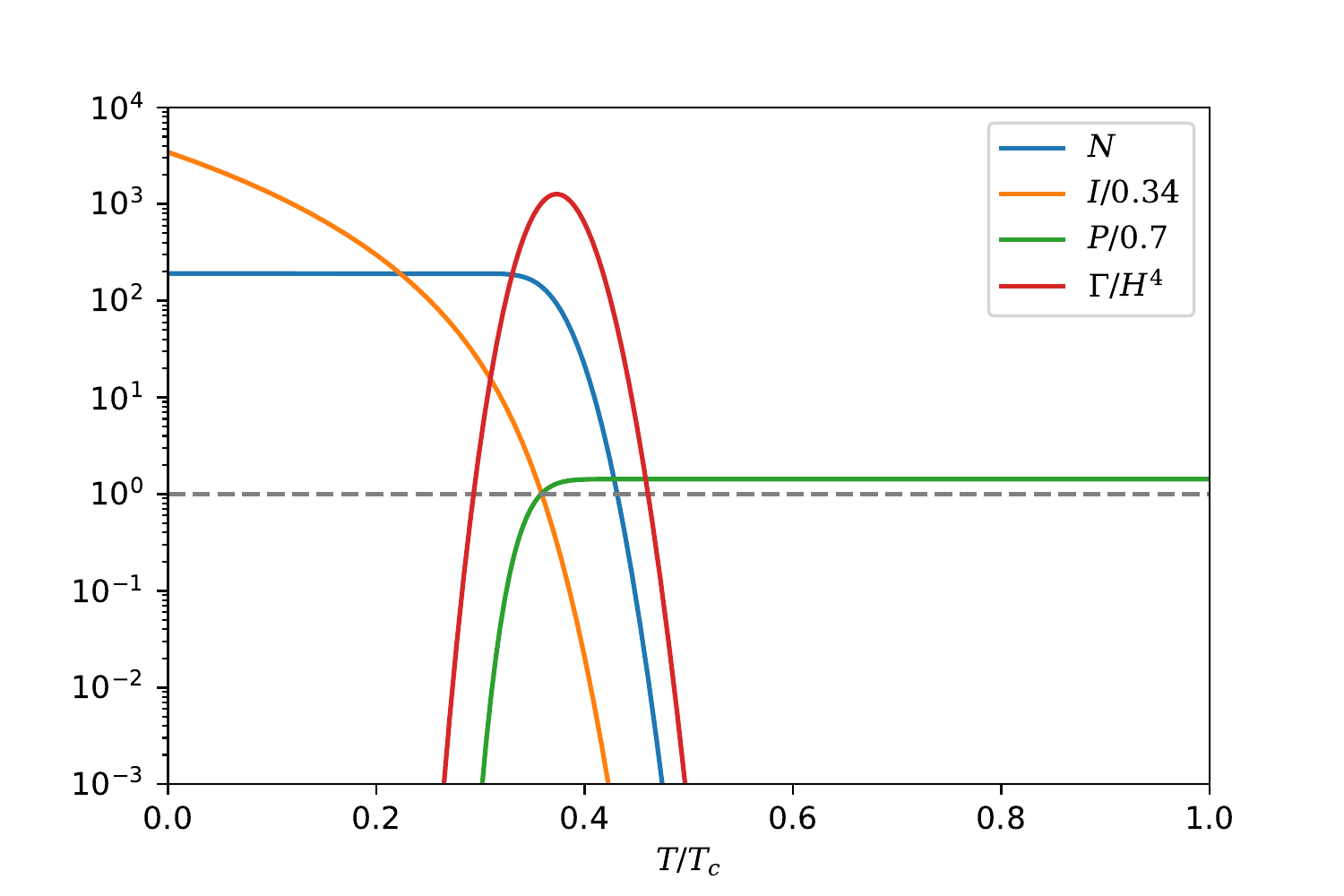}
		\end{minipage}
	}
	\centering
	\caption{Illustration for values of $\Gamma/H^4$ (red line), the number of bubbles $N$ (blue line) per horizon, the fraction of vacuum $I$ (orange line) converted to true vacuum, and the probability $P$ (green line) of finding a point in false vacuum as function of $T/T_c$, corresponding to a SFOPT with zero-temperature potential barrier.
	The intersection point of the blue with and the horizontal dashed line represent the nucleation temperature.
	The intersection points of the green and orange lines with the dashed line indicate the percolation temperature.
	}\label{PIN}
\end{figure}

From eq.~\eqref{corad}, eq.~\eqref{hubble} and the adiabatic time-temperature relation, the physical radius of bubbles $R(T,T^\prime)$ ($T^\prime > T$) can be calculated as following
\begin{equation}
R(T, T^\prime)=a(T^\prime)r(T,T^\prime)=v_ba(T^\prime)\int_{T}^{T^\prime}\frac{d\tilde{T}}{a(\tilde{T})H(\tilde{T})\tilde{T}}=\frac{v_b}{T^\prime}\int_{T}^{T^\prime}\frac{d\tilde{T}}{H(\tilde{T})}\,\,.
\end{equation}
And the fraction converted to the true vacuum $I(T)$ can be derived as
\begin{equation}
I(T) = \frac{4\pi v_b^3}{3}\int_{T}^{T_c}\frac{dT^\prime\Gamma(T^\prime)}{H(T^\prime){T^\prime}^4}\left(\int_{T}^{T^\prime}\frac{d\tilde{T}}{H(\tilde{T})}\right)^3\,\,.
\end{equation}
Then the percolation temperature $T_p$ can be directly derived from $I(T_p) = 0.34$ or equivalently $P(T_p) = 0.7$.
Figure~\ref{PTevo} shows the bubble formation and evolution at different temperatures, and it also denotes the characteristic shape of the effective potential with the decreasing of the temperature.
Figure~\ref{PIN} shows the typical evolution for the values of $\Gamma(T)/H(T)^4$, the number of bubbles per Hubble volume $N(T)$, the fraction of vacuum converted to true vacuum $I(T)$, and the probability of finding a point in false vacuum $P(T)$.
The intersection point of the blue line with the horizontal dashed line represent the nucleation temperature.
The intersection point of the green line with the horizontal dashed line (or the intersection point of the orange line with the dashed line) indicates the percolation temperature.
The minimum temperature $T_{\min}$ is defined as the temperature at which the bounce action minimizes itself.
The minimum temperature should be the temperature where the potential barrier disappears (i.e. the bounce action becomes zero) for the typical behaviors as shown in the left panel of figure~\ref{fg1}.
For the effective potential that still persists a potential barrier at zero-temperature, there exists a minimum of the action as shown in the right panel of figure~\ref{fg1}.
And the temperature at which the bounce action gets its minimum value, is the minimum temperature for this case.

For the radiation-dominated scenario, if a phase transition can achieve the percolation temperature, it indicates that the phase transition process can be completed.
This percolation criterion is not valid for a phase transition with ultra supercooling.
Since the ultra supercooling can introduce a vacuum-dominated situation and induce a inflationary stage for the false vacuum. Therefore, the percolation mentioned above may never be achieved.
In the strong and ultra supercooling cases, for successful completion of phase transition, it requires the physical volume of the false vacuum $\mathcal{V}_{\rm false}$ decreases around percolation.
Hence, the condition of the completion of phase transition for strong and ultra supercooling reads \cite{Turner:1992tz, Ellis:2018mja}
\begin{equation}
\frac{1}{\mathcal{V}_{\rm false}}\frac{d\mathcal{V}_{\rm false}}{dt} = 3H(t) - \frac{dI(t)}{dt} = H(T)\left(3 + T\frac{dI(T)}{dT}\right) < 0\,\,. \label{EC}
\end{equation}
If this is fulfilled at the percolation temperature $T_p$, we can conclude that the phase transition terminates successfully.

\begin{figure}[t]
	\centering
	\subfigure{
		\begin{minipage}[t]{1\linewidth}
			\centering
			\includegraphics[scale=0.8]{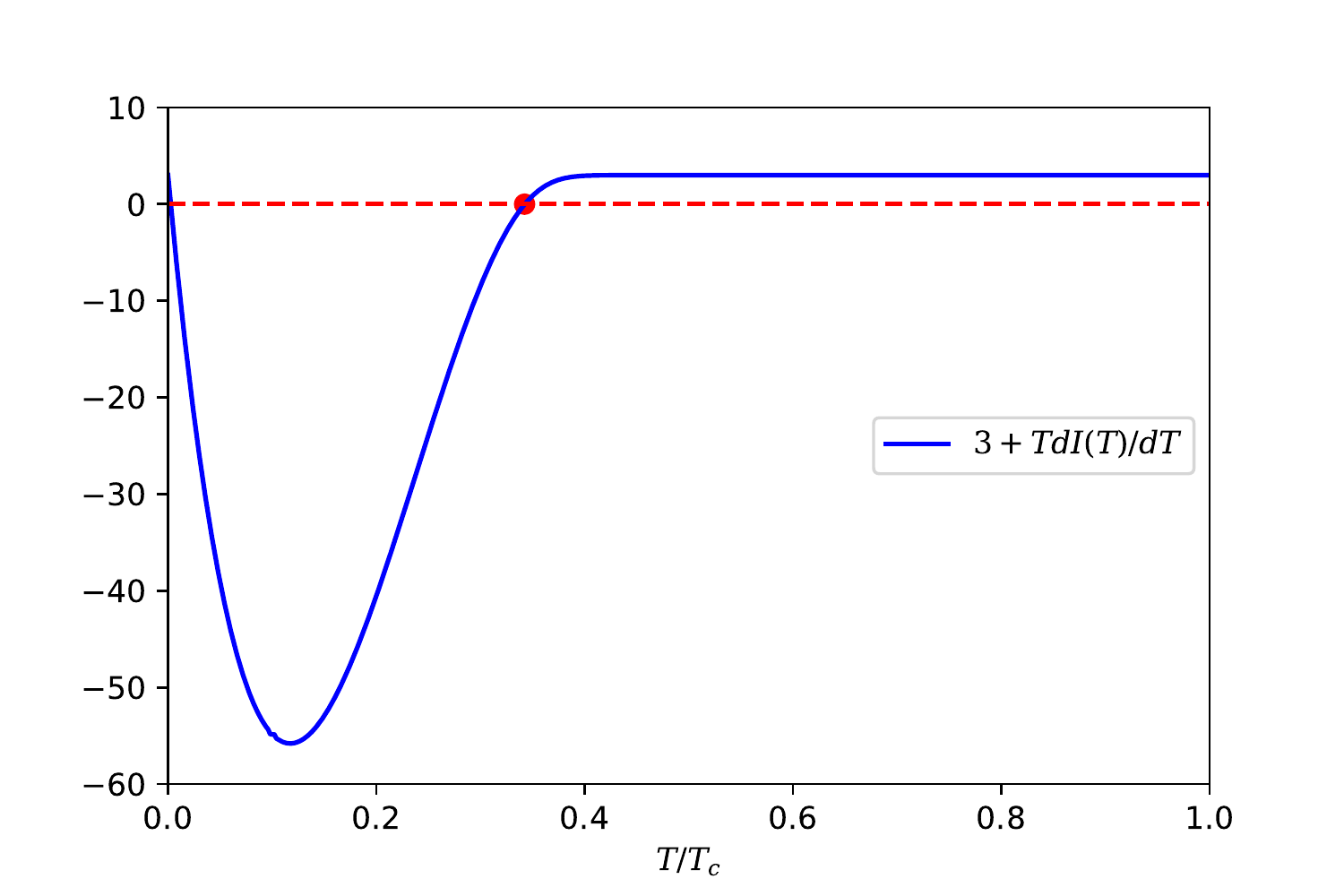}
		\end{minipage}
	}
	\centering
	\caption{The typical evolution of $3+TdI/dT$ for the ultra supercooling. The red dot indicates the maximum temperature $T_{\max}$.
	}\label{completion}
\end{figure}

Since the Hubble rate is always positive, the condition can only be satisfied when the values in the bracket of eq.~(\ref{EC}) become negative.
We show the typical behavior of $3+TdI/dT$ for the ultra supercooling in figure~\ref{completion}.
There is a temperature range with negative values.
With the decreasing of temperature, the value of $3+TdI/dT$ first decreases, and then increases.
$T_{\max}$ is then defined as the temperature when $3+TdI/dT$ first decreases to zero with the decreasing of temperature.
The red dot indicates the maximum temperature $T_{\max}$ in figure~\ref{completion}.
If the parameter space of a concrete model have a percolation temperature that is lower than the maximum temperature, the completion of phase transition is questionable.
And when the maximum temperature can not be derived from a concrete model, the phase transition can not terminate.

\subsection{Characteristic length scale}
Another important parameter is the characteristic length scale of the SFOPT, which is expected to be the scale carrying the majority of the energy released by phase transition.
And the length scale is an important parameter for the calculation of GWs produced by bulk fluid motion and bubble collision.
Usually, the mean bubble separation is regarded as the characteristic length scale.

The commonly used estimation of the mean bubble separation is based on the short-duration approximation of a SFOPT.
Hence, the bounce action can be expanded around some time $t_*$ as
\begin{equation}
S(t) \approx S(t_*) - \beta(t-t_*) + \mathrm{O}((t-t_*)^2)... ,
\end{equation}
where $\beta$ is a good approximation to the inverse time duration of a slight supercooling. %mild supercooling?
Therefore, we generally estimate the mean bubble separation as
\begin{equation}
R_* = \frac{(8\pi)^{1/3}}{\beta}v_b\,\,,\label{HRappro}
\end{equation}
and $\beta$ can be derived as following
\begin{equation}
\beta = -\frac{d}{dt}\left(\frac{S_3(T)}{T}\right)\Bigg|_{t = t_*} = H(T)T\frac{d}{dT}\left(\frac{S_3}{T}\right)\Bigg|_{T = T_*}\,\,.
\end{equation}
For convenience, we use the normalized parameter $\tilde{\beta} = \beta/H$ in the following discussions.

However, in the case of a SFOPT with strong and ultra supercoling, $\beta$ becomes order unity or even minus. The above approximation breaks down and the mean bubble separation should be calculated from the strict definition.
Explicitly, the mean bubble separation is defined as
\begin{equation}
R_* = n_b^{-1/3}\,\,, \label{HR}
\end{equation}
where $n_b$ is the bubble number density that can be written as \cite{Turner:1992tz}
\begin{equation}
n_b = \int_{t_c}^{t}dt^\prime\frac{a(t^\prime)^3}{a(t)^3}\Gamma(t^\prime)P(t^\prime) \,\,.
\end{equation}
Then we can get the number density distribution \cite{Turner:1992tz}
\begin{equation}
\frac{dn_b}{dR}(t, t') = \Gamma(t')\left(\frac{a(t')}{a(t)}\right)^4 \frac{P(t')}{v_b(t')}\,\,.
\end{equation}

According to the above quantities, there are other definitions for the characteristic length scale in literatures. Ref.~\cite{Kobakhidze:2017mru} proposes that the physical distance $R_{\rm major} = R(t_p, t_r)$, which is the distance of the bubbles expansion from the time $t_r$ to the percolation time, can be a good characteristic length scale.
Here, $t_r$ is defined as the time that maximizes the number distribution at percolation time, and derived by
\begin{equation}
\frac{d}{dt'}\frac{dn_b}{dR}(t_p,t')\Bigg|_{t'=t_r} = 0\,\,.
\end{equation}
Refs.~\cite{Ellis:2018mja,Huber:2007vva} argue that the dominant contribution to the GW comes from the bubbles with largest fraction of the released phase transition energy.
Therefore, the relevant characteristic length scale should be the bubble size $R_{\max}$, which is the scale that maximize the energy distribution
\begin{equation}
\mathcal{E}_b = R^3\frac{dn_b}{dR}   \,\,.
\end{equation}
The peak of the GW spectrum correspond to the thickness of the fluid shell \cite{Hindmarsh:2016lnk}.
Hence, The characteristic length scale is defined as $\bar{R} \sim (v_b - c_s)R_{\max}$ \cite{Ellis:2018mja}, which is another good definition for the characteristic length scale.

In this work, we use the strict definition of the mean bubble separation as a characteristic length scale for the phase transition, and compare the results with the commonly used approximation for the mean bubble separation.
For convenience, we use the characteristic length scale normalized by the Hubble scale in the following discussion.

\subsection{Energy budget}
For the calculation of GW from a SFOPT, the efficiency parameter and phase transition strength parameter are essential.
The former describes the fraction of vacuum energy released by a SFOPT, which is transferred into the bulk kinetic energy of the fluid.
The later is the ratio of the vacuum to the radiation energy density.
To proceed, the analyses should start from the hydrodynamical treatment ~\cite{Steinhardt:1981ct,Laine:1993ey,Ignatius:1993qn,Kamionkowski:1993fg,KurkiSuonio:1995pp,Espinosa:2010hh} of a system with thermal plasma and scalar field.
For consistence, we give a brief review of the hydrodynamics based on refs.~\cite{Kamionkowski:1993fg,Espinosa:2010hh}.
The energy-momentum tensor of the scalar field $\phi$ is
\begin{equation}
T_{\phi}^{\mu\nu}=\partial^{\mu}\phi\partial^{\nu}\phi - g^{\mu\nu}\left[\frac{1}{2}\partial_{\rho}\phi\partial^{\rho}\phi - V_{\rm eff}(\phi)\right] \,\,,
\end{equation}
where $V_{\rm eff}$ is the effective potential.
For the thermal plasma, which is assumed in local equilibrium and can be treated as a perfect fluid, the energy-momentum tensor is given by
\begin{equation}
T_{f}^{\mu\nu}= wU^{\mu}U^{\nu} - g^{\mu\nu}p \,\,,
\end{equation}
where $p$ is the pressure, $w=T(\partial p/\partial T)$ is the enthalpy density ($\omega = e + p$, $e$ is the energy density), and $U^{\mu}$ is the four-velocity of the fluid
\begin{equation}
U^{\mu} = \frac{(1,\vec{v})}{\sqrt{1 - |\vec{v}|^2}} = (\gamma,\gamma\vec{v})   \,\,.
\end{equation}
From the energy-momentum conservation at the bubble wall $\partial_{\mu}T^{\mu\nu} = \partial_{\mu}T_{\phi}^{\mu\nu} + \partial_{\mu}T_{f}^{\mu\nu}$, in the wall frame, we get the following matching conditions across the bubble wall (``+" and ``-" represents the quantities in the symmetric and broken phase, respectively)
\begin{equation}
w_+\gamma_+^2v_+^2 + p_+ = w_-\gamma_-^2v_-^2 + p_-, \quad w_+\gamma_+^2v_+ = w_-\gamma_-^2v_- \label{match} \,\,.
\end{equation}

For further discussions, we need to know the equation of state (EoS) of the plasma.
We usually use the conventional EoS of the bag model.
In the high-temperature symmetric phase
\begin{equation}
p_+ = \frac{1}{3}a_+T_+^4 - \epsilon \,\,, \quad e_+ = a_+T_+^4 + \epsilon \,\,,
\end{equation}
while in the low-temperature broken phase
\begin{equation}
p_- = \frac{1}{3}a_-T_-^4 \,\,, \quad e_- = a_-T_-^4 \,\,,
\end{equation}
where $a_{\pm} = \pi^2 g_{\rm eff}/30$,  $g_{\rm eff}$
%\footnote{We set $g_{\rm eff} = 100$ in following calculation}
is the number of light degree of freedom in the plasma in the symmetric and broken phase, and $\epsilon$ is the bag constant describing the difference in energy density and pressure across the bubble wall.

Based on the EoS of the bag model, the conventional definition of phase transition strength parameter is,
\begin{equation}
\alpha = \frac{\epsilon}{a_+T_+^4} \,\,.
\end{equation}
Then from eq.~\eqref{match}, one can get the relation
\begin{equation}
v_+ = \frac{1}{1+\alpha}\left[\left(\frac{v_-}{2} + \frac{1}{6v_-}\right) \pm \sqrt{\left(\frac{v_-}{2} + \frac{1}{6v_-}\right)^2 + \alpha^2 + \frac{2}{3}\alpha - \frac{1}{3}}\right] \label{v+} \,\,.
\end{equation}
This definition of $\alpha$ is widely used in most of literatures.
From a given particle physics model, the effective potential can be directly derived.
Since the free energy of scalar field is equivalent to the effective potential $V_{\rm eff}$, we can obtain  the energy density $e = V_{\rm eff} - TdV_{\rm eff}/dT$  and the pressure $p = -V_{\rm eff}$.
Then $\epsilon$ corresponds to the energy density difference between the symmetric phase and broken phase.
Therefore, the strength parameter can be written as
\begin{equation}\label{convenalpha}
\alpha = \frac{\Delta V_{\rm eff} - T\frac{\partial\Delta V_{\rm eff}}{\partial T}}{\rho_R} \,\,,
\end{equation}
where $\rho_R = a_+T_+^4 = \pi^2g_{\rm eff}T_+^4/30$.
In the calculation of GWs, $T_+$ should be the temperature of the plasma surrounding the vacuum bubble where GWs have been produced.
We emphasize that it is important to find the correct $T_+$ in the following calculations.

%Note that when particles get masses comparable to \tc{$T$,  it can cause deviation from the EoS of the bag model}.
%In many particle physics models, the free-energy is dominated by a large number of degrees of freedom.
%Hence, this deviation is small.
For some cases, the deviation from the bag EoS is large, we can still parameterize the plasma in terms of quantities that mimic the EoS of the bag model.
For the situation with large deviation from the bag EoS, we can define
\begin{equation}
a_{\pm} = \frac{3}{4T_{\pm}^3}\frac{\partial p}{\partial T}\Bigg|_{\pm}=\frac{3w_{\pm}}{4T_{\pm}^4} \,\,, \quad \epsilon_{\pm} = \frac{1}{4}(e_{\pm} - 3p_{\pm}) \,\,,
\end{equation}
then
\begin{equation}
p_{\pm} = \frac{1}{3}a_{\pm}T_{\pm}^4 - \epsilon_{\pm} \,\,, \quad e_{\pm} = a_{\pm}T_{\pm}^4 + \epsilon_{\pm} \,\,,
\end{equation}
where $a_{\pm}$ and $\epsilon_{\pm}$ (so-called trace anomaly) are temperature dependent quantities and should be interpreted carefully.
Now a more general definition (the alternative definition) for the strength parameter is 
\begin{equation}
\alpha^\prime = \frac{\epsilon_+ - \epsilon_-}{a_+T_+^4} = \frac{4}{3}\frac{\epsilon_+ - \epsilon_-}{w_+} \,\,.
\end{equation}
When we consider a specific particle physics model, the strength parameter can be computed as
\begin{equation}
\alpha^\prime = \frac{\Delta V_{\rm eff} - \frac{T}{4}\frac{\partial\Delta V_{\rm eff}}{\partial T}}{\rho_R} \,\,.
\end{equation}
And for this case, eq.~\eqref{v+} still applies, with replacing $\alpha$ by $\alpha^\prime$.
For simplicity, we still use $\rho_R = \pi^2g_{\rm eff}T_+^4/30$ for this case in this work.
Therefore, two different definitions of strength parameter introduce different vacuum energy, and further should affect the evolution of Hubble rate according to eq.~\eqref{hubble}.
The related phase transition quantities derived from the two different definitions are denoted as unprimed and primed from here and after.

\begin{figure}[t]
	\centering
	
	\subfigure{
		\begin{minipage}[t]{1\linewidth}
			\centering
			\includegraphics[scale=0.8]{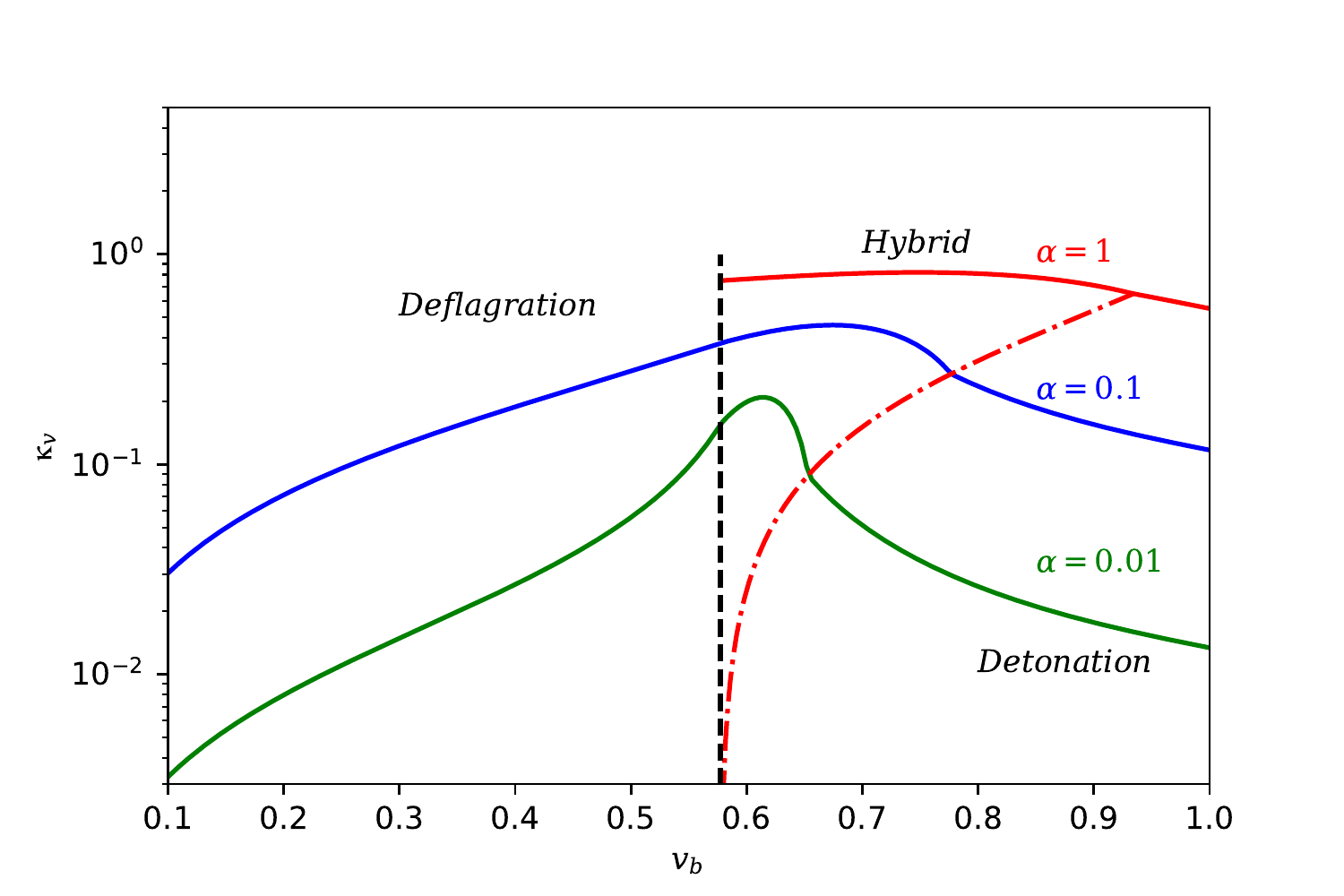}
		\end{minipage}%
	}%
	\centering
	\caption{The analytical fit of $\kappa_v$ \cite{Espinosa:2010hh} as a function of the bubble wall velocity $v_b$ and the phase transition strength parameter $\alpha$.
	The deflagration mode corresponds to the left region of dashed (black) line, the region between the dashed (black) line and dash-doted (red) line represents the hybrid mode, and the region below the dash-doted line denotes the detonation mode.
	}\label{kappa}
\end{figure}

To calculate the bulk kinetic energy stored in the plasma, one needs to know the velocity and enthalpy profile.
Then we should consider the energy-momentum conservation of the plasma that is away from the bubble wall, and this condition gives
\begin{equation}
\partial_{\mu}T^{\mu\nu} = U^{\nu}\partial_{\mu}(U^{\mu}w) + U^{\mu}w\partial_{\mu}U^{\nu} - \partial^{\nu}p = 0   \,\,.
\end{equation}
With the assumption of the spherical-symmetric bubble configuration, there is no natural length scales, and hence the above equations can be written in terms of $\xi = r/t$ where $r$ is the distance from the bubble center and $t$ is the time since nucleation.
In other words, the velocity and enthalpy profile are self-similar, being functions of $\xi$.
From the conservation equation, we obtain the equation of the velocity profile \cite{Kamionkowski:1993fg,Espinosa:2010hh}
\begin{equation}
\gamma^2(1-v\xi)\left[\left(\frac{\mu}{c_s}\right)^2 - 1\right]\frac{dv}{d\xi} = \frac{2v}{\xi}\label{velpf} \,\,,
\end{equation}
where $\mu(\xi,v) = (\xi-v)/(1-v\xi)$ and $c_s$ is the sound speed.
In the bag model $c_s = 1/\sqrt{3}$.
For more general EoS, $c_s = \sqrt{dp/de}$ depends on the temperature,
but it just get small deviation from $1/\sqrt{3}$.
Therefore, we choose $c_s= 1/\sqrt{3}$ for simplicity.
According to the velocity profile, the enthalpy profile can be derived from \cite{Kamionkowski:1993fg}
\begin{equation}
\frac{1}{w}\frac{dw}{dv} = \frac{4\gamma^2\mu}{3c_s^2} \,\,.
\end{equation}
The efficiency parameter, which is the ratio of the bulk kinetic energy to the vacuum energy, can then be calculated by
\begin{equation}
\kappa_v = \frac{3}{\epsilon v_b^3}\int w(\xi)v^2\gamma^2\xi^2d\xi \,\,.
\end{equation}
For the deviation of the bag model, we replace $\epsilon$ by $\epsilon_+ - \epsilon_-$.
The simulated and fitted results of efficiency parameter $\kappa_v$ are given as a function of $\alpha$ and $v_b$ in \cite{Espinosa:2010hh}.
In figure~\ref{kappa}, we show the analytical fit of $\kappa_v$ \cite{Espinosa:2010hh} as a function of the bubble wall velocity $v_b$ and the strength parameter $\alpha$.
The deflagration mode corresponds to the left region of the black dashed line, the region between the black dashed and red dash-doted line represents the hybrid mode, and the region below the dash-doted line denotes the detonation mode.
The definition of $\alpha$ is partly a matter of convention, and actually does not affect the fitted results for the efficiency parameter $\kappa_v$.
%However, recent study \cite{Cutting:2019zws} shows these results of the efficiency parameter for deflagration may not be valid for all parameter space of $v_b$ and $\alpha$, there may exist a large kinetic energy deficit in some parameter regions.
However, there may exist a large kinetic energy deficit in some parameter spaces, since ref.~\cite{Cutting:2019zws} shows the efficiency parameter for deflagration may not be valid for all parameter space of $\alpha$ and $v_b$.

\subsection{Bubble expansion mode}

\begin{figure}[t]
	\centering
	\subfigure{
		\begin{minipage}[t]{1\linewidth}
			\centering
			\includegraphics[scale=0.6]{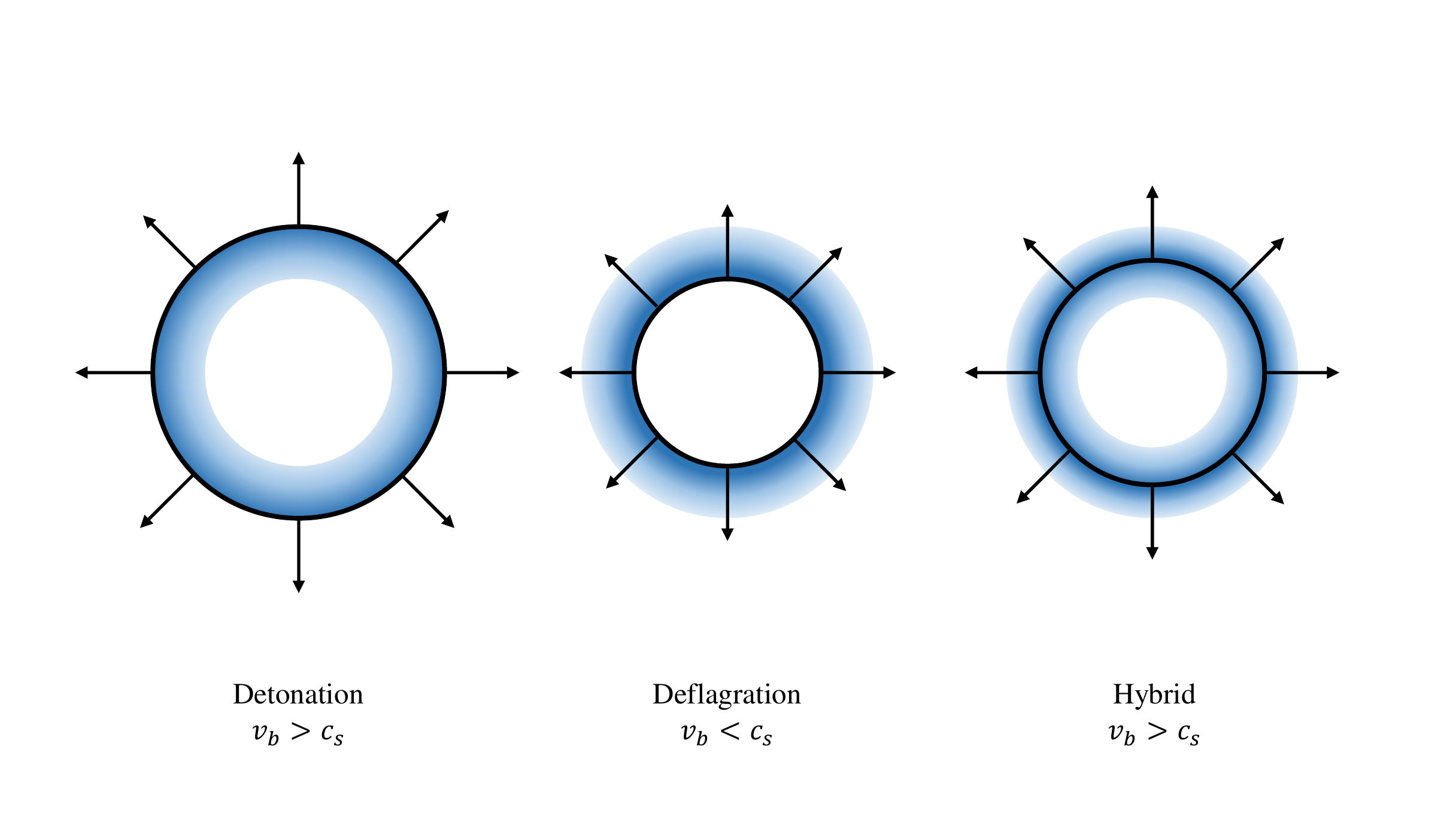}
			%\caption{fig2}
		\end{minipage}%
	}%
	\centering
	\caption{Schematic diagram for the different bubble dynamics. The left, middle, right diagram represents
		the dynamics of detonation, deflagration and hybrid, respectively. The black circle is the bubble wall. The color shaded region shows the
plasma with non-zero velocity in the universe frame.}\label{exmode}
\end{figure}

For a cosmological first-order phase transition, there exists six hydrodynamical modes \cite{KurkiSuonio:1995pp} with
different behaviors. Within them, there are three relatively stable hydrodynamical modes,
called detonation, deflagration, and hybrid (or supersonic deflagration).
We only focus on the detonation and deflagration cases in this work.
Here, we use $v_{\pm}$ and $\tilde{v}_{\pm}$ to denote the fluid velocity in the symmetric and broken phase for the bubble wall frame (the bubble wall is at rest) and the frame of Universe (the rest frame of the fluid far away from the bubble wall).
And they are related by the Lorentz transformation
\begin{equation}
\tilde{v}_{\pm} = \mu(v_b, v_{\pm})  \,\,.
\end{equation}

A schematic representation of detonation is depicted in the right panel of figure~\ref{exmode}.
For a detonation, the bubble wall moves at a supersonic speed $v_b > c_s$, and the fluid outside the wall is at rest in the Universe frame.
In the bubble wall frame, the fluid is moving into the wall with $v_+ = v_b$ and entering the broken phase, it slows down so that $v_- < v_+$ just behind the wall.
In the rest frame of the Universe, the fluid velocity just behind the bubble wall jumps to $\tilde{v}_- = \mu(v_b, v_-)$ and decreases gradually until it goes to zero smoothly at $\xi = c_s$. A rarefaction wave can be formed behind the wall.
The detonation solution is confined to a minimum value of $v_-$ and hence $v_b$, in the wall frame this indicate $v_-\ge c_s$.
Therefore, detonation can be divided into Chapman-Jouget detonation ($v_-=c_s$, $v_b = v_{CJ}$) and weak detonation ($v_->c_s$, $v_b > v_{CJ}$). The Chapman-Jouget speed $v_{CJ}$ is defined as \cite{Steinhardt:1981ct,Kamionkowski:1993fg}
\begin{equation}
v_{CJ} = \frac{\sqrt{\alpha(2+3\alpha)} + 1}{\sqrt{3}(1+\alpha)} \,\,,
\end{equation}
which is the mostly used in literature for the discussions of the bubble wall velocity.
However, the Chapman-Jouget detonation is problematic for the cosmological phase transition.

For the deflagration mode, the pictorial representation is depicted in the middle panel of figure~\ref{exmode}.
Compare with detonations,  the fluid of deflagration inside the bubble wall is at rest.
In the Universe frame, $\tilde{v}_- = 0$, and hence $v_-=v_b$.
The fluid velocity of wall frame that is larger behind the wall than in front, $v_->v_+$.
As one move out the bubble wall, the fluid velocity decreases, until a shock-front is encountered at $\xi_{\rm sh}$, and eventually would become zero outside of the shock-front.
In the frame of the shock-front, $\tilde{v}_+^{\rm sh}\tilde{v}_-^{\rm sh}=1/3$; the fluid is at rest outside the shock-front, hence $\tilde{v}_+^{\rm sh}=\xi_{\rm sh}$.
We can obtain the relation that $\xi_{\rm sh}\mu(\xi_{\rm sh}, \tilde{v}_+^{\rm sh})=1/3$ in the Universe frame.

%The fluid velocity just in front of the bubble wall

The hybrid (supersonic deflagration) could be treated as the superposition of detonation and deflagration.
The pictorial representation is depicted in the left panel of figure~\ref{exmode}.
For the hybrid solution, the bubble wall is followed by a rarefaction wave of Jouguet type and is proceed by a shock-front.
In the wall frame, the fluid velocity of the broken phase is $v_-=c_s$.
As the bubble wall velocity increases ($v_b\rightarrow v_{CJ}$), the compression wave in front of the bubble wall becomes thinner, and eventually disappears.
figure~\ref{kappa} shows the bubble velocity interval for the different expansion mode with different values of strength parameter.

We can obtain the following picture, the bubble expansion mode is in general deflagration for subsonic wall velocity ($v_b < c_s$).
As the wall velocity increases, a rarefaction wave is developed behind the bubble wall.
The shock becomes thinner until it completely vanishes and the bubble expansion proceeds by a Jouguet detonation with the bubble wall velocity is $v_b = v_{CJ}$.
Then the bubble expansion proceeds with the so-called detonation with $v_b > v_{CJ}$.

\begin{figure}[t]
	\centering
	\subfigure{
		\begin{minipage}[t]{1\linewidth}
			\centering
			\includegraphics[scale=0.5]{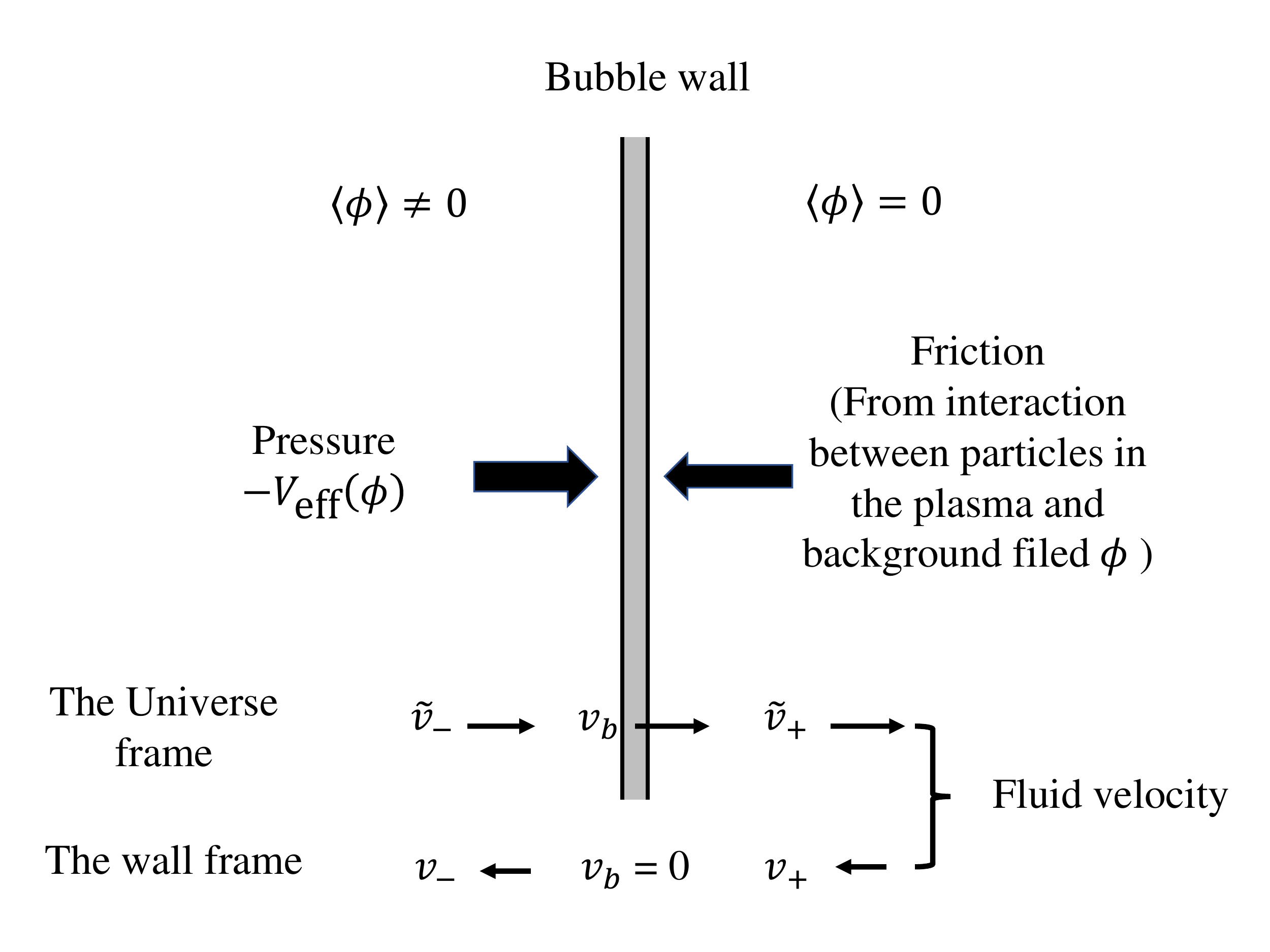}
			%\caption{fig2}
		\end{minipage}%
	}%
	\centering
	\caption{Schematic diagram for the forces acting on the wall and the fluid velocity in different reference of frame. In the bubble wall frame, the bubble velocity is zero and the fluid moves into the bubble center with velocity $v_{\pm}$. The bubble velocity is non-zero in the Universe frame and this velocity is generally used in the calculation of GW signals. The fluid moves outward the bubbles with velocity $\tilde{v}_{\pm}$ in the Universe frame.
    The vacuum energy released during the phase transition induces the driving force to put the bubble wall outward, while interactions with the particles in the plasma produce friction force.}\label{vb}
\end{figure}

Until now, the bubble wall velocity is given as a free parameter.
Successful electroweak baryogenesis favors small bubble wall velocity (more precisely, the relative velocity between the bubble wall and the front plasma) for sufficient diffusion process.\footnote{There are some exceptions that
large bubble wall velocity is possible for electroweak baryogenesis~\cite{No:2011fi,Cline:2020jre}.}
For phase transition GW signals, larger bubble wall velocity usually produce stronger GW signals.
Thus, it is important to calculate the exact bubble wall velocity for a given phase transition model.
However, to obtain the bubble wall velocity in a concrete model is still a difficult work, since it is related to the hydrodynamics and the microscopic properties of the a plasma during the phase transition. It is also model dependent.
But previous studies \cite{Moore:1995si,Megevand:2009gh,Huber:2013kj,Konstandin:2014zta,Espinosa:2010hh,Dorsch:2018pat} give a general picture for the calculation of the bubble wall velocity, we give a brief comment on it in the following.

In figure~\ref{vb}, we show the schematic diagram for the forces acting on the wall and the fluid velocity in different reference of frame. In the bubble wall frame, the bubble velocity is zero and the fluid moves into the bubble center with velocity $v_{\pm}$. The bubble velocity is non-zero in the Universe frame and this velocity is generally used in the calculation of GWs and baryogenesis. The fluid moves outward the bubbles with velocity $\tilde{v}_{\pm}$ in the Universe frame.
There is a driving force generated by the released vacuum energy during the phase transition.
With the expansion of the bubble, the interaction between the walls and the surrounding particles of the plasma create friction to prevent the growing of bubbles.
Generally, the friction should be eventually equal to the driving force and the bubble wall get a terminal velocity.
The friction generated by the particle species that change mass
during the phase transition requires a knowledge not only
about the scalar sector of the theory (order parameter field),  but also about the particles that cause the friction.
Detonation and deflagration is the case with a steady terminal velocity.
In ref.~\cite{Dorsch:2018pat}, the bubble wall velocity in the dimension-six effective model and the quartic toy model is
calculated from the first principle under some ideal approximations.
For extremely SFOPT, there is a situation which the friction can never catch up the driving force, and hence the bubble wall keep accelerating without limitation $v_b\rightarrow1$.
This expansion mode is so-call run-away.
However, this is based on the leading-order calculation of the friction.
The friction is independent of the $\gamma$ \cite{Bodeker:2009qy}~\footnote{When $v_b \rightarrow 1$, $\gamma \rightarrow \infty$} and hence is bounded.
When the higher-order contributions to the friction is considered, the friction have been proved to proportional to $\gamma$ \cite{Bodeker:2017cim}.
Therefore the accelerating should be stopped eventually and the bubble wall gets a steady velocity.
For simplicity, in this work the run-away situation and hybrid are beyond our consideration.
We only focus on the detonation and deflagration mode.

\section{Gravitational wave signals}
\label{GW}

\begin{figure}[t]
	\centering
	\subfigure{
		\begin{minipage}[t]{1\linewidth}
			\centering
			\includegraphics[scale=0.6]{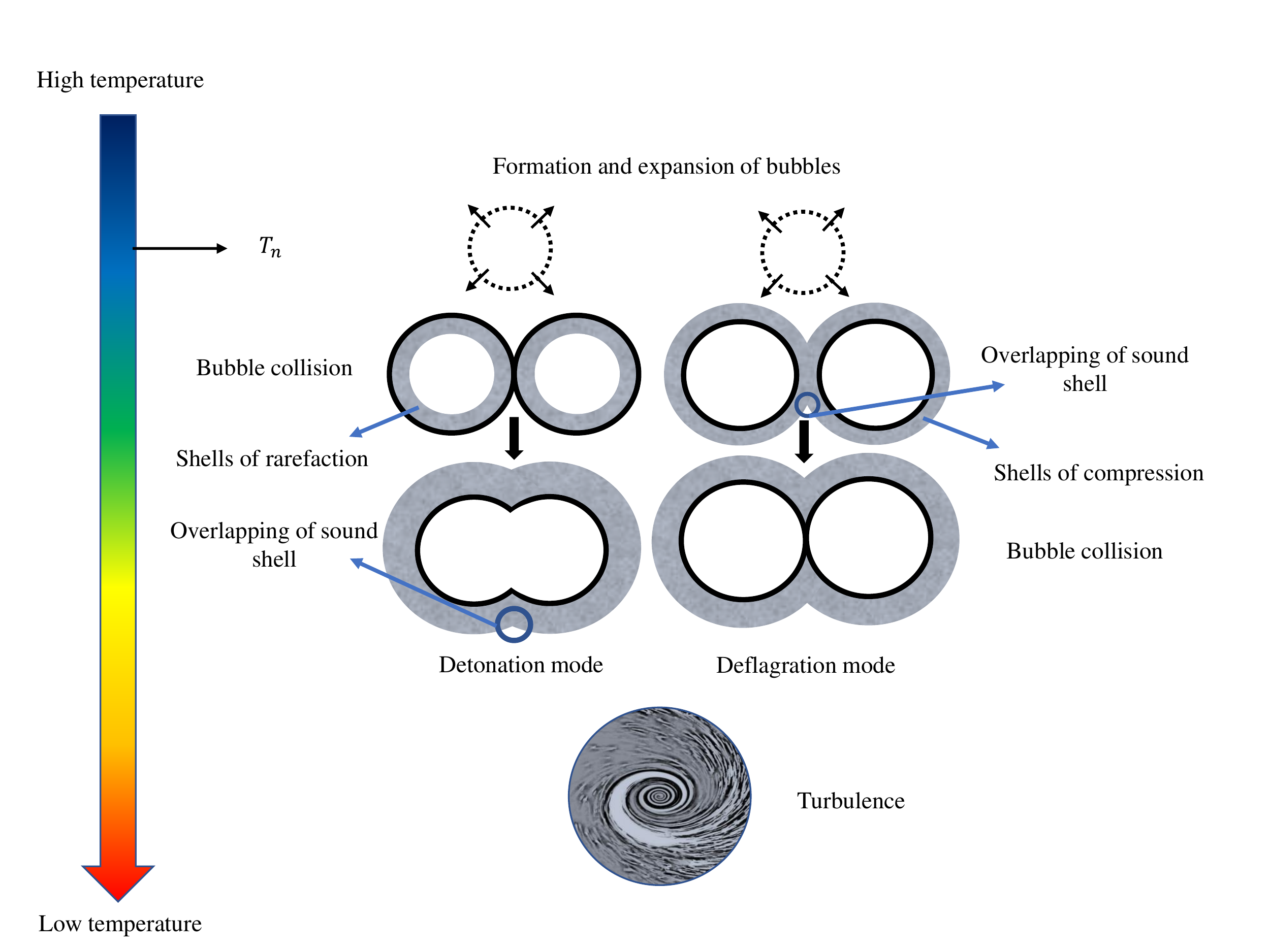}
			%\caption{fig2}
		\end{minipage}%
	}%
	\centering
	\caption{Schematic diagram for the three GW production mechanisms.
		For different bubble expansion mode, the occurrence of bubble collision and sound wave has different order.
		For detonation, bubble collision start first, and the overlapping of sound shell happens after the collisions.
		However, the overlapping of sound shell is earlier than the collision for the deflagration mode.
		With further evolution, the lower panel shows the turbulence occurs during the propagation of the shells.}\label{GWprop}
\end{figure}

From cosmological perturbation theory, the two degrees of freedom of the tensor mode of the metric perturbations $h_{ij}$ are the only radiative modes that correspond to the GWs.
Hence GWs can be represented by a tensor perturbation $h_{ij}$ of the Fredmann Robertson-Walker metric as
\begin{equation}
ds^2 = -dt^2 + a^2(t)(\delta_{ij} + h_{ij})dx^idx^j \,\,,
\end{equation}
where tensor perturbation $h_{ij}$ is transverse and traceless
\begin{equation}
\partial_ih_{ij} = h_{ii} = 0  \,\,.
\end{equation}
%\tc{Only Eqs.45 and 46 is  not enough if you want to discuss this point.}
In Fourier space, the equation of motion of the perturbation derived from Einstein field equation is
\begin{equation}
\ddot{h}_{ij}(\mathbf{k}, t) + 3H\dot{h}_{ij}(\mathbf{k}, t) + \frac{k^2}{a^2}h_{ij}(\mathbf{k}, t) = 16\pi G \Pi_{ij}^{TT}(\mathbf{k}, t) \,\,,
\end{equation}
where $G$ is the Newton constant, $t$ denotes the physical time, a dot represents the derivative with respect to t, and $\Pi_{ij}^{TT}$ is the transverse-traceless part of the shear stress $\Pi_{ij}$.
The shear stress is given by $a^2\Pi_{ij} = T_{ij} - a^2p(\delta_{ij} + h_{ij})$, where $T_{ij}$ is the spatial components of the energy-momentum tensor.

Once GWs have been generated in the early Universe, they propagate freely and are redshifted by the expansion of the universe.
Hence, the GW energy density today is
\begin{equation}
\rho_{GW} = \frac{\langle \dot{h}_{ij} \dot{h}_{ij} \rangle}{32\pi G} = \int\frac{df}{f}\frac{d\rho_{GW}}{d\log f} \,\,,
\end{equation}
where $\langle \rangle$ denotes ensemble average and $f$ is the observed GWs frequency at current time.
The main properties of GWs are described by the power spectrum, which is defined as the spectrum of energy density per logarithmic frequency interval
\begin{equation}
h^2\Omega_{GW}(f) = \frac{h^2}{\rho_c}\frac{d\rho_{GW}}{d\log f} \,\,,
\end{equation}
where $\rho_c = 3H_0^2/(8\pi G)$ is the critical energy density today. Here, $H_0 = 100h \rm km/s/Mpc$ is the Hubble rate today, and $h = 0.72$.
Actually, we do not consider the current Hubble tension between $h=74.03\pm 1.42$ from Ia Supernovae  (SH0ES)~\cite{Riess:2019cxk}  and $h=67.4\pm 0.5$ from CMB (planck)~\cite{Aghanim:2018eyx}.

Based on the recent study on the GW signals from the cosmological SFOPT, there are three mechanism that can generate tensor perturbation, the bubble collision, sound wave, and turbulence.
In figure~\ref{GWprop}, we give the schematic diagram for the three GW production mechanisms.
\begin{enumerate}
	\item \emph{Bubble collision}.
As shown in the figure~\ref{GWprop}, at nucleation temperature bubbles have been produced then expand until they collide to each other.
The GW production depend on the shape of the uncollided bubble wall rather than the evolution of the scalar field in the region of the collied bubbles.
This is the key idea of the ``envelope approximation'' \cite{Kosowsky:1991ua,Kosowsky:1992rz,Kosowsky:1992vn}.
Since the time duration is short and the efficiency of converting vacuum energy into the scaler field gradient energy is low, the contribution to the production of GW from this source is sub-dominant unless the bubble wall runs away or a vacuum transition.
	\item \emph{Sound wave}.
	Based on the ``sound shell'' model \cite{Hindmarsh:2016lnk,Hindmarsh:2019phv}, the expansion of bubbles generate compression and rarefaction waves.
	They are the shells of compression and rarefaction (sound pulses) around the bubbles and continue to propagate after the wall that drive them has disappeared.
	For detonation expansion mode, after the collision of bubbles these shells (shells of rarefaction) continue to propagate outward into the false vacuum.
	As the shell of fluid from different bubbles overlap, GW can be produced.
	For deflagration expansion mode, the overlapping of the shells (shells of compression) might start before the collision of bubble walls \cite{Weir:2017wfa}.
	Since the sound wave persists long after the bubble collision, the difference of the production of GW introduced by different expansion mode is negligible.
	\item \emph{Turbulence}. With the continued propagating of the fluid, a turbulent stage may also be triggered, then GW can be generated by turbulence mechanism. There are three major factors to determine the contribution of turbulence mechanism to the total GW signals. The first one is the fraction of kinetic energy converted into vortical motion.
    The second one is the relevant time scale.
    The last one is the different descriptions on the turbulence decay.
    Different assumptions on the turbulence lead to different predictions on the GW spectra.
    More dedicated work is needed to unravel the spectral shape and amplitude of the GW signals from turbulence.
\end{enumerate}
The three sources generally coexist, although peaking at different length and on different time scale, the contribution to the total GW can be derived by combining them together
\begin{equation}
	\Omega_{\rm GW} = \Omega_{\rm co} + \Omega_{\rm sw} + \Omega_{\rm turb} \,\,.
\end{equation}
As mentioned in last section, we do not consider the run-away case in this work, so the total contribution to the GW spectrum mainly comes from the sound wave and turbulence.

\subsection{Bubble collision}
For the collision of bubble walls, the numerical simulation based on ``envelope approximation'', gives the formula of the GW spectrum as \cite{Huber:2008hg, Caprini:2015zlo}
\begin{equation}
h^2\Omega_{\rm co}(f) \simeq 1.67\times10^{-5}(\frac{H_*R_*}{(8\pi)^{1/3}})^2\left(\frac{\kappa_{\phi}\alpha}{1 + \alpha}\right)^2\left(\frac{100}{g_{\star}}\right)^{1/3}\frac{0.11v_b}{0.42 + v_b^2}\frac{3.8(f/f_{\rm co})^{2.8}}{1 + 2.8(f/f_{\rm co})^{3.8}} \,\,,
\end{equation}
where $g_*$ is the total number of degrees of freedom and $H_*$ is the Hubble rate at the temperature $T_*$ when GWs are produced.
The peak frequency is
\begin{equation}
f_{\rm co} \simeq 1.65\times10^{-5}\text{Hz}\frac{(8\pi)^{1/3}}{H_*R_*}\left(\frac{0.62v_b}{1.8 - 0.1v_b + v_b^2}\right)\left(\frac{T_*}{100\rm GeV}\right)\left(\frac{g_*}{100}\right)^{1/6} \,\,.
\end{equation}
The coefficient $\kappa_{\phi}$ denotes the fraction of vacuum energy converted into the scaler field gradient energy.
For thermal phase transition, we can expect $\kappa_{\phi}$ to be extremely small, hence is negligible.
On the other hand, for vacuum transition and runaway bubble expansion mode, all of the vacuum energy transfers into the bubble wall.
Then $\kappa_{\phi}$ must be of order one, the GWs are all sourced by the collision of bubble walls in principal.

\subsection{Sound wave}
Form recent numerical simulations \cite{Hindmarsh:2013xza,Hindmarsh:2015qta,Hindmarsh:2017gnf}, sound wave is a more significant and long-lasting source of the GW, which are produced by overlapping of the expanding sound shells in the fluid.
Based on these simulations, there are models \cite{Hindmarsh:2016lnk,Hindmarsh:2019phv,Konstandin:2017sat} built to explain the mechanism of the generation of GW from sound wave.
The more successful model is the sound shell model, which agrees with the numerical simulation result well for detonation mode, but less well for deflagration mode.
Other models \cite{Konstandin:2017sat} focuses on the dynamics of the expanding compression waves in real space, fail to account for the shape of the power spectrum around the peak.
And they predict there are small signals at low frequencies which is not included in the sound shell model.

Here, we give the simulated GW spectrum of the sound wave:
\begin{equation}
h^2\Omega_{\rm sw}(f) \simeq 1.64\times10^{-6}(H_*\tau_{\rm sw})(H_*R_*)\left(\frac{\kappa_v\alpha}{1 + \alpha}\right)^2\left(\frac{100}{g_*}\right)^{1/3}(f/f_{\rm sw})^3\left(\frac{7}{4 + 3(f/f_{\rm sw})^2}\right)^{7/2},\label{swf}
\end{equation}
with the peak frequency
\begin{equation}
f_{\rm sw} \simeq 2.6 \times10^{-5}\text{Hz}\frac{1}{H_*R_*}\left(\frac{T_*}{100 \rm GeV}\right)\left(\frac{g_*}{100}\right)^{1/6} \,\,,
\end{equation}
and $\tau_{\rm sw}$ is the duration of the sound wave source,
\begin{equation}
\tau_{\rm sw} = \min\left[\frac{1}{H_*}, \frac{R_*}{\overline{U}_f}\right]   \,\,.
\end{equation}
The root-mean-square (RMS) fluid velocity can be approximated as \cite{Hindmarsh:2017gnf, Caprini:2019egz, Ellis:2019oqb}
\begin{equation}
\overline{U}_f^2 \approx \frac{3}{4}\frac{\kappa_v\alpha}{1+\alpha} \,\,.
\end{equation}
The term $H_*\tau_{\rm sw}$ accounts for the GW amplitude for sound wave suppressed by a factor of $H_*R_*/\overline{U}_f$, if the sound wave source can not last more than a Hubble time.
The efficiency parameter $\kappa_v$ is the fraction of vacuum energy transfered into the fluid bulk kinetic energy.
We show its analytical fit in figure~\ref{kappa}, and the analytic formulae of efficiency parameter can be find in ref.~\cite{Espinosa:2010hh}.
Note, the simulated formulae is based on the strength of the phase transition with $\alpha < 1$.
However, for a SFOPT with ultra supercooling $\alpha > 1$, its extrapolation need further investigation.
We use the formulae for both $\alpha > 1$ and $\alpha < 1$ in this work before the valid simulation results for $\alpha > 1$ are obtained.

\subsection{Turbulence}
For the turbulence mechanism,  the first numerical simulations of the GW power spectrum by magnetohydrodynamic turbulence was obtained in ref.~\cite{Pol:2019yex}. However, due to the complex physical process,  the GW spectra from turbulence remain to be further investigated.
There are indications that turbulence are less efficient than sound wave on the production of GW from ref.~\cite{Pol:2019yex}.
The time scale of the shock formation \cite{Hindmarsh:2017gnf}, which is related to the onset of turbulence, is given by
\begin{equation}
\tau_{\rm sh} \sim L_f/\overline{U}_f   \,\,,
\end{equation}
where $L_f$ is the characteristic length scale of the fluid flows, and we can approximate the scale as the mean bubble separation $R_*$.
Hence, if $H_*R_*/\overline{U}_f < 1$, the shocks can develop during one Hubble time, and we can take the contribution of turbulence into consideration.
Here, we use the analytical results from the modelling of Kolmogorov-type turbulence \cite{Caprini:2009yp,Caprini:2015zlo}
\begin{equation}
h^2\Omega_{\rm turb}(f) \simeq 1.14\times10^{-4}H_*R_*\left(\frac{\kappa_{\rm turb}\alpha}{1 + \alpha}\right)^{3/2}\left(\frac{100}{g_*}\right)^{1/3}\frac{(f/f_{\rm turb})^3}{(1 + f/f_{\rm turb})^{11/3}(1 + 8\pi f/H_*)} \,\,,
\end{equation}
where $\kappa_{\rm turb} = \tilde{\epsilon}\kappa_v$ is the efficiency of conversion of vacuum energy into turbulent flow.
Here, $\tilde{\epsilon}$ represents the fraction of bulk motion which is turbulent.
According to recent simulations \cite{Hindmarsh:2015qta}, at most $5-10\%$ of bulk motion is converted into vorticity.
We set $\tilde{\epsilon}=0.1$ in the following calculation.
The peak frequency of turbulence $f_{\rm turb}$ is
\begin{equation}
f_{\rm turb} \simeq 7.91\times10^{-5}\text{Hz}\frac{1}{H_*R_*}\left(\frac{T_*}{100 \rm GeV}\right)\left(\frac{g_*}{100}\right)^{1/6},
\end{equation}
and $H_*$ is the Hubble rate at $T_*$
\begin{equation}
H_* = 1.65 \times 10^{-5}\text{Hz}\left(\frac{T_*}{100 \rm GeV}\right)\left(\frac{g_*}{100}\right)^{1/6} \,\,.
\end{equation}
The above formulae of the GW spectrum produced by turbulence is controversial \cite{Kosowsky:2001xp,Gogoberidze:2007an,Niksa:2018ofa,Caprini:2019egz} and further numerical results are needed.
In the following section, we use $HR_*$ as the abbreviation of $H_*R_*$.

\subsection{Signal-to-noise ratio}
To quantify the detectability of the GW signals from a given model with SFOPT, it is necessary to calculate the signal-to-noise ratio (SNR) as the following
\begin{equation}\label{snr}
\text{SNR}=\sqrt{\mathcal{T} \int^{f_{\rm max}}_{f_{\rm min}} df \left( \frac{h^2 \Omega_{GW}}{h^2  \Omega_{\rm sens}}\right)^2}\,\,. \nonumber
\end{equation}
$\mathcal{T}$ is the total duration time of the experiment mission.
We choose the minimal data-taking time of LISA, namely, $\mathcal{T}\simeq 9.46\times10^7s$
based on refs.~\cite{Caprini:2019egz,LISA:documents}. It corresponds to
four-years duration time of the mission and a duty cycle of $75\%$.
$h^2\Omega_{\rm sens}$ corresponds to the expected sensitivity~\cite{Schmitz:2020syl} of a given experiment configuration.
According to ref.~\cite{Caprini:2015zlo}, we choose a threshold value $\rm SNR_{thre} =10$, which is not easy to be quantified.
A signal can be claimed to be detectable only if SNR is larger than
a threshold value $\rm SNR_{thr}$ of a detector.
In the following discussions of the three typical models, we present the corresponding SNR for each model.

\section{Benchmark models}
\label{models}
Based on the previous detailed discussions and clarifications, we recalculated the phase transition dynamics
and the corresponding GW signals in three benchmark models, the dimension-six effective model, quartic toy
model and logarithm model.
For the dimension-six effective model, the parameter spaces are allowed by the current data as shown in our previous study \cite{Huang:2015izx,Huang:2016odd}. For the quartic toy model and logarithm model, to avoid the collider constraints, the Higgs field is not the preferred order-parameter scalar field.
However, in order to clearly show the qualitative  properties of the phase transition dynamics, we take the Higgs as an example in our numerical calculations.
These three models represent the three typical SFOPT types, which are classified in ref.~\cite{Chung:2012vg}.
By using the correct characteristic temperature and the correct GW formulae discussed above, we get more reliable
GW spectra, especially for the strong and ultra supercooling case.
The discussions on these benchmark models can also be used to other models.

\subsection{Dimension-six effective model}
We study the benchmark model in SM effective field theory with the Higgs sextic term, which can provide a new Higgs potential and a SFOPT
~\cite{Zhang:1992fs,Grojean:2004xa,Huang:2015izx,Huang:2016odd,Cao:2017oez,Croon:2018erz,Croon:2019rqu} as the following tree-level potential
\begin{equation}
V(\phi)=\frac{\mu^2}{2}\phi^2 + \frac{\lambda}{4}\phi^4 + \frac{\kappa}{8\Lambda^2}\phi^6 \,\,.
\end{equation}
From the perspective of SM effective
field theory, this effective scenario can represent general properties of many models, like
singlet, doublet, triplet extended Higgs model and composite Higgs model as discussed in our previous work~\cite{Cao:2017oez}.
The Higgs sextic operator can be obtained by integrating out the heavy degree of freedom when
the heavy particle masses are much larger than the cutoff scale of the effective theory.
Besides the Higgs sextic operator, other dimension-six operators can be obtained simultaneously
and contribute to the electroweak precise observables when we match the renormalizable model to
the SM effectve operators. In our previous study, we find this dimension-six effective model still works well
to realize a SFOPT and satisfy the current elecroweak precise measurements after considering the constraints from all
the possible dimension-six effective operators.
Since our study in this work focuses on the clarification of the different characteristic temperatures,
we only keep the Higgs sextic term among all the dimension-six effective operators and the leading thermal loop corrections.
Other dimension-six operators and loop corrections at zero temperature could give some minor
modifications to our results, but not change the qualitative properties.
Based on the above assumptions, we begin our discussions with the following finite-temperature
effective potential
\begin{equation}
V_{\rm eff}(\phi,T)\approx\frac{\mu^2 + cT^2}{2}\phi^2 + \frac{\lambda}{4}\phi^4 + \frac{\kappa}{8\Lambda^2}\phi^6  \,\,,
\end{equation}
where $\Lambda/\sqrt{\kappa}$ is the effective cutoff scale and $c$ is the thermal correction
\begin{equation}\label{cdim6}
c=\frac{1}{16}(g^{\prime 2}+3 g^2+4 y_t^2+4 \frac{m_h^2}{v^2}-12 \frac{\kappa v^2}{\Lambda^2}) \,\,.
\end{equation}
Since the effective cutoff $\Lambda/\sqrt{\kappa}$ depends on the ratio of $\Lambda$  and $\sqrt{\kappa}$, we can keep the effective field theory valid up to our interested energy scales by rescaling $\Lambda$ and $\kappa$ simultaneously.
For larger $\kappa$, we have larger energy scale.
$g^{\prime}$ and $g$ are the $U(1)$ and $SU(2)$ gauge coupling, respectively.
$y_t$ is the top quark Yukawa coupling. $m_h=125$ GeV is the Higgs mass and  vacuum expectation value (VEV) $v=246$~GeV.
The model parameters can be expressed in terms of the SM Lagrangian parameters
\begin{equation}
\lambda = \lambda_{SM}\left(1 - \frac{\Lambda_{\rm max}^2}{\Lambda^2}\right)\,\,,
\end{equation}
\begin{equation}
\mu^2 = \mu_{SM}^2\left(-1 + \frac{\Lambda_{\rm max}^2}{2\Lambda^2}\right)\,\,,
\end{equation}
where $\Lambda_{\rm max}=\sqrt{3\kappa}v^2/m_h$.
In this model, we can derive its critical temperature as
\begin{equation}
T_c = \frac{\sqrt{\lambda^2\Lambda^2 - 4\kappa\mu^2}}{2\sqrt{c\kappa}}\,\,,
\end{equation}
and the washout parameter
\begin{equation}
\frac{\phi_c}{T_c} = \frac{2\Lambda\sqrt{-c\lambda}}{\sqrt{\lambda^2\Lambda^2 - 4\kappa\mu^2}}\,\,.
\end{equation}
The true minimum is
\begin{equation}
\phi_{\rm true} = \sqrt{\frac{-2\lambda\Lambda^2 + 2\Lambda\sqrt{\lambda^2\Lambda^2 - 3\kappa(\mu^2 + cT^2)}}{3\kappa}}\,\,.
\end{equation}

\begin{figure}[t]
	\centering
	
	\subfigure{
		\begin{minipage}[t]{1\linewidth}
			\centering
			\includegraphics[scale=0.8]{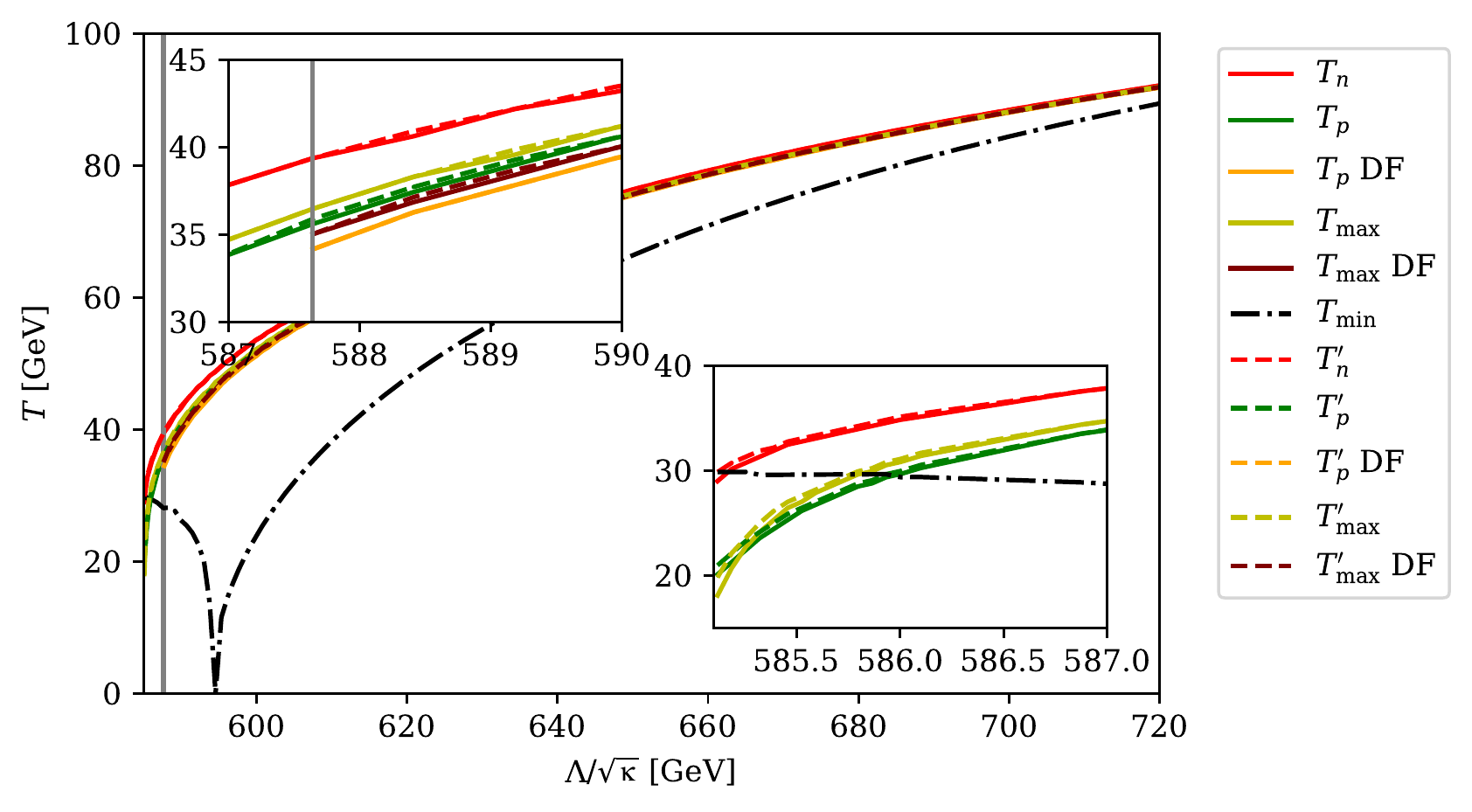}
		\end{minipage}%
	}%
	\centering
	\caption{
		The characteristic temperatures of the dimension-six effective model.
		The colored solid lines represent various temperatures for the conventional definition of the phase transition strength parameter.
		The colored dashed lines represent temperatures that are derived for the strength parameter defined by trace anomaly.
		The black dash-dotted line denotes the minimum temperature.
		The vertical gray line indicates the cutoff scale $\Lambda/\sqrt{\kappa}$ that can produce a SFOPT with $\alpha=\alpha^\prime=1$ at percolation temperature.
		The orange and maroon lines (both solid and dashed) show that different bubble wall velocities only have negligible modification to the percolation and the maximum temperature for the cutoff scale that generates a SFOPT $\alpha<1$ and $\alpha^\prime<1$ at percolation temperature.
		DF is the abbreviation of deflagration ($v_b = 0.3$).	
	}\label{ctdim6}
\end{figure}

\begin{figure}[t]
	\centering
	
	\subfigure{
		\begin{minipage}[t]{1\linewidth}
			\centering
			\includegraphics[scale=0.8]{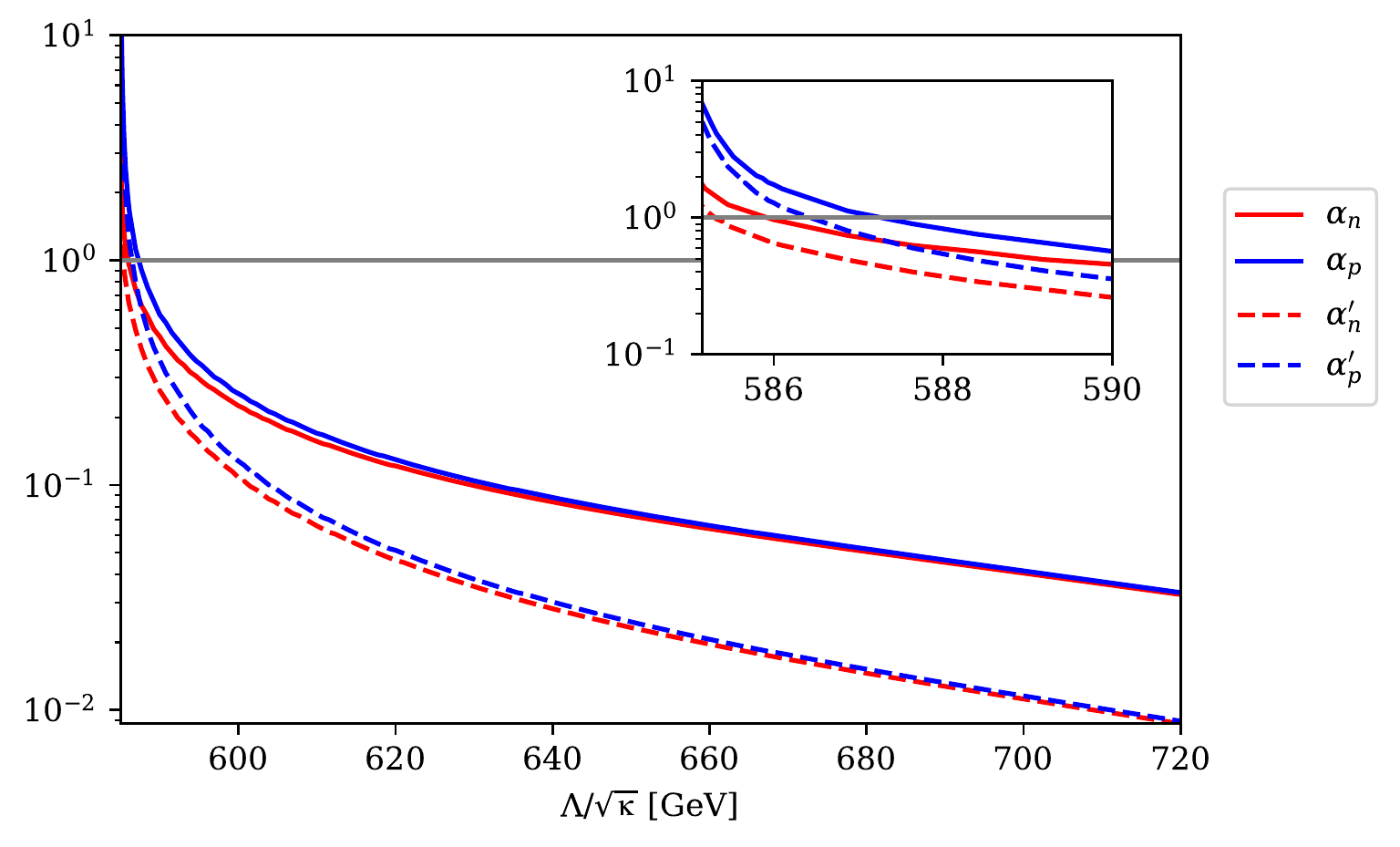}
		\end{minipage}%
	}%
	\centering
	\caption{The phase transition strength as a function of the effective cutoff scale $\Lambda/\sqrt{\kappa}$.  Solid lines (red and blue) represent the values with conventional definition at nucleation and percolation temperature. Dashed lines (red and blue) denote the values with alternative definition at nucleation and percolation temperature. The horizontal gray line indicates that the phase transition strength parameters are equal to one.}\label{stdim6}
\end{figure}

\begin{figure}[t]
	\centering
	
	\subfigure{
		\begin{minipage}[t]{1\linewidth}
			\centering
			\includegraphics[scale=0.8]{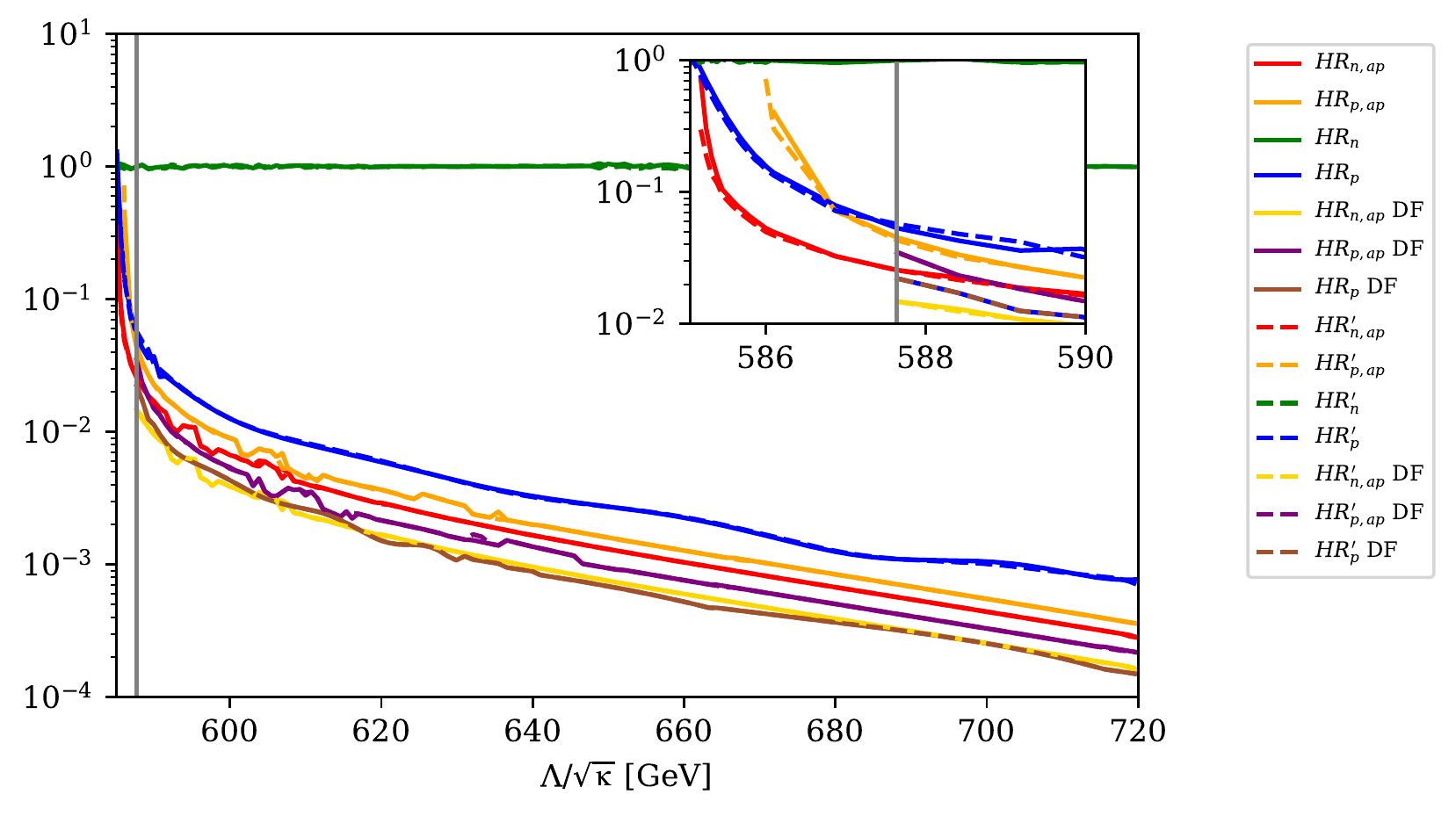}
		\end{minipage}%
	}%
	\centering
	\caption{The characteristic length scale of the dimension-six effective model. $HR_{*,ap}$ (red, orange, yellow, and purple solid lines), which is derived by the approximation, and $HR_*$ (green, blue, and brown solid lines), which is calculated by the first principle eq.~\eqref{HR}, indicate the characteristic length scale at the nucleation and percolation temperature.
	The colored dashed lines denote the values of characteristic length scale at different temperatures for different definitions of the phase transition strength.
	The vertical gray line indicates the cutoff scale $\Lambda/\sqrt{\kappa}$ that can produce a SFOPT with $\alpha_p=\alpha_p'=1$.
	}\label{hrdim6}
\end{figure}

Based on the detailed discussions and formulae therein,
we can perform analytical and numerical calculations.
We show the different characteristic temperatures
of the dimension-six effective model in figure~\ref{ctdim6}.
The colored solid lines represent various temperatures for the conventional definition of the phase transition strength parameter ($T_n$ for detonation, $T_p$ for detonation, $T_{p}$ for deflagration, $T_{\rm max}$ for detonation, and $T_{\rm max}$  for deflagration).
The colored dashed lines represent temperatures that are derived for the strength parameter defined by trace anomaly.
Different definitions give very slightly modification to these characteristic temperatures.
Basically, we can see the temperatures denoted by colored lines decrease with decreasing of the cutoff scale. Lower cutoff scale usually has lower nucleation temperature and
percolation temperature.
The black dash-dotted line shows the minimum temperature.
The minimum temperature first decreases with the cutoff scale, then increases with the decreasing cutoff scale after it drops to zero.
This behavior is induced by two different typical evolution of bounce action shown in figure~\ref{fg1}.
When the cutoff is not small enough, there no potential barrier exist at zero temperature, and hence the minimum temperature is the temperature at which the potential barrier disappear.
But with the decreasing of the cutoff scale, this temperature becomes zero at some specific parameter point.
After that point,  the barrier begins to show up at zero temperature.
And the bounce action become more flat with decreasing cutoff scale, hence the minimum of action appears at higher temperature.
Since the minimum temperature is only related to the evolution of the bounce action, it should not be affected by different definitions of strength parameter.
The vertical gray line indicates the cutoff scale $\Lambda/\sqrt{\kappa}$ which can  produce a ultra supercooling with $\alpha=\alpha^\prime=1$ at percolation temperature.\footnote{The corresponding cutoff scales that can generate a SFOPT with $\alpha = 1$ and $\alpha' = 1$ at percolation temperature are different. However, these cutoff scales are very close. And we take the same value here and after.}
DF is the abbreviation of deflagration ($v_b = 0.3$ as default benchmark value).
Different bubble wall velocities give negligible modifications to the percolation temperature and the maximum temperature (the green, orange, silver, and purple lines, both solid and dashed) for the cutoff scales which can generate a SFOPT with $\alpha<1$ and $\alpha^\prime<1$ at percolation temperature.
We can see that for a given definition of the phase transition strength, the hierarchy between the nucleation temperature $T_n$ and the percolation temperature $T_p$ becomes obvious for the small cutoff scale $\Lambda/\sqrt{\kappa}$.
This small cutoff scale region corresponds to the strong and ultra supercooling cases where the percolation temperature is obviously smaller than the nucleation temperature.
It is crucial to choose the correct characteristic temperature to calculate the GW spectra in the following. The percolation temperature is a more appropriate temperature.
For example, when $\Lambda/\sqrt{\kappa}=585.1$ GeV, the resulting percolation temperature $T_p\sim 20$ GeV,
which is much smaller than the nucleation temperature $T_n$ (about 30 GeV).
For the cutoff scale which gives $T_{\rm max} < T_p$, the completion of phase transition is questionable.
And the cutoff scale that is lower than 585.1 GeV is excluded, since they can not obtain a maximum temperature.
Hence the completion of phase transition is not guaranteed. 
After considering more precise loop corrections of the effective potential, the allowed cutoff scale can slightly
shift to lower value~\cite{Ellis:2018mja}.

After we get the characteristic temperature, it is straight forward to obtain the corresponding phase transition strength at different temperatures.
The obvious hierarchy between the nucleation temperature $T_n$ and percolation temperature $T_p$ makes $\alpha_p$ obviously larger than $\alpha_n$ as shown in figure~\ref{stdim6}.
This means we should use $T_p$ and $\alpha_p$ to obtain more reliable and stronger GW spectra when
the cutoff scale is lower, namely, for the strong supercooling and ultra supercooling cases.
In figure~\ref{stdim6}, we show the different definitions of phase transition strength as a function of the cutoff scale $\Lambda/\sqrt{\kappa}$.
Solid lines represent the values with conventional definition at the nucleation and percolation temperature.
Dashed lines denote the values with alternative definition at the nucleation and percolation temperature.
Lower cutoff scales usually give larger phase transition strength.
For the same cutoff value and the same temperature, the phase transition strength of
the conventional definition is obviously larger than the value of the alternative definition.
With the decreasing of the cutoff scale $\Lambda/\sqrt{\kappa}$, the type of the SFOPT changes from slight supercooling, to
mild supercooling, strong supercooling, and finally to ultra supercooling.
We can see that the lower the cutoff scale, the stronger the supercooling.
In the strong supercooling and ultra supercooling case, $T_n-T_p$ (or $T'_n-T'_p$) and the corresponding $\alpha_p-\alpha_n$ (or $\alpha'_p-\alpha'_n$) becomes distinctively large.
Strong supercooling and ultra supercooling are favored by the GW detectors since they can produce obviously stronger GW signals.
And they are also motivated by supercooling dark matter scenario.
For example, if the electroweak phase transition is ultra supercooling and
the percolation temperature might be lower than the QCD phase transition scale~\cite{Iso:2017uuu},
the QCD phase transition can become a SFOPT and form quark nugget as the dark matter candidate~\cite{Witten:1984rs}.

Figure~\ref{hrdim6} shows the characteristic length scale of the dimension-six effective model.
$HR_{*,ap}$ (red, orange, yellow, and purple solid lines), which is derived by the approximation eq.~\eqref{HRappro}, and $HR_*$ (green, blue, and brown solid lines), which is calculated by first principle, indicate the characteristic length at nucleation and percolation temperature.
The dashed lines denote the values of characteristic length at different temperatures for alternative definition of phase transition strength.
For different definitions, the characteristic length scale only gets a negligible modification.
The yellow, purple and brown lines (both solid and dashed) show that the deflagration expansion mode ($v_b = 0.3$) gives negligible modifications to the characteristic length scale for regime of a SFOPT with $\alpha_p < 1$ for both definition.
The vertical gray line indicates the cutoff scale $\Lambda/\sqrt{\kappa}$ that can produce a SFOPT with $\alpha_p=\alpha_p'=1$.
In this figure, we can see the commonly used approximation of the characteristic length scale (the yellow and purple lines, both solid and dashed) fit the strict calculation result (the brown lines, both solid and dashed) very well for the deflagration mode.
However, the approximated length scale (the red and orange lines, both solid and dashed) for the detonation mode fit the strict calculation (the blue lines, both solid and dashed) less well.\footnote{We give the approximated result for the cutoff scales which produce a SFOPT with $\beta > 1$ at the nucleation and percolation temperature.}
%\tc{This approximation are still applicable for the }
Here, the normalized characteristic length scale $HR_n$ and $HR_n'$ with the strict calculation by eq.~\eqref{HR} is around order 1.
And this is reasonable, since nucleation temperature is defined as one bubble is nucleated at one Hubble volume, the mean bubble separation should be one Hubble radius at nucleation temperature.
From this point of view, taking the nucleation temperature as the temperature at which the GWs are produced is not a good approximation.
Actually, the percolation temperature $T_p$ is a better approximation for the temperature at which GWs are produced.
Therefore, for the GW spectra derived from the strict calculations of $HR$ and $HR'$, it is better to use the values of phase transition parameters at the percolation temperature.

\begin{figure}[t]
	\centering
	
	\subfigure{
		\begin{minipage}[t]{0.5\linewidth}
			\centering
			\includegraphics[scale=0.5]{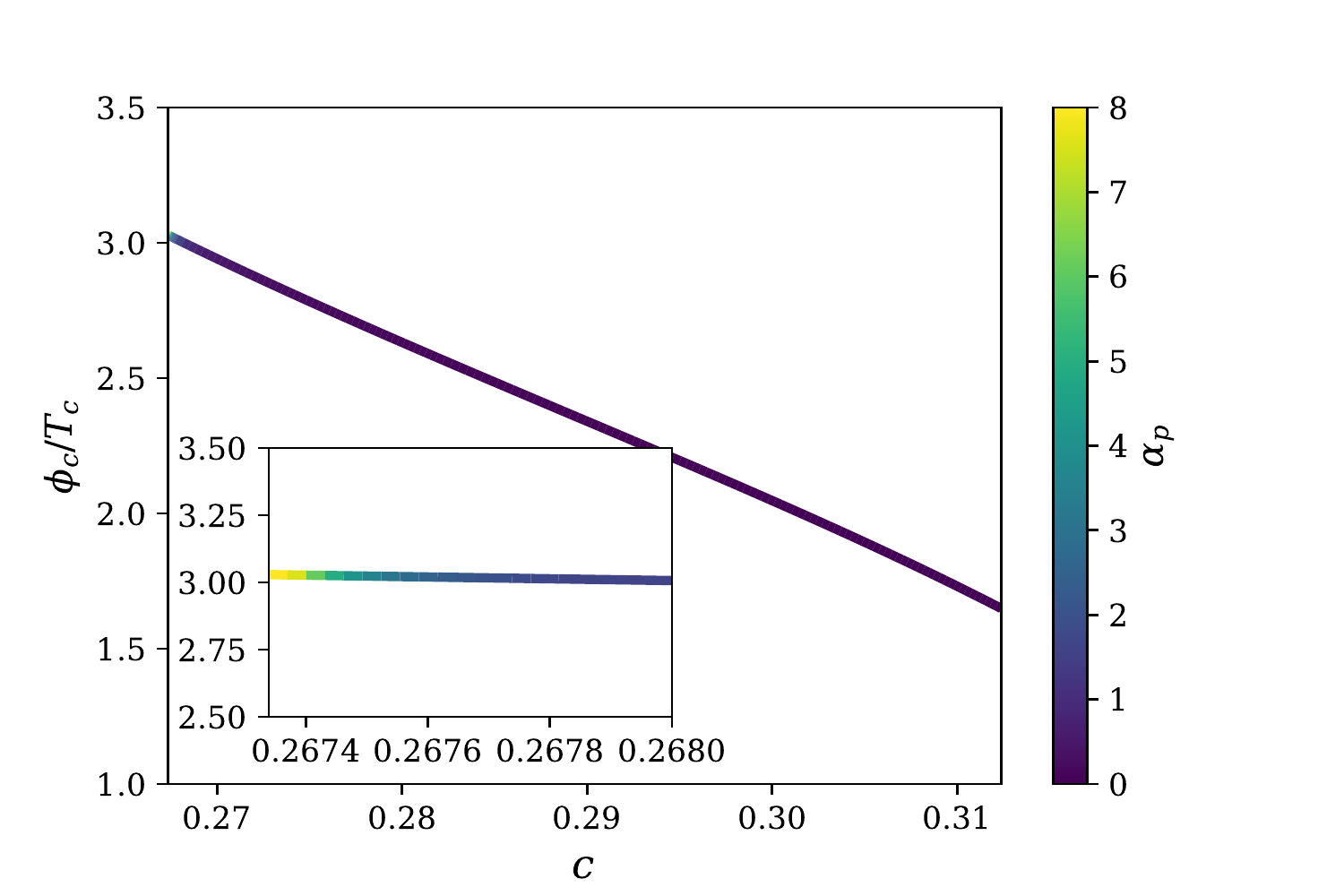}
		\end{minipage}%
	}%
	\subfigure{
		\begin{minipage}[t]{0.5\linewidth}
			\centering
			\includegraphics[scale=0.5]{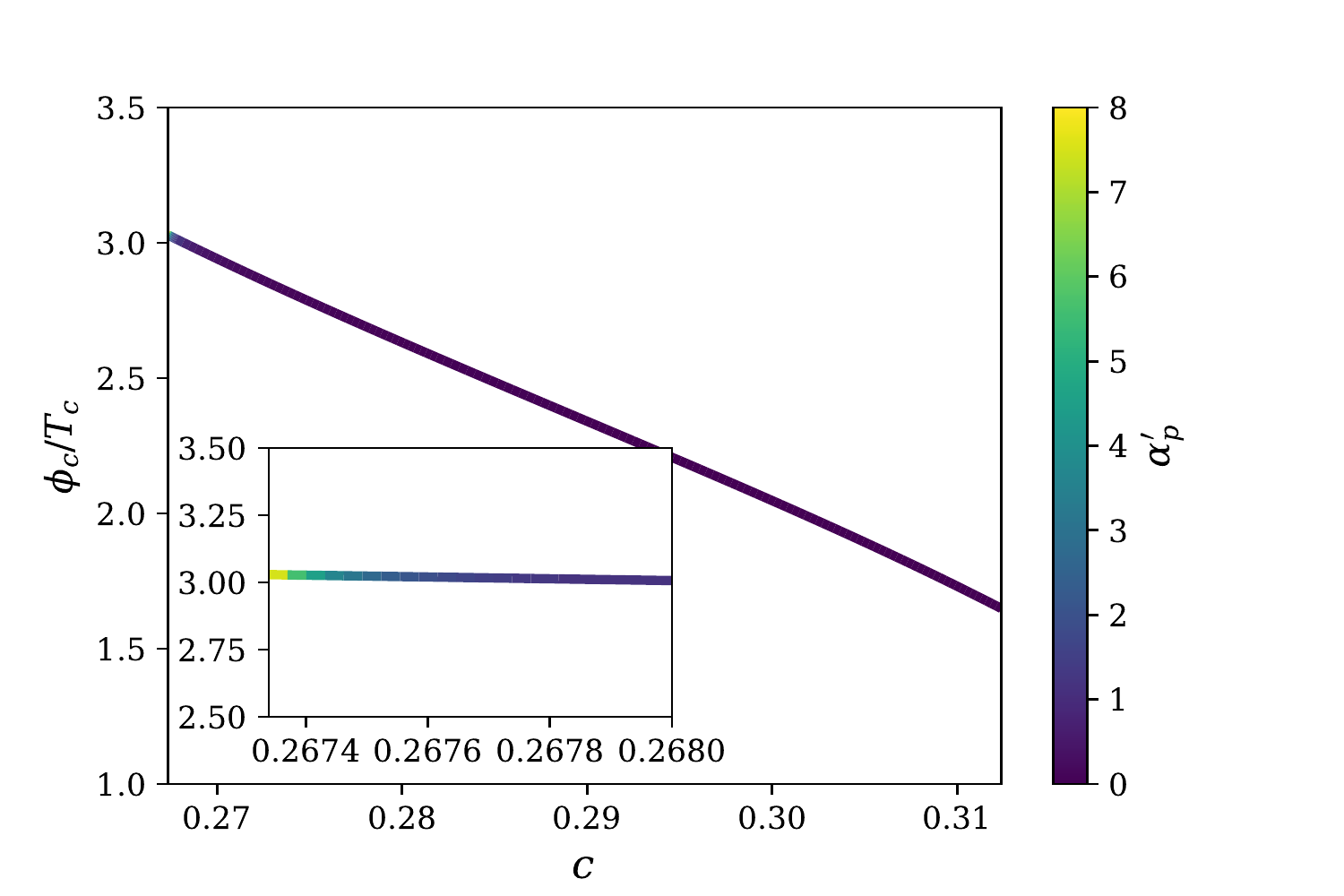}
			%\caption{fig2}
		\end{minipage}%
	}%
	\centering
	\caption{The relation between the washout parameter $\phi_c/T_c$, the thermal correction $c$, and the phase transition strength parameter in the dimension-six effective model. The multicolored lines show the parameter space with smaller $c$ and larger $\phi_c/T_c$ can generate a ultra supercooling case. The left figure is for $\alpha_p$, and the right figure is for $\alpha^{\prime}_p$.}\label{vtdim6}
\end{figure}

Furthermore, we show the relation between the washout parameter $\phi_c/T_c$, the thermal correction $c$, and the phase transition strength in the dimension-six effective model in figure~\ref{vtdim6}.
The left panel is for the conventional definition $\alpha_p$, and the right panel is for the alternative definition $\alpha^{\prime}_p$.
For both definitions of the phase transition strength parameter, the multicolored lines show the parameter with smaller $c$ and larger $\phi_c/T_c$ can generate a SFOPT with ultra supercooling.
In general, larger washout parameter gives larger phase transition strength.
For example, for the ultra supercooling case, the washout parameter $\phi_c/T_c$ should be larger than 2 in the dimension-six effective model.

In tables.~\ref{tb1} and \ref{tb2}, we show five benchmark sets of the dimension-six effective model with the conventional definition $\alpha$ and alternative definition $\alpha^{\prime}$, respectively.
For the same cutoff scale, the phase transition strength parameters in table~\ref{tb1} are obviously larger than those in table~\ref{tb2}.
We can see that for different definitions, it gives different values of phase transition strength.
In general, the conventional definition gives larger phase transition strength compared to the alternative definition.
In each definition, the value of phase transition strength is larger at $T_p$ than the one at $T_n$.
It is reasonable since $\alpha \propto 1/T^4$ and $T_p$ is obviously smaller than $T_n$.
And the time duration of the phase transition obtain a larger value at $T_p$ for the same definition and cutoff scale.
In the first row of each table, the parameters calculated at the percolation temperature ($T_p$, $\alpha_p$ and $HR_p$) depend on the bubble wall velocity.
This behavior is natural since the bubble wall velocity is one of the factors that determine how fast the bubble walls can percolate.
Thus, in the calculations, we should carefully choose the definition and the characteristic temperature. %select or other word?

\renewcommand\arraystretch{1.28}
\begin{table}[t]\small%
	\centering
	\begin{tabular}{|cccccccccc|}
		\hline
		& $\Lambda/\sqrt{\kappa}$ [GeV] & $T_n$ [GeV] & $T_p$ [GeV] & $\alpha_n$ & $\alpha_p$ & $\tilde{\beta}_n$ & $\tilde{\beta}_p$ & $HR_p$ & $v_b$\\
		\hline
		\multirow{2}*{$BP_1$}& \multirow{2}*{589.181}&42.175 &38.976&0.495&0.656&156.091&108.552&0.0360&1\\
       		&&42.175&37.812&0.495&0.731&156.091&91.670&0.0125&0.3\\
		$BP_2$&586.870&37.524&33.483&0.741&1.121&89.671&40.560&0.0794&1\\
		$BP_3$&586.1&35.123&30.229&0.937&1.624&58.798&7.251&0.140&1\\
		$BP_4$&585.864&34.232&28.767&1.027&1.949&47.771&-6.264&0.190&1\\
		$BP_5$&585.254&30.716&22.391&1.523&4.992&9.587&-69.824&0.710&1\\
		\hline
	\end{tabular}
    \caption{
    	The important phase transition parameters of the dimension-six effective model for the five benchmark sets with the conventional definition $\alpha$.
    }\label{tb1}
\end{table}

\begin{table}[t]\small%
	\centering
	\begin{tabular}{|cccccccccc|}
		\hline
		& $\Lambda/\sqrt{\kappa}$ [GeV] & $T_n^\prime$ [GeV] & $T_p^\prime$ [GeV] & $\alpha_n^\prime$ & $\alpha_p^\prime$ & $\tilde{\beta}_n^\prime$ & $\tilde{\beta}_p^\prime$ & $HR_p^\prime$ & $v_b$\\
		\hline
		\multirow{2}*{$BP_1^\prime$}& \multirow{2}*{589.181}&42.175&39.266&0.299&0.410&156.091&108.001&0.0420&1\\
		&&42.175&37.812&0.299&0.484&156.091&91.670&0.0115&0.3\\
		$BP_2^\prime$&586.870&37.524&33.483&0.491&0.803&89.671&40.560&0.0794&1\\
		$BP_3^\prime$&586.1&35.411&30.517&0.627&1.183&62.376&9.739&0.134&1\\
		$BP_4^\prime$&585.864&34.520&29.054&0.699&1.453&51.396&-3.869&0.176&1\\
		$BP_5^\prime$&585.254&31.291&23.253&1.057&3.652&15.507&-54.590&0.633&1\\
		\hline
	\end{tabular}
    \caption{The important phase transition parameters of the dimension-six effective model for the five benchmark sets with the alternative definition $\alpha'$.}\label{tb2}
\end{table}

\begin{figure}[t]
	\centering
	
	\subfigure{
		\begin{minipage}[t]{0.5\linewidth}
			\centering
			\includegraphics[scale=0.5]{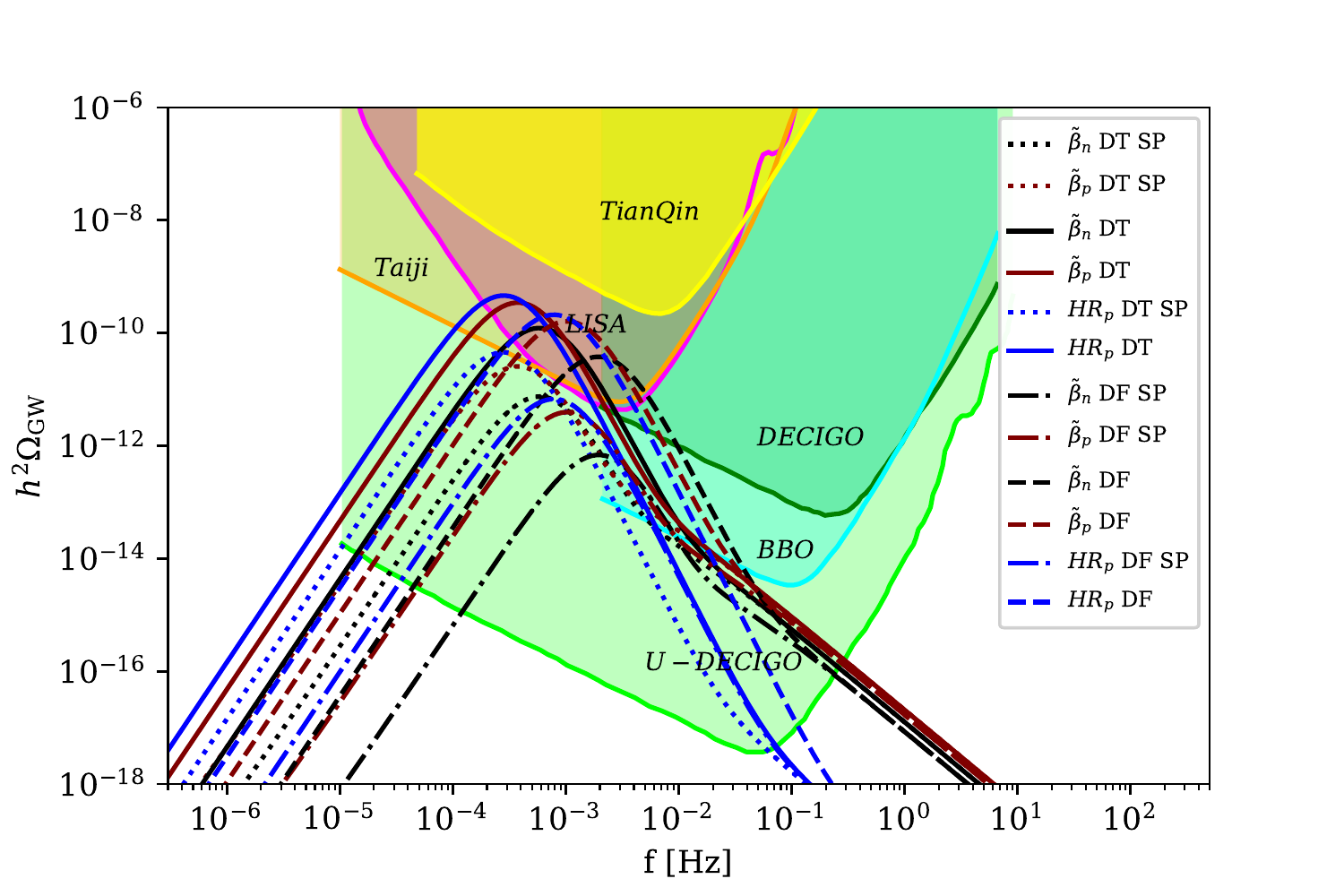}
		\end{minipage}%
	}%
	\subfigure{
		\begin{minipage}[t]{0.5\linewidth}
			\centering
			\includegraphics[scale=0.5]{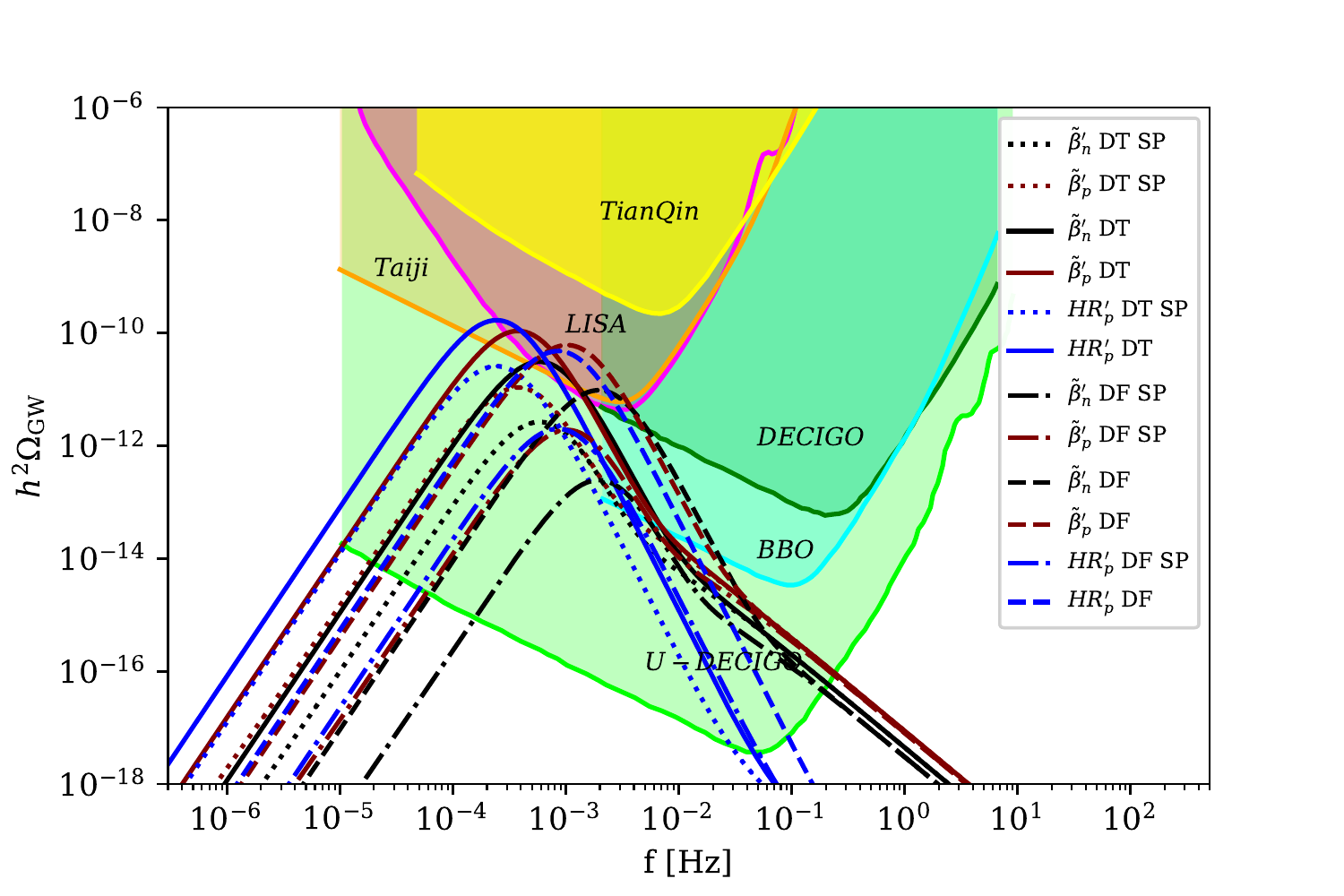}
			%\caption{fig2}
		\end{minipage}%
	}%
	\quad
	\subfigure{
		\begin{minipage}[t]{0.5\linewidth}
			\centering
			\includegraphics[scale=0.5]{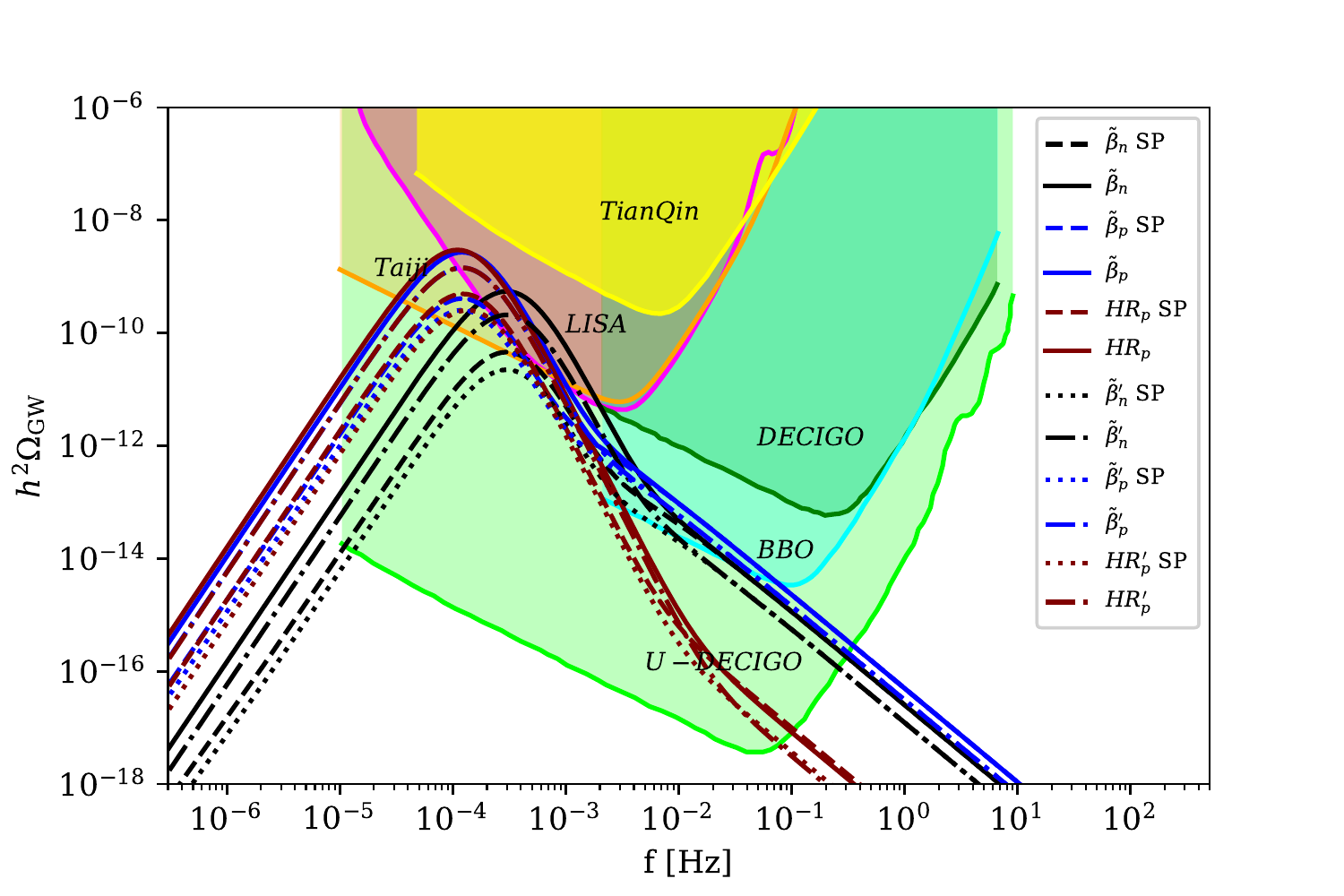}
		\end{minipage}
	}%
	\subfigure{
		\begin{minipage}[t]{0.5\linewidth}
			\centering
			\includegraphics[scale=0.5]{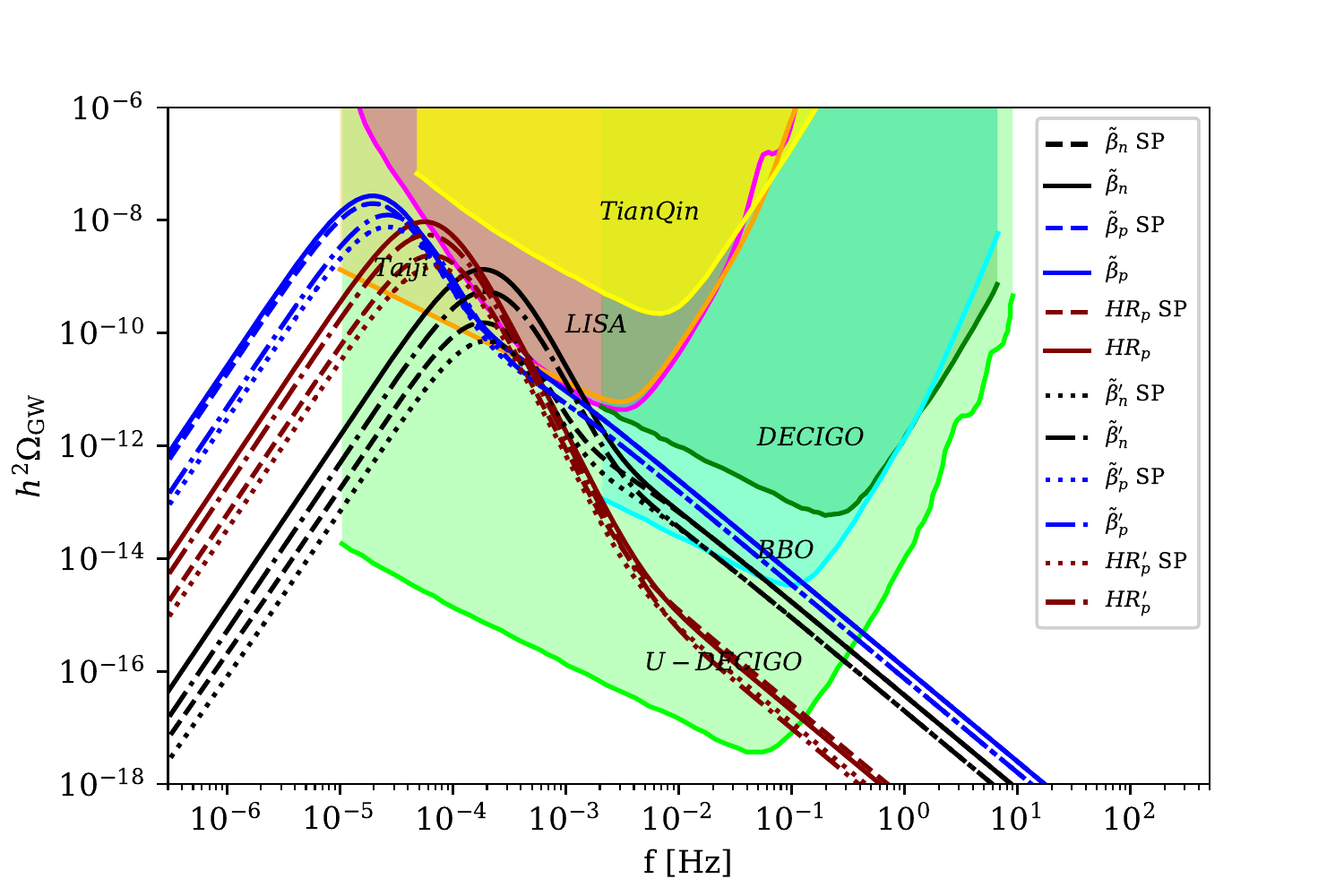}
		\end{minipage}
	}%
	\quad
	\subfigure{
		\begin{minipage}[t]{0.5\linewidth}
			\centering
			\includegraphics[scale=0.5]{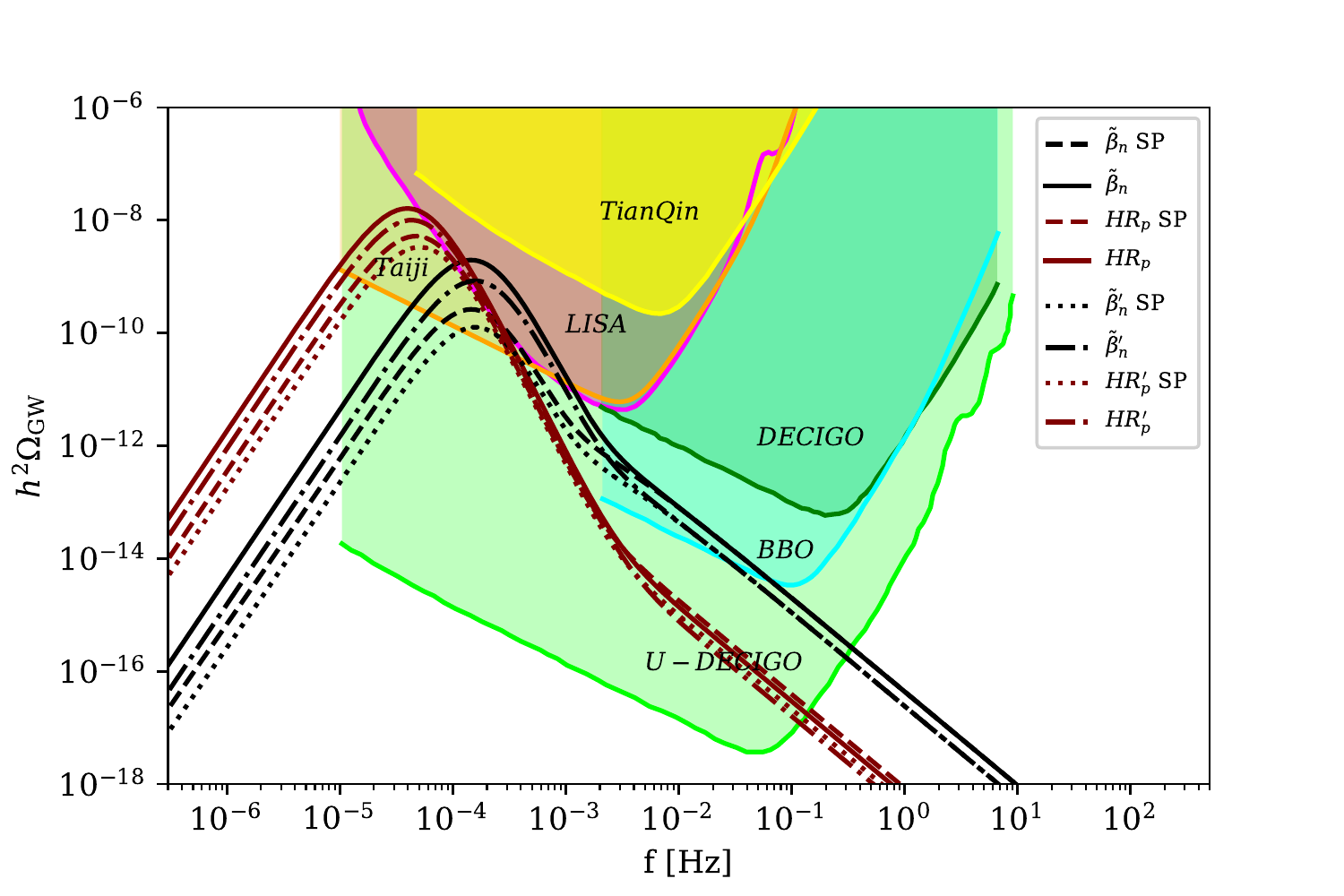}
		\end{minipage}
	}%
	\subfigure{
		\begin{minipage}[t]{0.5\linewidth}
			\centering
			\includegraphics[scale=0.5]{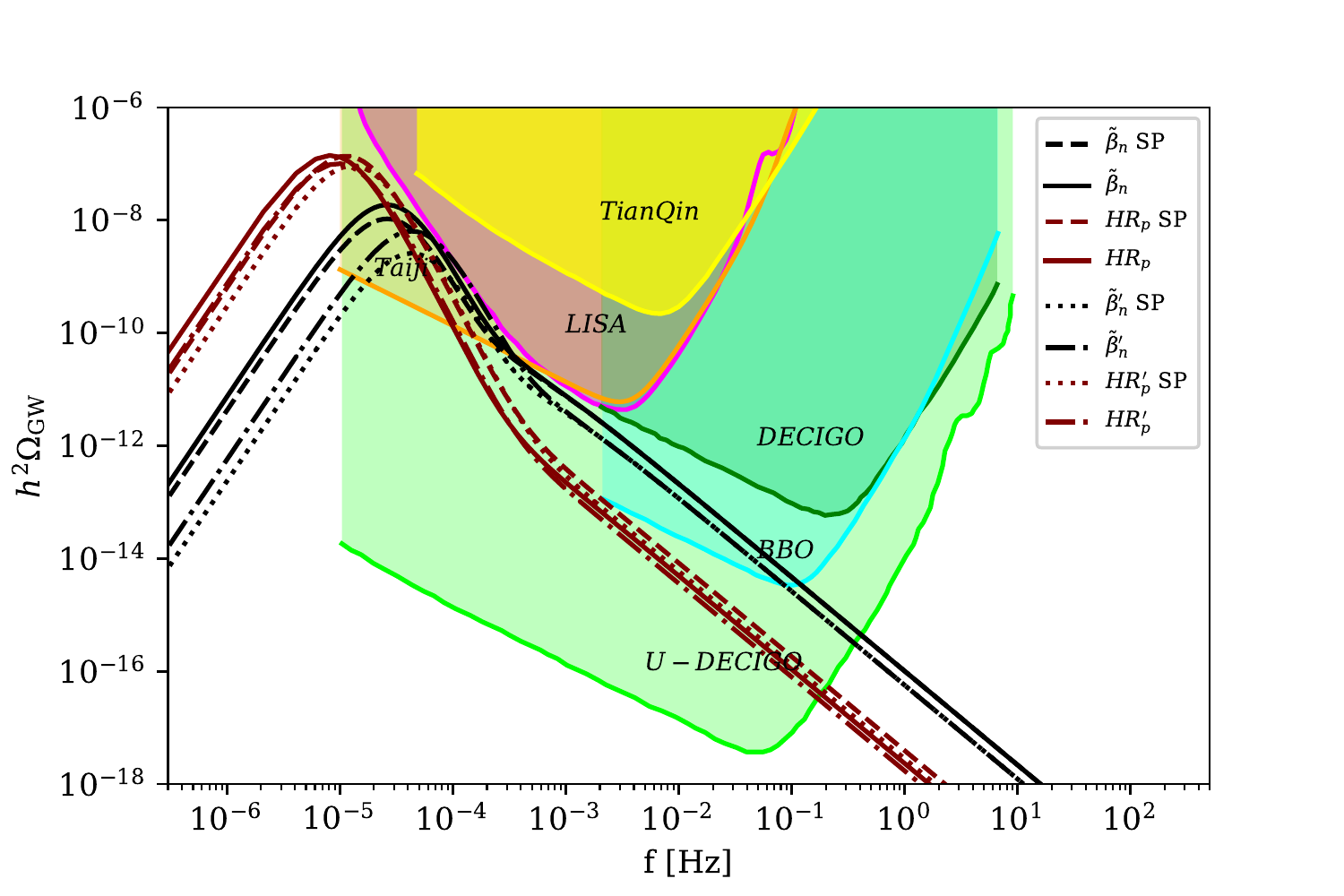}
		\end{minipage}
	}%
	\centering
	\caption{The GW spectra of the dimension-six effective model for different benchmark sets.
		The colored shaded regions represent the expected sensitivity of the GW interferometers LISA, DECIGO, U-DECIGO, BBO, Taiji, and TianQin, respectively.
		$\tilde{\beta}$ and $\tilde{\beta}^\prime$ denote the spectra that are derived by the approximated length scale at different temperatures for different definitions of phase transition strength.
		$HR$ and $HR^\prime$ represent the spectra that are obtained by the strict calculation of the mean bubble separation at the percolation temperature for different definitions of phase transition strength.
		DT and DF represent $v_b = 1$ and $v_b = 0.3$ respectively.
		SP denotes the GW spectra with the suppressed contribution of sound wave.
		The upper left plot and upper right plot show the GW spectra of $BP_1$ and $BP_1^\prime$.
		The middle left and right plot denote the GW spectra of $BP_2$ and $BP_2^\prime$ and the GW spectra of $BP_3$ and $BP_3^\prime$.
		The bottom left and right plot show the GW spectra of $BP_4$ and $BP_4^\prime$ and the GW spectra of $BP_5$ and $BP_5^\prime$ respectively.
	}
	\label{gwdim6}
\end{figure}

According to the phase transition parameters listed in Tables.~\ref{tb1} and \ref{tb2}, we compute the GW spectra of the dimension-six effective model for different benchmark sets and show them in figure~\ref{gwdim6}.
The color shaded regions represent the expected sensitivity of the GW interferometers LISA, DECIGO, U-DECIGO, BBO, Taiji, and TianQin, respectively.
$\tilde{\beta}$ and $\tilde{\beta}^\prime$ denote the spectra that are derived by the approximation of characteristic length scale at the nucleation and percolation temperature for different definitions of phase transition strength.
$HR$ and $HR^\prime$ represent the spectra which are obtained by the strict calculation of the mean bubble separation at the percolation temperature for different definitions of phase transition strength.
DT and DF are the abbreviation of detonation ($v_b = 1$) and deflagration ($v_b = 0.3$), respectively.
SP denotes the GW spectra with suppressed sound wave contribution as discussed in the previous section of sound wave mechansim.
The upper left plot and upper right plot show the GW spectra of $BP_1$ and $BP_1^\prime$.
The middle left plot denotes the GW spectra of $BP_2$ and $BP_2^\prime$.
The middle right plot denotes the GW spectra of $BP_3$ and $BP_3^\prime$.
The bottom left plot shows the GW spectra of $BP_4$ and $BP_4^\prime$.
The bottom right plot represents the GW spectra of $BP_5$ and $BP_5^\prime$.
From tables.~\ref{tb1} and \ref{tb2}, for some cutoff scales we find $\beta$ becomes negative at the percolation temperature, and hence the approximation in eq.~\eqref{HRappro} is failed at the percolation temperature, but is still applicable at the nucleation temperature.
Therefore, we show the resulting GW spectra with the approximation used at the nucleation temperature and the strict calculation of mean bubble separation in the bottom left and right of figure~\ref{gwdim6}.
We can see distinct differences of the GW spectra for different calculation methods.
Generally, the GW spectra from the $HR_p$ and detonation is the strongest compared to other cases and is within the sensitivity of LISA, Taiji, BBO, and U-DECIGO.
However, for $BP_3$ and $BP_3'$ the GW spectra from $\tilde{\beta}$ and $\tilde{\beta}'$ is stronger than other.
Since for the corresponding cutoff scale, the approximation gives an overestimated length scale at the percolation temperature as shown in figure~\ref{hrdim6}.
After considering the suppression effect of sound wave, some GW spectra of benchmark sets (e.g., $BP_1$ and $BP_1'$) become marginal for LISA, Taiji, DECIGO, but is still available for BBO and U-DECIGO.
To determine whether LISA can detect the GW signal from this model, it is essential to use the correct temperature, length scale and formulae.

\begin{figure}[t]
	\centering
	
	\subfigure{
		\begin{minipage}[t]{1\linewidth}
			\centering
			\includegraphics[scale=0.8]{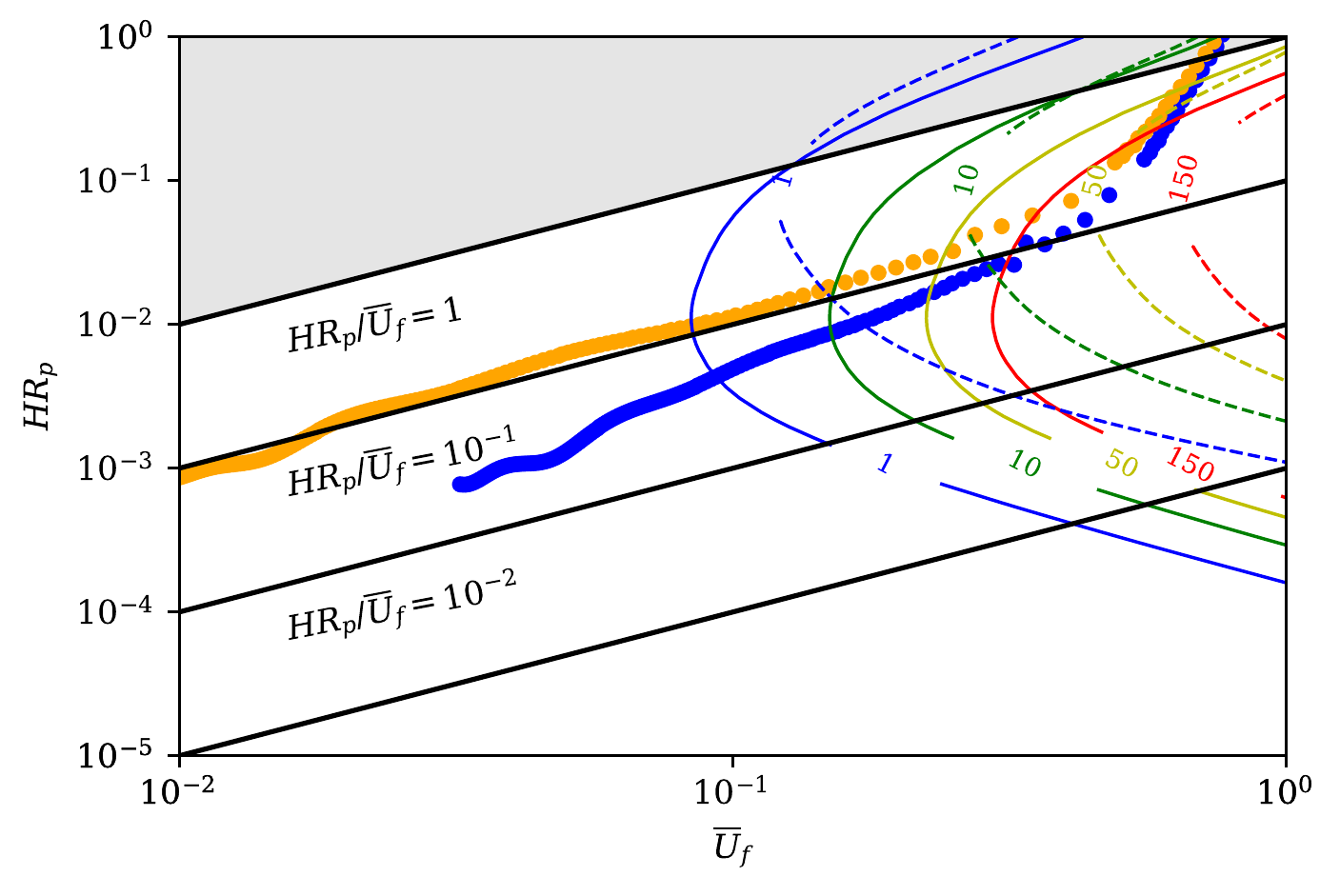}
		\end{minipage}%
	}%
	\centering
	\caption{Signal-to-noise ratio for the dimension-six effective model suppressed (colored dashed contours) and unsuppressed (colored solid contours) power spectra for LISA configuration and $T_p = 50$ GeV.
		The black solid lines show the magnitude of $HR_p/\overline{U}_f$.
		The blue and orange dots are parameter points of the dimension-six effective model with strict calculation of the length scale $HR_p$ and $\overline{U}_f$, which are derived from the conventional and alternative definition of phase transition strength respectively, for the detonation mode ($v_b =1$).}\label{snrdim6}
\end{figure}

In figure~\ref{snrdim6}, we show the SNR for the dimension-six effective model suppressed (colored dashed lines) and unsuppressed (colored solid lines) power spectra for LISA configuration and $T_p = 50$ GeV.
The black solid lines show the magnitude of $HR_p/\overline{U}_f$.
The blue and orange dots are selected parameter points of the dimension-six effective model, with strict calculation of the length scale $HR_p$ and $\overline{U}_f$ which are derived form the conventional and alternative definition of phase transition strength, respectively.
The stronger phase transition strength corresponds to the larger characteristic length and $\overline{U}_f$.
Since we assume $\rm SNR_{thr} = 10$, for the parameters with same definition of phase transition strength, we find the detectability for suppressed and unsuppressed GW signal are different.
We can also see different definitions of phase transition strength should also affect the detectability.
The stronger the phase transition, the less influence on the detectability.
It is worth noticing that figure~\ref{snrdim6} is just for illustration that different definitions of phase transition strength should give different SNR, and the suppression effect can also affect the detection ability of some specific parameters.
To precisely study the SNR of a parameter point for a given model, one should calculate it with the strict phase transition parameters derived from this point.

From the numerical results, we can find that for very low cutoff scale, the phase transition strength $\alpha$
or $\alpha^{\prime}$ can be larger than 1, namely, the ultra supercooling can occurs at very low cutoff scale.
For both strong supercooling and ultra supercooling cases,
there are obvious differences for the phase transition strength at the nucleation and percolation temperature.
In these cases, it is not correct to use the nucleation temperature $T_n$ to calculate the phase transition strength parameter.
We should use the percolation temperature $T_p$ to define all the important parameters and calculate the GW spectra.
For the phase transition duration parameter $\beta$, it becomes smaller with the decreasing of the cutoff scale.
The phase transition duration should be larger when the cutoff scale becomes very small.
And there are also distinctive differences between different definitions.
Therefore, in this dimension-six effective model, the SFOPT phase transition can be slight supercooling,
mild supercooling, strong supercooling and ultra supercooling. We should use the correct physical quantity to
calculate the GW spectra, especially for the strong supercooling and ultra supercooling.

\subsection{Quartic toy model}

In the previous model, the heavy particles are integrated out. Now, we consider a different typical model where the new particles are not so heavy and they can not be simply integrated out.
Instead, there are thermal or non-thermal contributions to the cubic term.
From the standard finite-temperature field theory, in order to accommodate the tree-level effect, we modify the parameter of the cubic term to make a cubic term that can exist at zero temperature, and give the following quartic toy model
\begin{equation}\label{1prime}
V_{\rm eff}(\phi,T)\approx \frac{\mu^2 + cT^2}{2}\phi^2 - (ET + A)\phi^3 + \frac{\lambda}{4}\phi^4 \,\,.
\end{equation}
It is just a toy model, which could be obtained from the approximation of some multi-field phase transition models.
We set the zero-temperature VEV $\phi_{\rm true}=v$,\footnote{For more general case $v\ne246 \rm GeV$. We just choose $v=246$~GeV as an example.} then we can derive the formula of the temperature when the potential barrier disappears,
\begin{equation}
T_o^2 = -\frac{\mu^2}{c} = \frac{\lambda - 3A/v}{c}v^2,
\end{equation}
and the critical temperature is given by
\begin{equation}
T_c = \frac{2}{\lambda c - 2E^2}\left[AE + \frac{\sqrt{\lambda c(2A^2 + (\lambda c - 2E^2)T_o^2)}}{2}\right]\,\,.
\end{equation}
The washout parameter is
\begin{equation}
\frac{\phi_c}{T_c} = \frac{2(E + A/T_c)}{\lambda}\,\,.
\end{equation}
The true minimum can be obtained as
\begin{equation}
\phi_{\rm true} = \frac{3(ET+A)}{2\lambda}\left[1 + \sqrt{1 - \frac{4\lambda (\mu^2 + cT^2)}{9(ET + A)^2}}\right]\,\,.
\end{equation}

This toy model is convenient to perform some semi-analytic calculations, which can help us to
clearly understand the underlying physics.
In this work, we use a semi-analytic method to study this toy model mentioned above.
The approximated analytic form of the action for this toy model is given by \cite{Adams:1993zs}
\begin{equation}
\frac{S_3}{T}=\frac{\pi a8\sqrt{2}}{81T\tilde{\lambda}^{3/2}}(2-\delta)^{-2}(\gamma_1\delta + \gamma_2\delta^2 + \gamma_3\delta^3)\,\,,
\end{equation}
where $a=EA+T$, $b=(\mu^2 + cT^2)/2$, $\tilde{\lambda}= \frac{\lambda}{4}$,  $\delta=8 \tilde{\lambda}b/a^2$, $\gamma_1 = 8.2938$, $\gamma_2 = -5.5330$, and $\gamma_3 = 0.8180$.
It is worth noticing that this formula is only applicable for $0 < \delta <2$.
We can also use the analytic expression for the bounce action that derived by Linde \cite{Dine:1992wr}:
\begin{equation}
\frac{S_3}{T} = \frac{13.72(\mu^2 + cT^2)^{3/2}}{2^{3/2}T(ET + A)^2}f\left(\frac{(\mu^2 +cT^2)\lambda}{2(ET + A)^2}\right)\,\,,
\end{equation}
where $f(x) = 1 + \frac{x}{4}\left[1 + \frac{2.4}{1-x} + \frac{0.26}{(1-x)^2}\right]$.
Note this analytic formula is valid for $0<x<1$.
%that means it can give a constraint on the parameters of the model.
Both the analytic formula of Linde~\cite{Dine:1992wr} and the semi-analytic formula of Adams~\cite{Adams:1993zs} can fit numerical calculation very well.
The above analytic formula is valid for thermal tunneling process, and the following semi-analytic formula derived by Adams is available for the quantum tunneling process
\begin{equation}
S_4 = \frac{\pi^2}{3\tilde{\lambda}}(2-\delta)^{-3}\left[\alpha_1\delta + \alpha_2\delta^2 + \alpha_3\delta^3\right]\,\,,
\end{equation}
where $\alpha_1 = 13.832$, $\alpha_2 = -10.819$, and $\alpha_3 = 2.0765$.

\subsubsection{$A \ne 0$ }
\begin{figure}[t]
	\centering
	
	\subfigure{
		\begin{minipage}[t]{1\linewidth}
			\centering
			\includegraphics[scale=0.8]{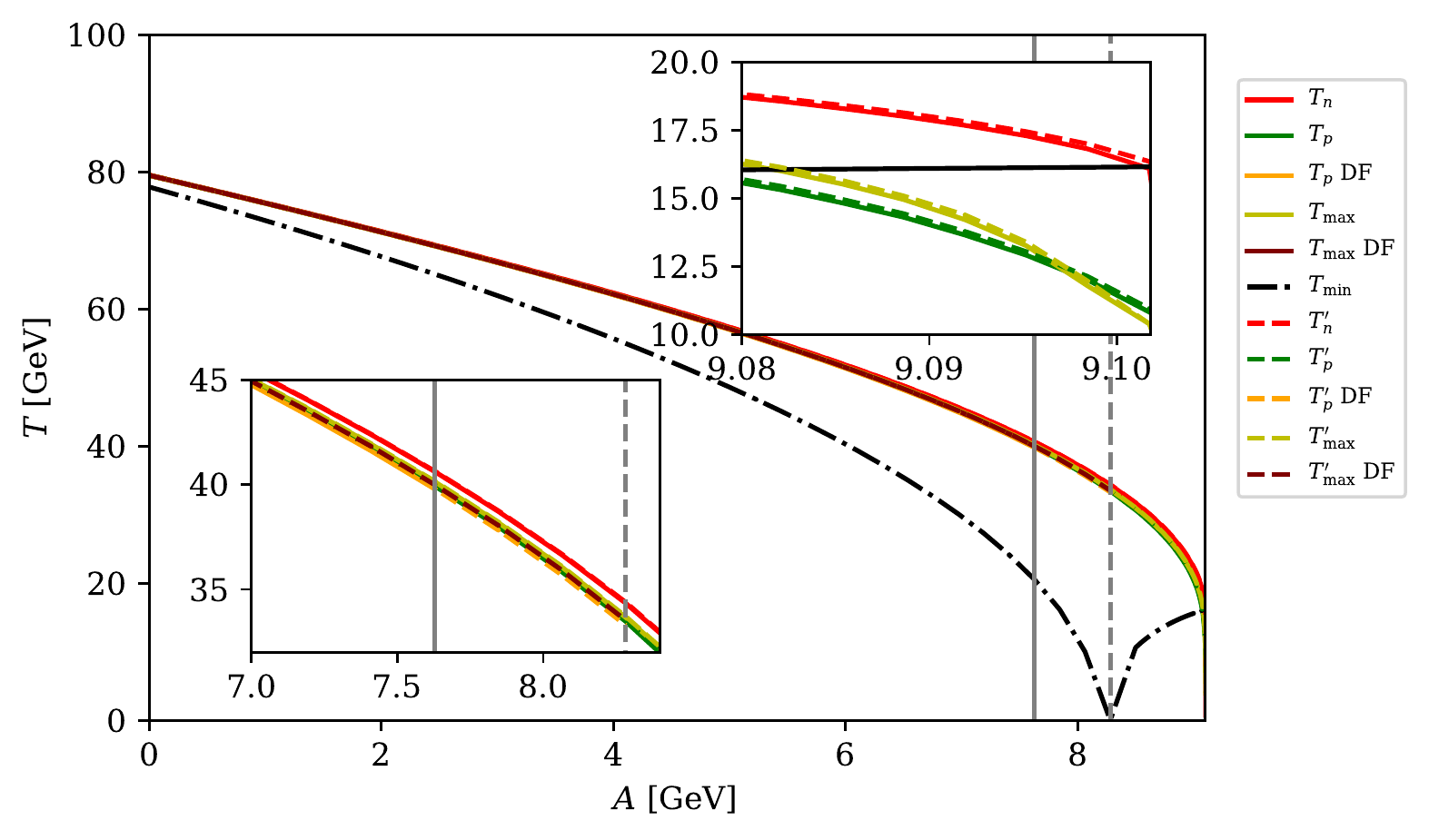}
		\end{minipage}%
	}%
	\centering
	\caption{The characteristic temperature of the quartic toy model ($A\ne0$).
The colored solid lines represent various characteristic temperatures for the conventional definition $\alpha$.
The colored dashed lines denote temperatures that are derived for the alternative definition $\alpha'$.
The black dash-dotted line denotes the minimum temperature.
The vertical gray solid and dashed lines indicate values of $A$ that produce a SFOPT with $\alpha_p=1$ and $\alpha_p^\prime=1$, respectively.
The orange and maroon lines shows different bubble wall velocities give negligible correction to the percolation and the maximum temperature for the regime that generates a SFOPT with $\alpha<1$ and $\alpha^\prime<1$ at the percolation temperature.
		}\label{cte3}
\end{figure}

The quartic toy model with $A \ne 0$ may comes from the approximation of the multifield phase transition models~\cite{Wang:2019pet,Wan:2018udw,Huang:2015bta,Huang:2018aja,Brdar:2019qut,Alves:2019igs}.
In figure~\ref{cte3}, we show the characteristic temperature of the quartic toy model ($A\ne0$).
The solid lines represent various characteristic temperatures for the conventional definition $\alpha$.
The dashed lines denote temperatures that are derived for the alternative definition $\alpha^{\prime}$.
The black dash-dotted line denotes the minimum temperature.
The vertical solid and dashed lines (gray) indicate values of $A$ which can produce a SFOPT with $\alpha_p=1$ and $\alpha_p^\prime=1$, respectively.
The orange and maroon lines shows that different bubble wall velocity can not give significant modifications to the percolation and the maximum temperature for the regime that generate a SFOPT with $\alpha<1$ and $\alpha^\prime<1$.
From figure~\ref{cte3}, we can see that the hierarchy between the nucleation temperature and percolation temperature becomes larger with the increasing of $A$ for the same definition of phase transition strength.
This behavior may indicate that the tree-level barrier could increase the temperature hierarchy and enhance the supercooling.
We should use $T_p$ as more reliable temperature in the following calculations.
As the same for dimension-six effective model, we only show the parameter space of $A$ that can give a maximum temperature ,i.e., the parameter space that can guarantee the completion of phase transition.

\begin{figure}[t]
	\centering
	
	\subfigure{
		\begin{minipage}[t]{1\linewidth}
			\centering
			\includegraphics[scale=0.8]{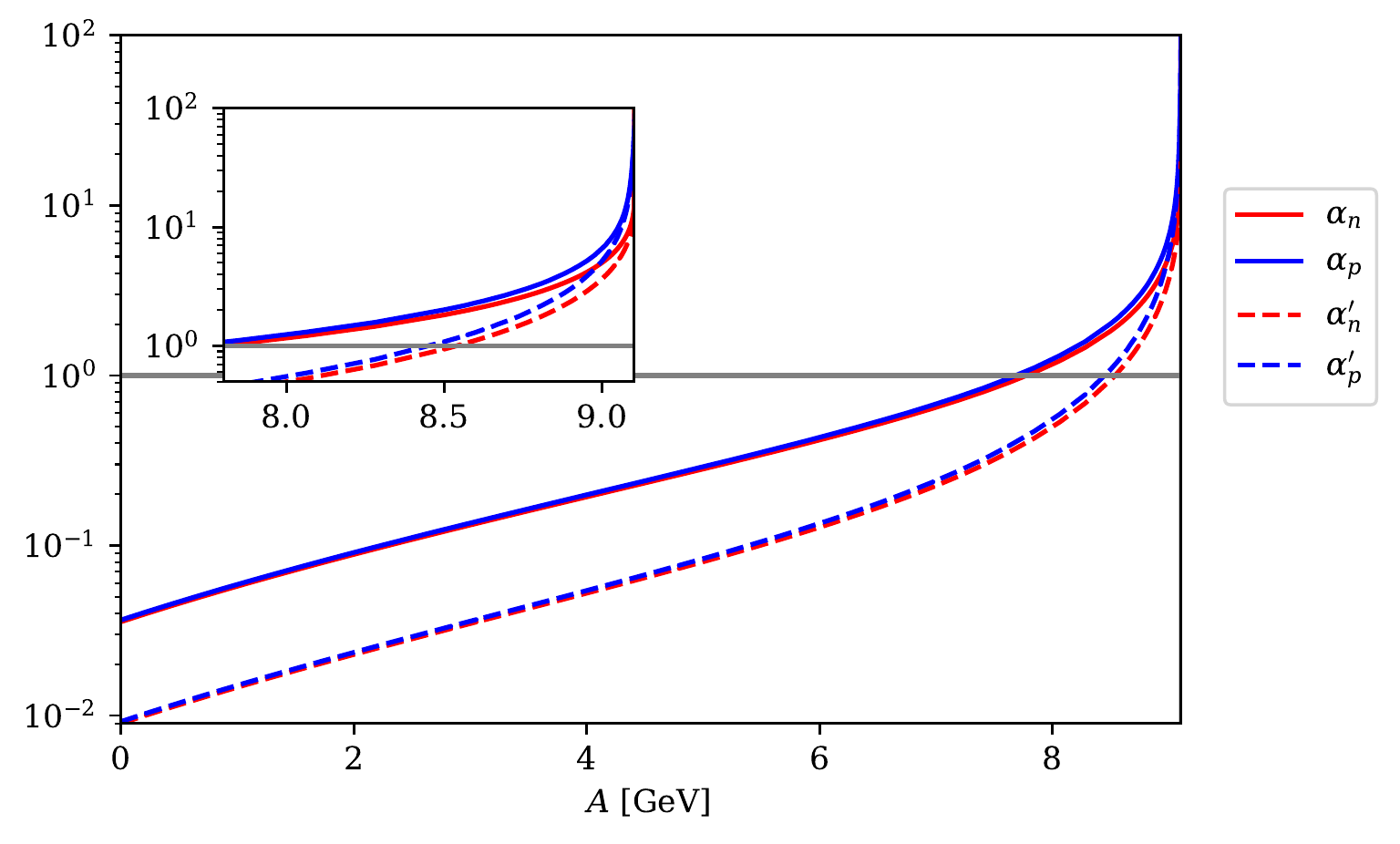}
		\end{minipage}%
	}%
	\centering
	\caption{The phase transition strength as a function of $A$ at the nucleation and percolation temperature for the quartic toy model ($A\ne0$) different definitions.
Solid lines represent the values with conventional definition.
Dashed lines denote the values with alternative definition.
The horizontal gray line indicates that the phase transition strength parameters are equal to one.}\label{ste3}
\end{figure}

We show the phase transition strength as a function of $A$ at the nucleation and percolation temperature for different definitions in figure~\ref{ste3}.
Solid lines represent the values with conventional definition.
Dashed lines denote the values with alternative definition.
Similar to the temperature behavior in figure~\ref{cte3},
the hierarchy between $\alpha$ and $\alpha^{\prime}$ increases with the increasing of $A$.
For $A>7.8$ GeV, we have the ultra supercooling case for the conventional definition of the strength parameter, namely, $\alpha_p>1$.
And in this ultra supercooling case, $\alpha_p$ is obviously larger than $\alpha_n$.
The alternative definition gives the similar qualitative behavior but weaker phase transition strength, and the ultra supercooling only becomes possible for $A>8.5$ GeV.

\begin{figure}[t]
	\centering
	
	\subfigure{
		\begin{minipage}[t]{1\linewidth}
			\centering
			\includegraphics[scale=0.8]{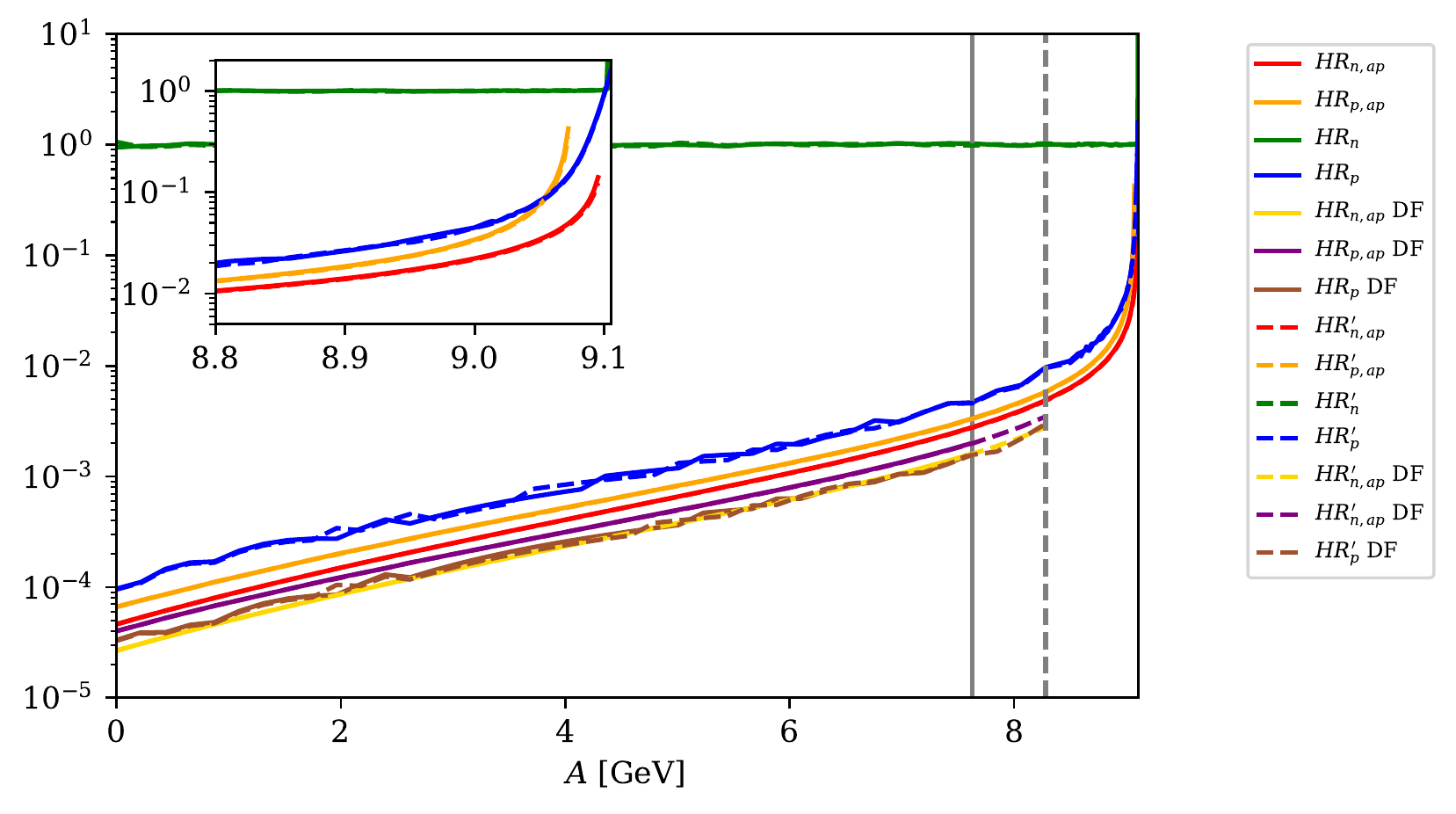}
		\end{minipage}%
	}%
	\centering
	\caption{
		The characteristic length scale of the quartic toy model ($A\ne0$). $HR_{*,ap}$ (red, orange, yellow, and purple solid lines), which is derived by the approximation, and $HR_*$ (green, blue, and brown solid lines), which is calculated by the first principle, denote the characteristic length at nucleation and percolation temperature.
The colored dashed lines denote the values of characteristic length scale at different temperatures for the alternative definition $\alpha'$.
The vertical gray solid and dashed lines indicate values of $A$ that produce a SFOPT with $\alpha_p=1$ and $\alpha_p^\prime=1$, respectively.}\label{hre3}
\end{figure}

In figure~\ref{hre3}, we show the characteristic length scale of the quartic toy model ($A\ne0$). $HR_{*,ap}$ (red, orange, yellow, and purple solid lines), which is derived by the approximation, and $HR_*$ (green, blue, and brown solid lines), which is calculated by the first principle, denote the characteristic length at the nucleation and percolation temperature.
The dashed lines denote the values of characteristic length at different temperatures for different definitions of phase transition strength.
Similar to the dimension-six effective model, here, different definitions of phase transition strength give a negligible modification to the characteristic length.
The yellow, purple and brown lines show the deflagration expansion mode ($v_b = 0.3$) gives negligible modifications to the length scale for regime of a SFOPT with $\alpha_p < 1$ and $\alpha_p' < 1$.
The vertical gray solid and dashed lines are the separation of the regions of a SFOPT with ultra supercooling and strong supercooling at the percolation temperature for both definitions.
As shown in this figure, the approximated characteristic length scale (red, orange,yellow,and purple lines, both solid and dashed) fits the result (blue and brown lines, both solid and dashed) derived by strict calculation for the deflagration case well, but fits less well for the detonation cases.

\begin{figure}[t]
	\centering
	
	\subfigure{
		\begin{minipage}[t]{0.5\linewidth}
			\centering
			\includegraphics[scale=0.5]{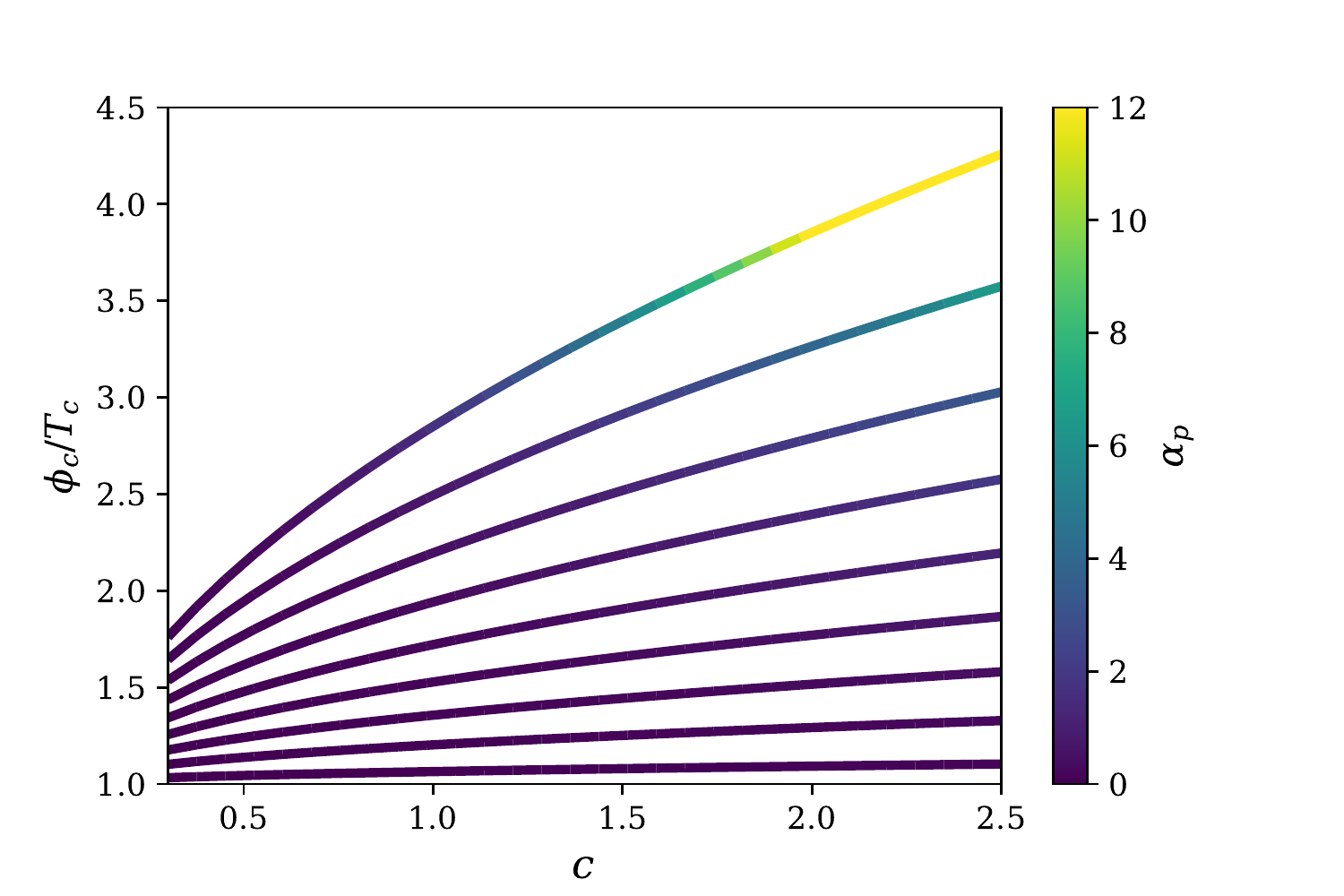}
		\end{minipage}%
	}%
	\subfigure{
		\begin{minipage}[t]{0.5\linewidth}
			\centering
			\includegraphics[scale=0.5]{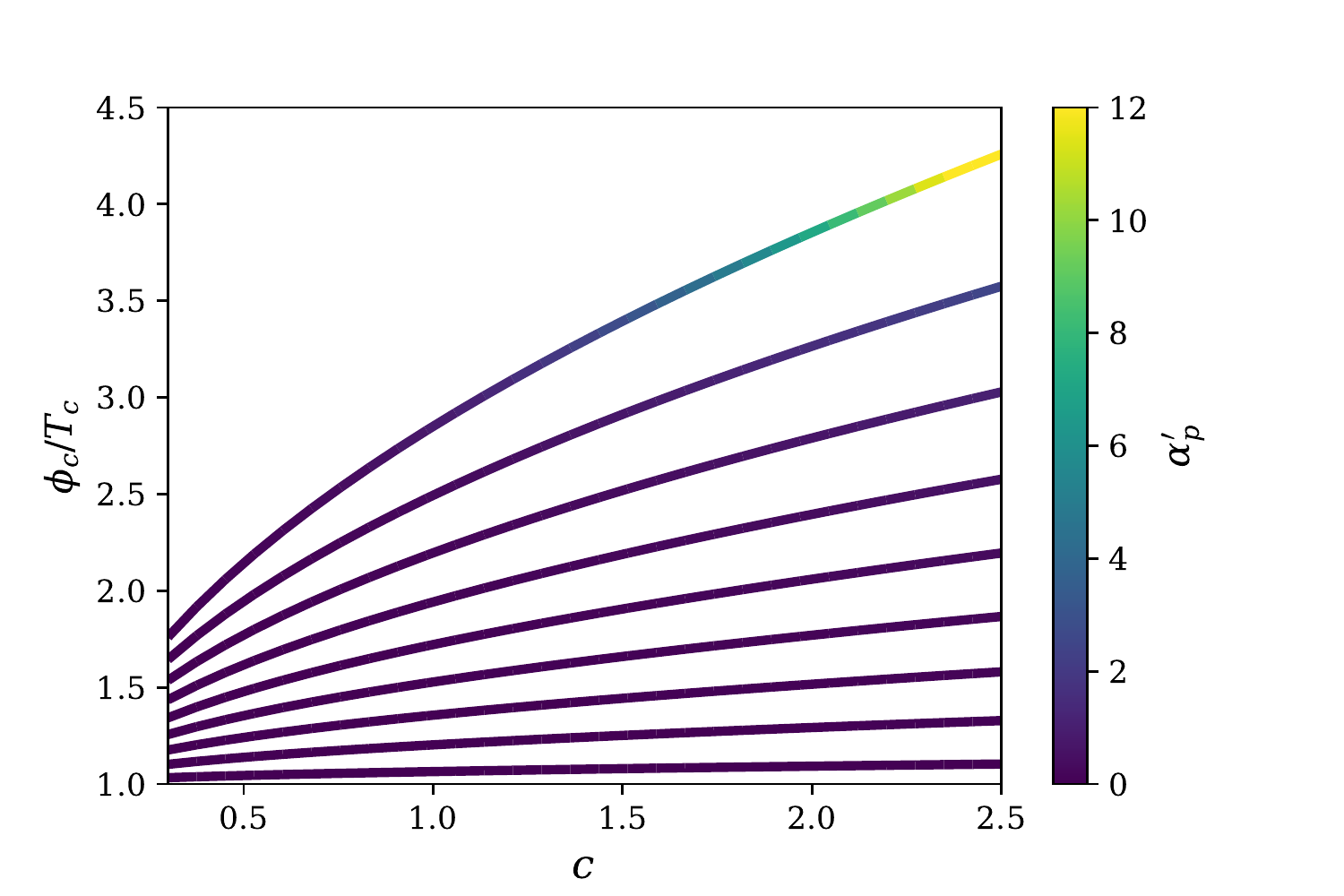}
			%\caption{fig2}
		\end{minipage}%
	}%
	\centering
	\caption{The relation between the washout parameter $\phi_c/T_c$, the thermal coupling $c$, and the phase transition
strength parameter for the quartic toy model ($A\ne0$).
The multicolored lines (from bottom to top) for $A=0.5,1.5,2.5,3.5,4.5,5.5,6.5,7.5,8.5$.}\label{vte3}
\end{figure}

In figure~\ref{vte3}, we show the relation between the washout parameter $\phi_c/T_c$, the thermal coupling $c$ and the phase transition strength for the quartic toy model.
The multicolored lines represent $A=0.5,1.5,2.5,3.5,4.5,5.5,6.5,7.5,8.5$ from bottom to top ($E$ and $\lambda$ are set as 0.05 and 0.1).
As shown in both panel, to generate a phase transition with ultra supercooling for both definitions, both the thermal coupling and the washout parameter should be large enough.

We list the phase transition parameters of the quartic toy model ($A\ne0$) for five benchmark sets with the conventional definition $\alpha$ in table~\ref{tb3}.
In table~\ref{tb4}, we show the phase transition parameters of  for the five benchmark sets with the alternative definition of the phase transition strength.
Here, for both cases we set $c$, $E$, and $\lambda$ as 0.5, 0.05, and 0.1, respectively.
In each table with the same definition of the phase transition strength, $T_p$ ($T_p'$) is smaller than
$T_n$ ($T_n'$), and $\alpha_p$ ($\alpha_p'$) is larger than $\alpha_n$ ($\alpha_n'$).

\begin{table}[t]\small%
	\centering
	\begin{tabular}{|cccccccccc|}
		\hline
		& $A$ [GeV] & $T_n$ [GeV] & $T_p$ [GeV] & $\alpha_n$ & $\alpha_p$ & $\tilde{\beta}_n$ & $\tilde{\beta}_p$ & $HR_p$ & $v_b$\\
		\hline
		\multirow{2}*{$BP_6$}& \multirow{2}*{7.410}&42.383 &41.766&0.802&0.844&1221.420&1018.770&0.00458&1\\
		&&42.383&41.607&0.802&0.855&1221.420&974.561&0.00131&0.3\\
		$BP_7$&7.7&40.003&39.309&0.950&1.007&995.824&831.650&0.00515&1\\
		$BP_8$&8.282&34.337&33.472&1.461&1.581&599.404&502.952&0.00970&1\\
		$BP_9$&8.5&31.694&30.684&1.824&2.014&462.593&383.227&0.0111&1\\
		$BP_{10}$&9.081&18.623&15.439&8.870&16.797&47.111&-10.117&0.218&1\\
		\hline
	\end{tabular}
    \caption{The phase transition parameters of the quartic toy model ($A\ne0$. $c$, $E$, and $\lambda$ are set as 0.5, 0.05, and 0.1, respectively.) for five benchmark sets with the conventional definition $\alpha$.
    }\label{tb3}
\end{table}

\begin{table}[t]\small%
	\centering
	\begin{tabular}{|cccccccccc|}
		\hline
		& $A$ [GeV] & $T_n^\prime$ [GeV] & $T_p^\prime$ [GeV] & $\alpha_n^\prime$ & $\alpha_p^\prime$ & $\tilde{\beta}_n^\prime$ & $\tilde{\beta}_p^\prime$ & $HR_p^\prime$ & $v_b$\\
		\hline
		\multirow{2}*{$BP_6^\prime$}& \multirow{2}*{7.410}&42.402&41.792&0.298&0.322&1228.850&1026.417&0.00458&1\\
		&&42.402&41.633&0.298&0.328&1228.850&981.736&0.00131&0.3\\
		$BP_7^\prime$&7.7&40.033&39.345&0.376&0.410&1003.97&839.21&0.00524&1\\
		$BP_8^\prime$&8.282&34.384&33.519&0.686&0.772&605.326&507.615&0.00958&1\\
		$BP_9^\prime$&8.5&31.740&30.730&0.939&1.084&466.690&386.435&0.0106&1\\
		$BP_{10}^\prime$&9.081&18.743&15.569&7.344&15.285&49.521&-7.961&0.212&1\\
		\hline
	\end{tabular}
    \caption{The phase transition parameters of the quartic toy model ($A\ne0$. $c$, $E$, and $\lambda$ are sit as 0.5, 0.05, and 0.1, respectively.) for five benchmark sets with the alternative definition $\alpha'$.
    }\label{tb4}
\end{table}

\begin{figure}[t]
	\centering
	
	\subfigure{
		\begin{minipage}[t]{0.5\linewidth}
			\centering
			\includegraphics[scale=0.5]{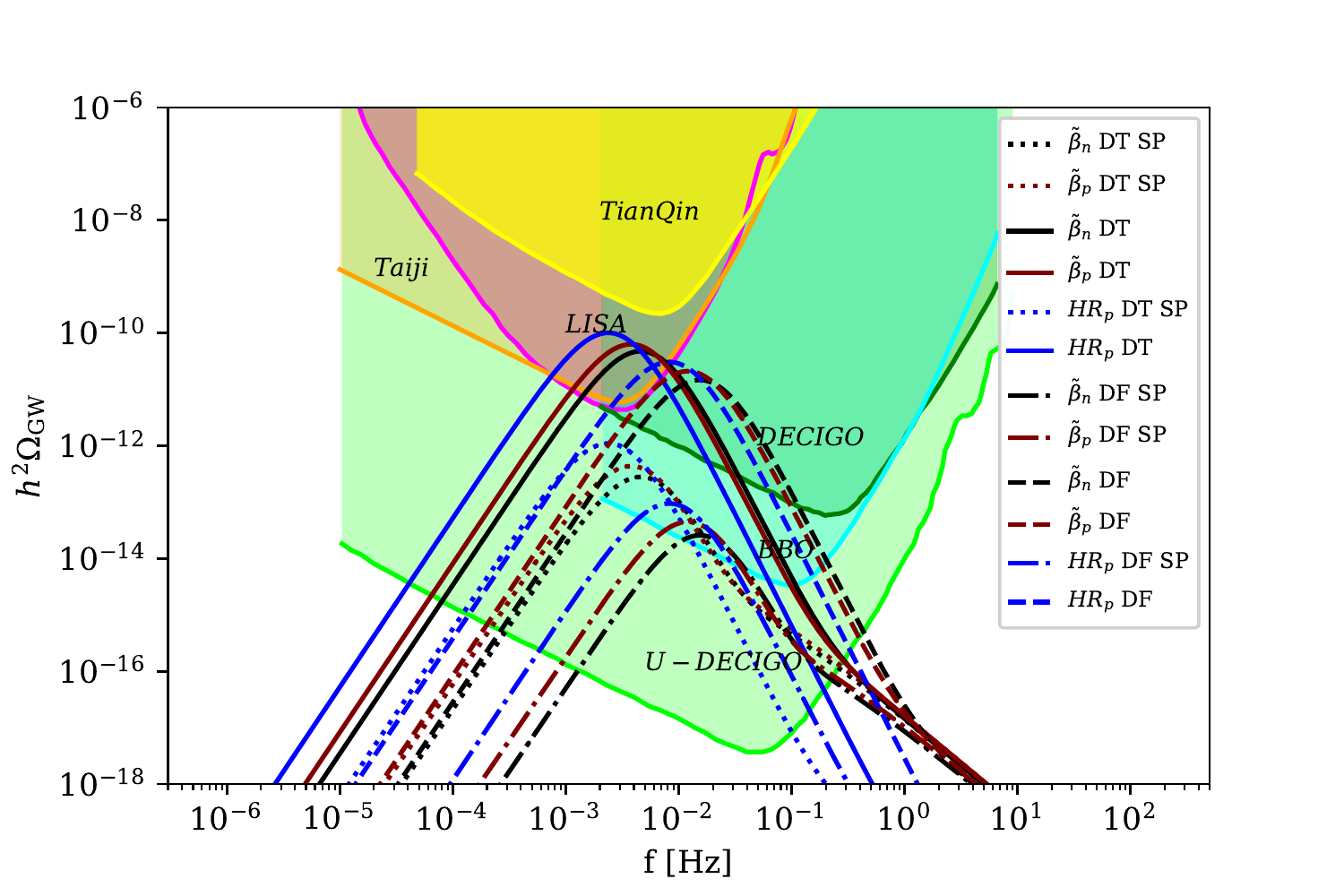}
		\end{minipage}%
	}%
	\subfigure{
		\begin{minipage}[t]{0.5\linewidth}
			\centering
			\includegraphics[scale=0.5]{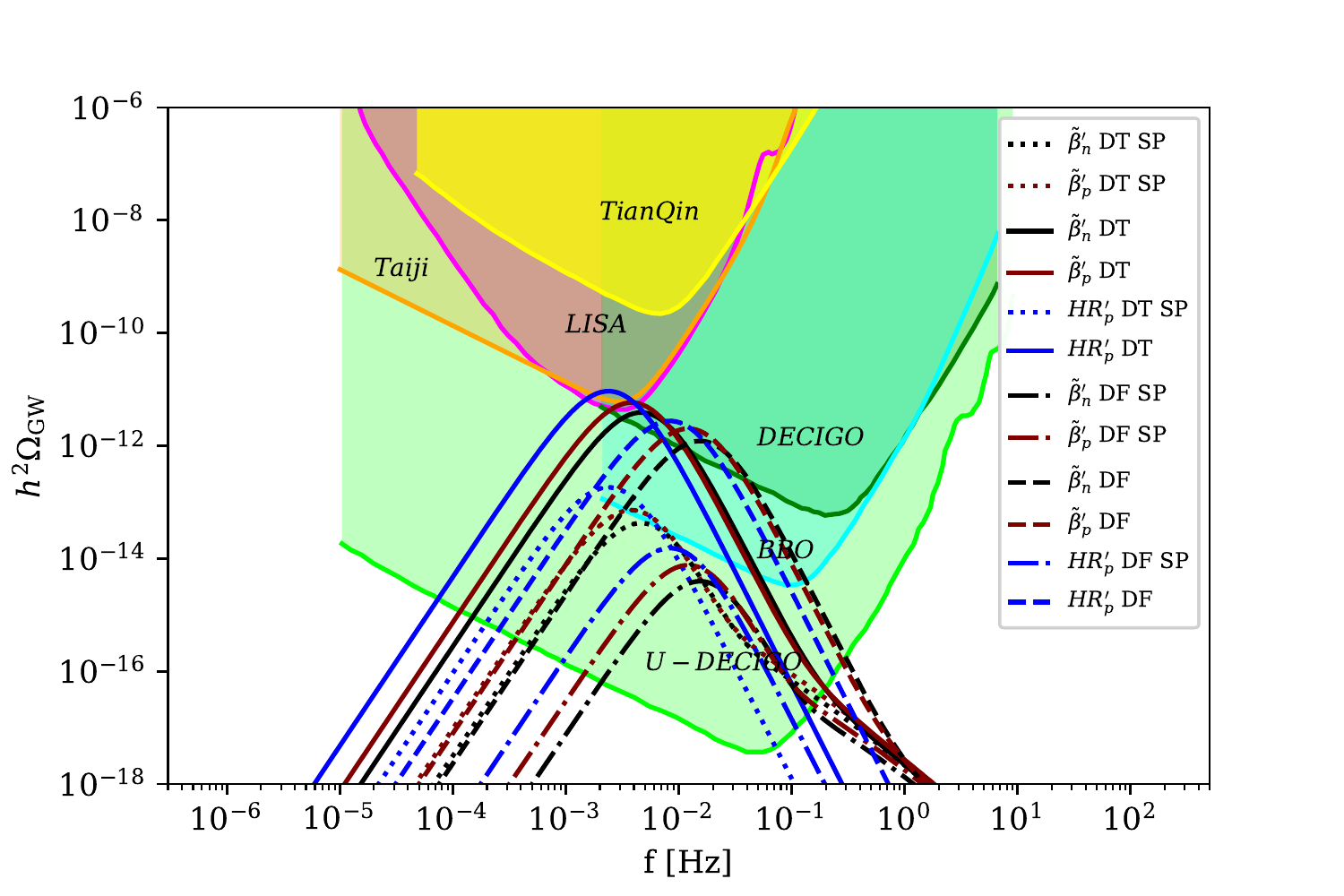}
			%\caption{fig2}
		\end{minipage}%
	}%
	\quad
	\subfigure{
		\begin{minipage}[t]{0.5\linewidth}
			\centering
			\includegraphics[scale=0.5]{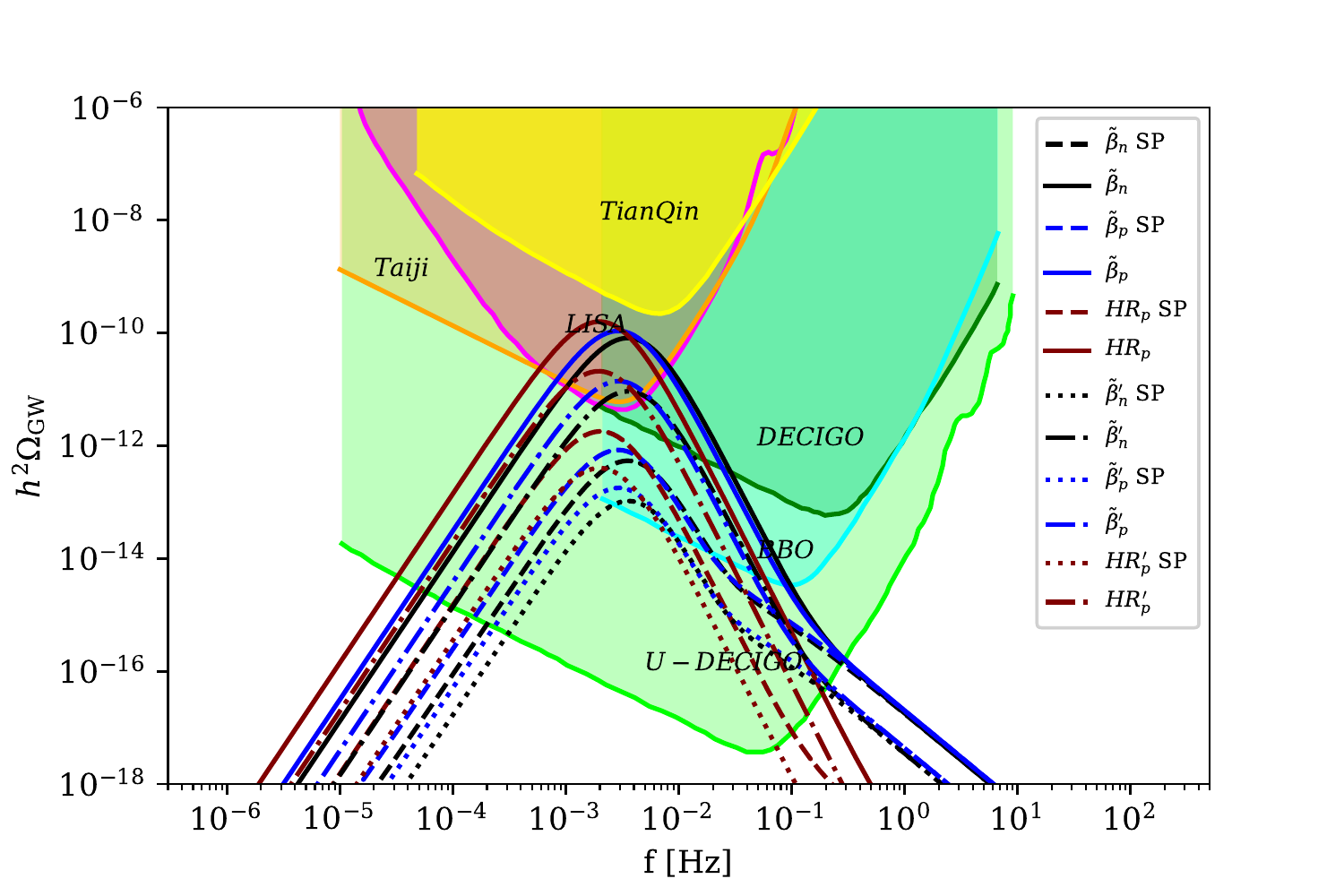}
		\end{minipage}
	}%
	\subfigure{
		\begin{minipage}[t]{0.5\linewidth}
			\centering
			\includegraphics[scale=0.5]{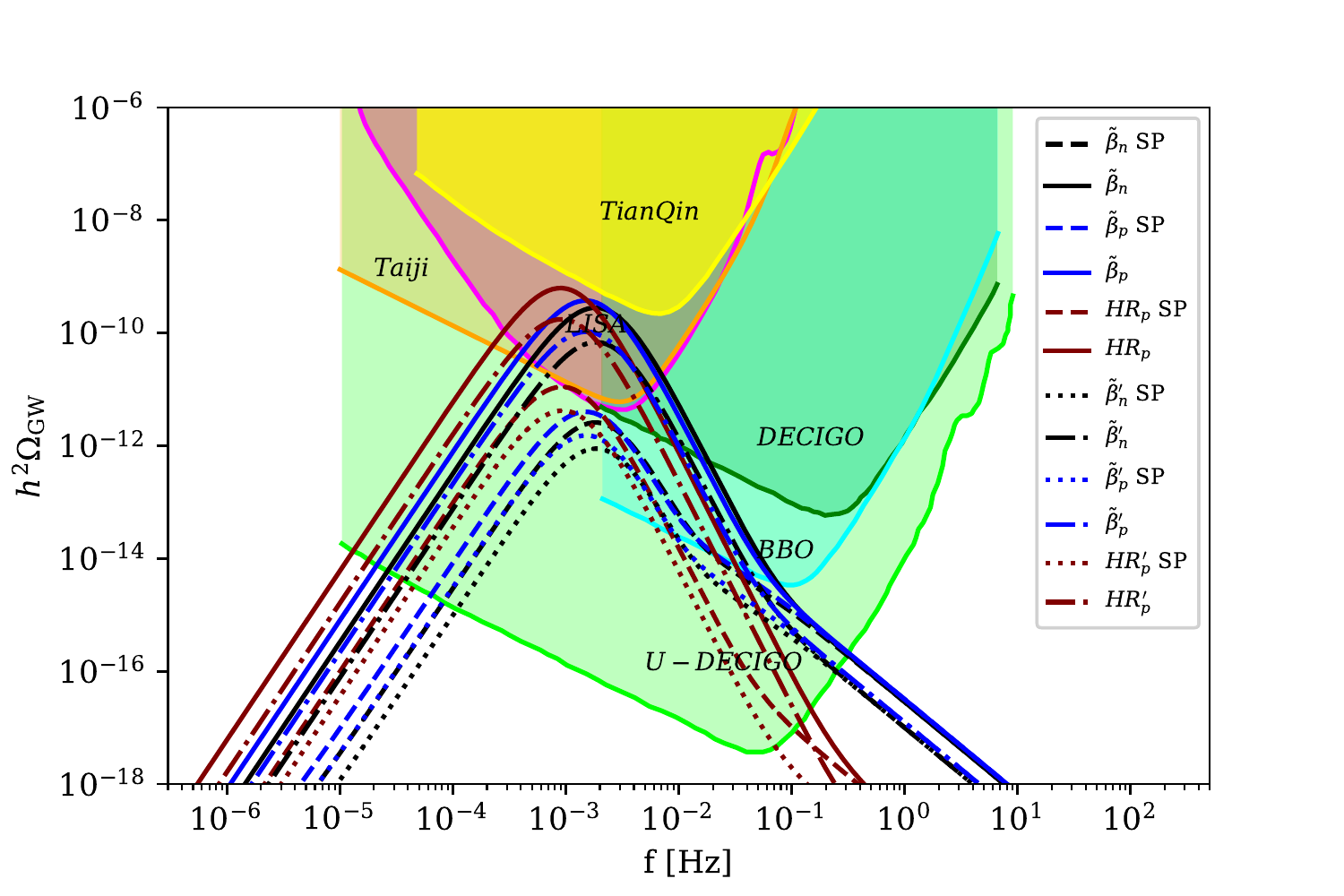}
		\end{minipage}
	}%
	\quad
	\subfigure{
		\begin{minipage}[t]{0.5\linewidth}
			\centering
			\includegraphics[scale=0.5]{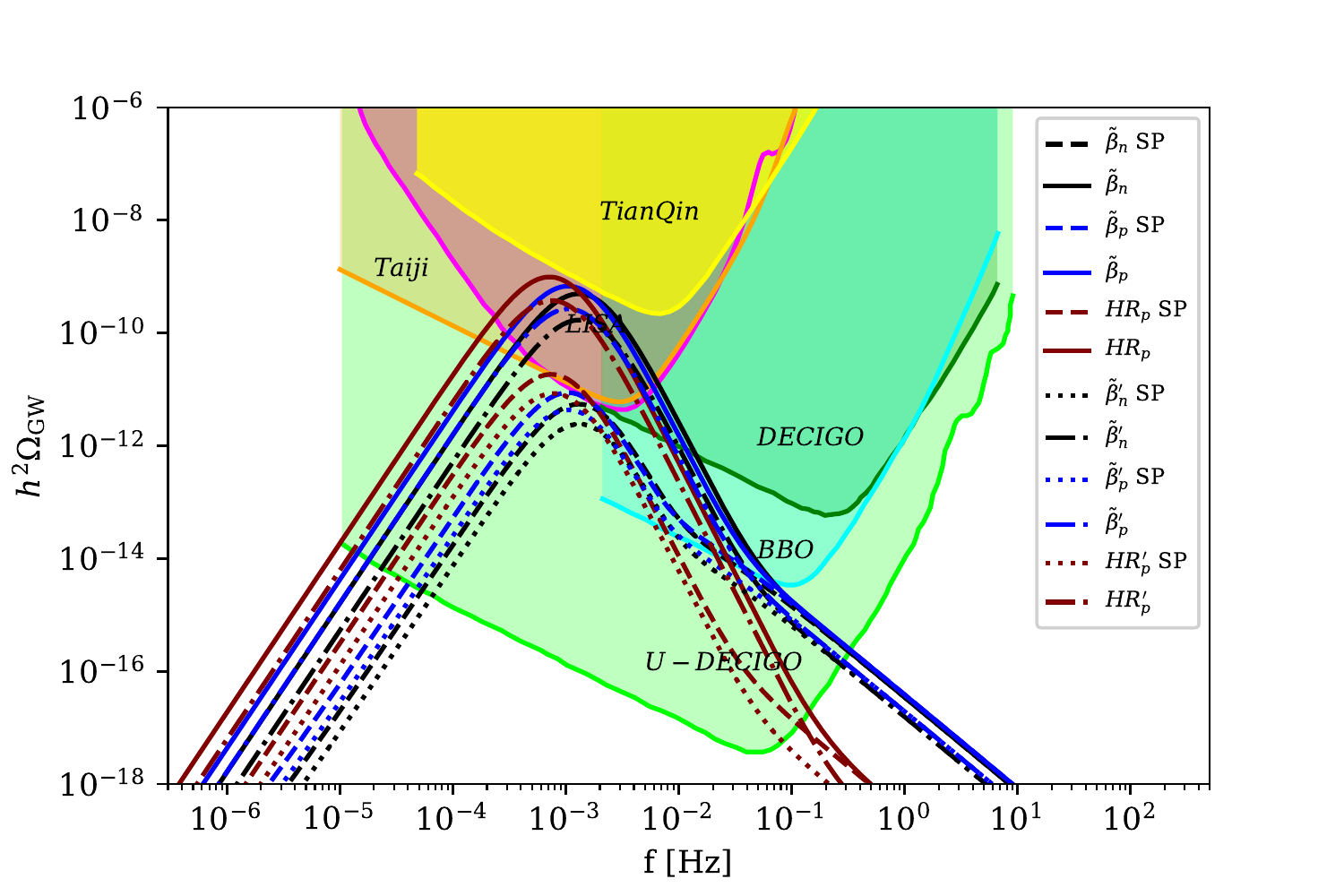}
		\end{minipage}
	}%
	\subfigure{
		\begin{minipage}[t]{0.5\linewidth}
			\centering
			\includegraphics[scale=0.5]{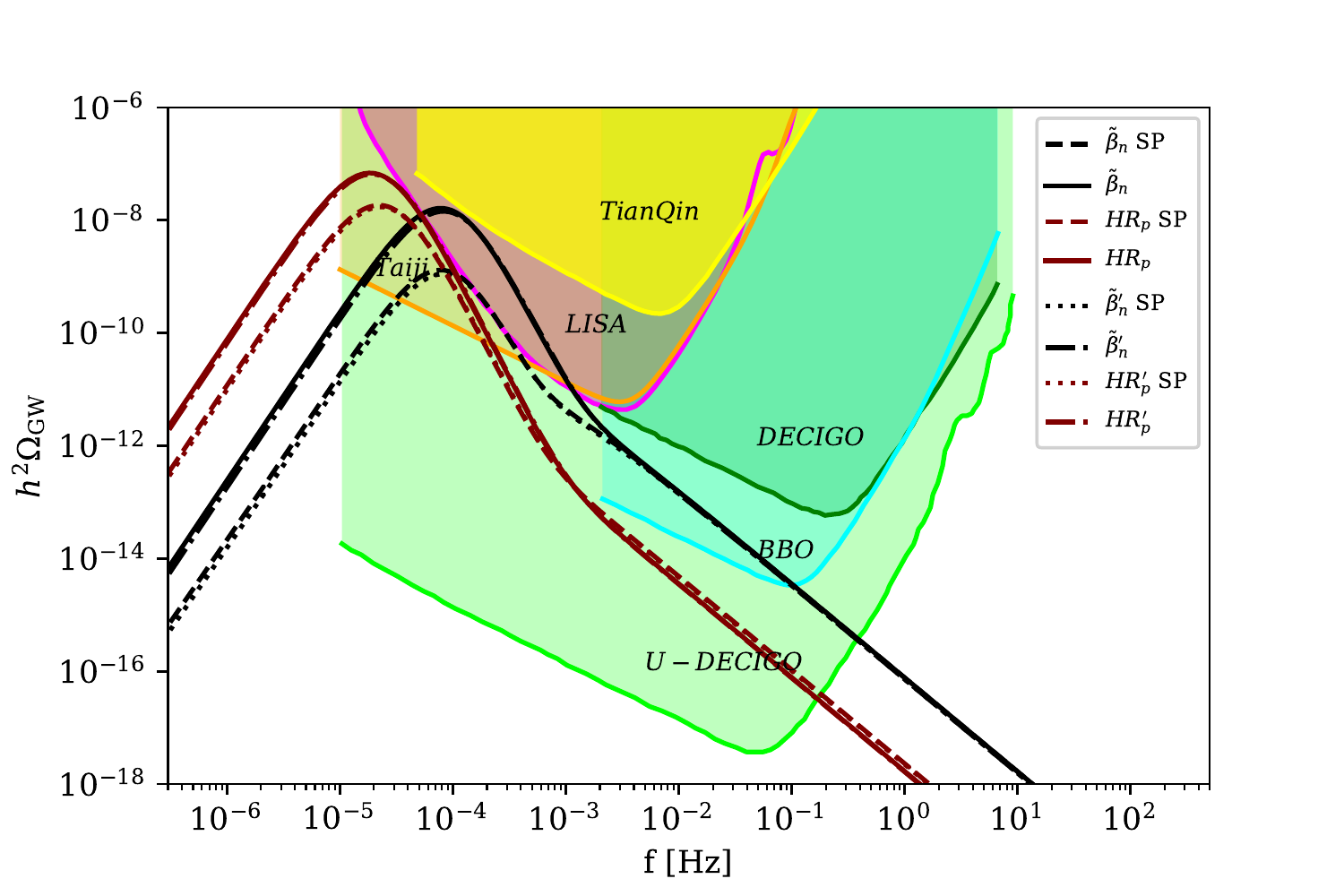}
		\end{minipage}
	}%
	\centering
	\caption{The GW spectra of the quartic toy model with $A\ne0$ for different benchmark sets.
The upper left plot and upper right plot show the GW spectrum of $BP_6$ and $BP_6^\prime$.
The middle left plot denotes the GW spectra of $BP_7$ and $BP_7^\prime$.
The middle right plot denotes the GW spectra of $BP_8$ and $BP_8^\prime$.
The bottom left plot shows the GW spectra of $BP_9$ and $BP_9^\prime$.
The bottom left plot represents the GW spectra of $BP_{10}$ and $BP_{10}^\prime$.}\label{gwe3}
\end{figure}

In figure~\ref{gwe3}, we show the GW spectra of the quartic toy model with $A\ne0$ for different benchmark sets given in table~\ref{tb3} and table~\ref{tb4}.
The color shaded regions represent the expected sensitivity of the GW interferometers LISA, DECIGO, U-DECIGO, BBO, Taiji, and TianQin, respectively.
$\tilde{\beta}$ and $\tilde{\beta}^\prime$ denote the spectra that are derived by the approximated characteristic length scale at the nucleation and percolation temperature for different definitions of strength parameter.
$HR$ and $HR^\prime$ represent the spectra that are obtained by the strict calculation of the mean bubble separation at nucleation and percolation temperature for different definitions of strength parameter.
The upper left plot and upper right plot show the GW spectra of $BP_6$ and $BP_6^\prime$.
The middle left plot denotes the GW spectra of $BP_7$ and $BP_7^\prime$.
The middle right plot denotes the GW spectra of $BP_8$ and $BP_8^\prime$.
The bottom left plot shows the GW spectra of $BP_9$ and $BP_9^\prime$.
The bottom left plot represents the GW spectra of $BP_{10}$ and $BP_{10}^\prime$.
From Tables.~\ref{tb3} and \ref{tb4}, we find $\tilde{\beta}$ become negative at percolation temperature for $BP_{10}$ and $BP_{10}^\prime$.
The approximation of eq.~(\ref{HRappro}) is not applicable for this situation, but is still valid at the nucleation temperature.
Therefore, we show the resulting GW spectra with the approximation used at nucleation temperature and the strict calculation of mean bubble separation in the bottom right of figure~\ref{gwe3}.
As shown from the GW spectra of these benchmark sets, the suppression effect of the sound wave has a strong influence on the amplitude.
Hence, this suppression effect makes the parameters (e.g., $BP_6$, $BP_6'$) that was suppose to be detected by some specific GW experiment configuration (e.g., LISA, Taiji, etc.) become marginal.

\begin{figure}[t]
	\centering
	
	\subfigure{
		\begin{minipage}[t]{1\linewidth}
			\centering
			\includegraphics[scale=0.8]{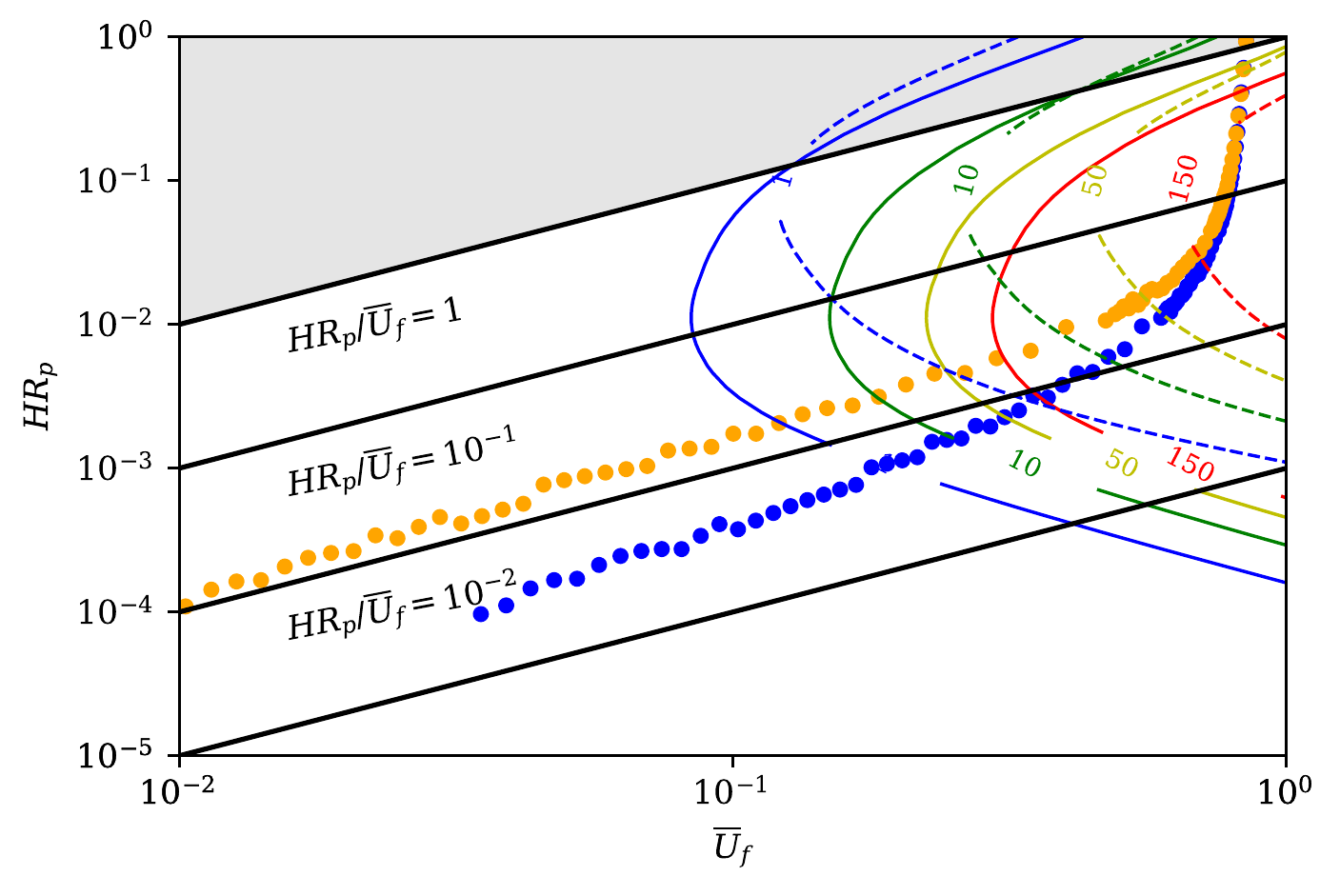}
		\end{minipage}%
	}%
	\caption{SNR for the quartic toy model ($A\ne0$) suppressed (colored dashed contours) and unsuppressed (colored solid contours) power spectra with LISA configuration and  $T_p = 50 $~GeV. The black solid lines show the magnitude of $HR_p/\overline{U}_f$.
The blue and orange dots are parameter points of the quartic toy model with strict calculation of the length scale $HR_p$ and $\overline{U}_f$, which are derived by the conventional and alternative definition of phase transition strength, respectively.}\label{snren0}
\end{figure}

In figure~\ref{snren0}, we show
SNR for the quartic toy model ($A\ne0$) suppressed (colored dashed lines) and unsuppressed (colored solid lines) power spectra with LISA configuration and  $T_p = 50 $~GeV.
The black solid lines show the magnitude of $HR_p/\overline{U}_f$.
The blue and orange dots are the selected parameter points of the quartic toy model with strict calculation of the length scale $HR_p$ and $\overline{U}_f$, which are derived by the conventional and alternative definition of phase transition strength, respectively.
For stronger phase transition, we can also obtain larger $HR_p$ and $\overline{U}_f$.
And for the allowed parameter space, the long-lasting sound wave ($HR_p/\overline{U}_f>1$) is very hard to be generated.

Compared to previous dimension-six effective model, the temperature hierarchy is not so large, but the
phase transition strength hierarchy is still obvious.
And it is much easier to realize ultra supercooling and generate more stronger phase transition in the allowed parameter space for this model.

\subsubsection{$A = 0$}

A large classes of particle physics models can be approximated to the form of effective potential given by eq.~\eqref{1prime} with $A=0$~\cite{Chung:2012vg,Huang:2017rzf,Huang:2017laj,Mohamadnejad:2019vzg,Huang:2017kzu,Huang:2019riv,Dev:2019njv}, such as the SM, the inert singlet, inert doublet model, minimal supersymmetry model and some hidden phase transition models.
In this case, $E$ usually comes from the cubic term of the thermal loop functions since
for the bosons, the thermal corrections contain terms like
\begin{equation}
J_b  \supset  \frac{-T}{12\pi}  m_b^2(\phi,T)^{3/2}\,\,.
\end{equation}
If the order parameter is the Higgs field, the value of $E$ should be small.
The first reason is that the couplings of new particles to Higgs boson are strongly constrained by current collider data.
The Higgs portal couplings should be small and then makes $E$ small.
The second reason is that the $E$ comes from the loop thermal effects.
Hence $E$ has a loop-factor suppression.
However, if the order parameter is not Higgs field, there are no such strong constraints and $E$ can be much larger, such as some hidden phase transition models from different motivations~\cite{Huang:2017laj,Huang:2017kzu}.

\begin{figure}[t]
	\centering
	
	\subfigure{
		\begin{minipage}[t]{1\linewidth}
			\centering
			\includegraphics[scale=0.8]{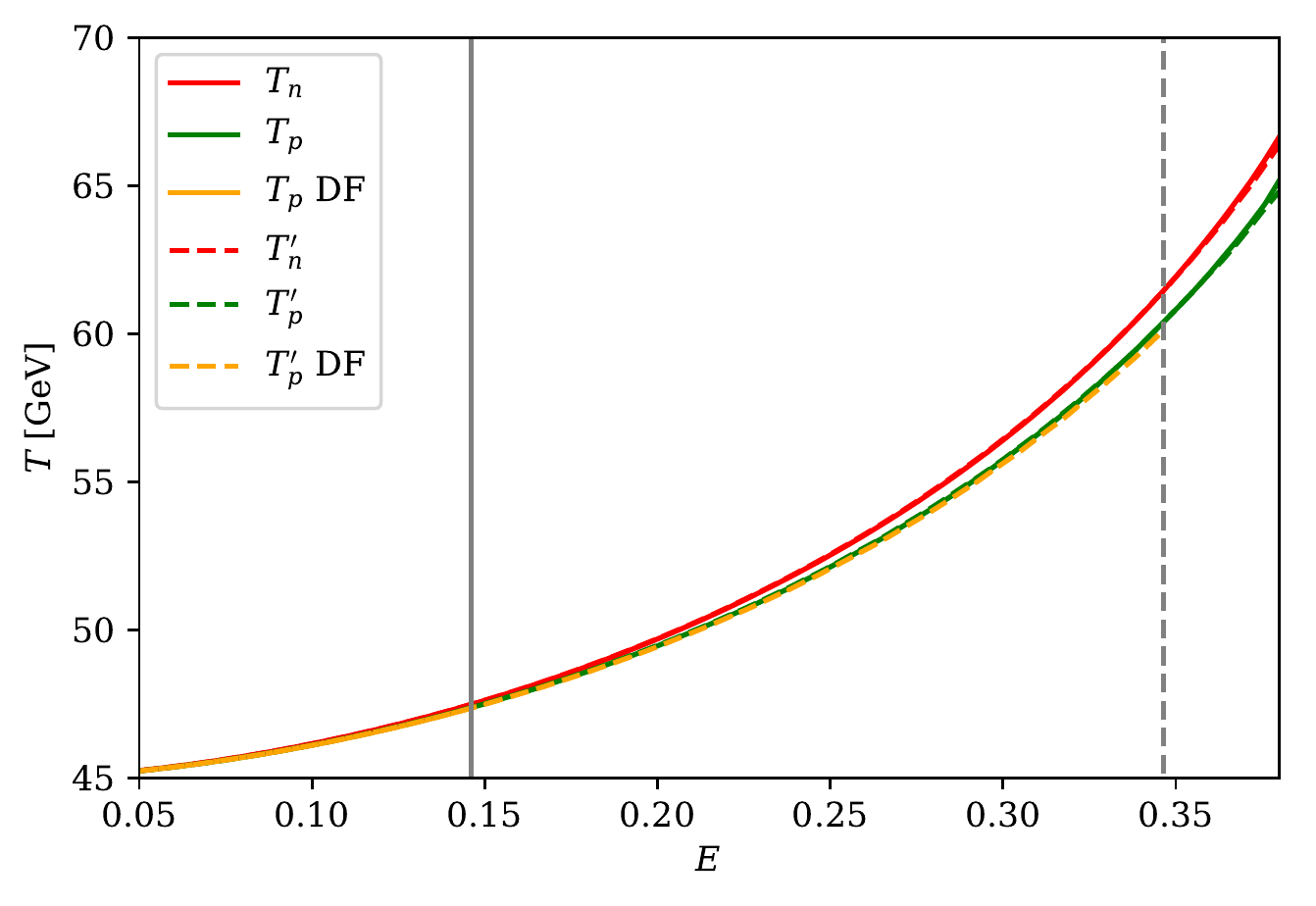}
		\end{minipage}%
	}%
	\centering
	\caption{The characteristic temperature of the quartic toy model ($A=0$). The colored solid lines represent various characteristic temperatures for the conventional definition $\alpha$.
	The colored dashed lines denote temperatures that is derived for the alternative definition $\alpha'$. The vertical gray solid and dashed lines indicate values of $E$ that produce a SFOPT with $\alpha_p=1$ and $\alpha_p^\prime=1$, respectively.}\label{cte3n}
\end{figure}

In figure~\ref{cte3n}, we show the characteristic temperature of the quartic toy model ($A=0$) for both definition of transition strength parameter (solid lines for conventional definition and dashed lines for alternative definition).
The vertical gray solid and dashed lines indicate values of $E$ that produce a SFOPT with $\alpha_p=1$ and $\alpha_p^\prime=1$, respectively.
Since the potential barrier is induced by the thermal effect and should disappear at some specific temperature, the phase transition must be completed.
We can see that larger value of $E$ induces larger hierarchy between the nucleation temperature $T_n$ and percolation temperature $T_p$.
However, this hierarchy is not significant compare to the previous two models.
Different bubble wall velocities and definitions of $\alpha $ just give negligible corrections to the characteristic temperature.

\begin{figure}[t]
	\centering
	
	\subfigure{
		\begin{minipage}[t]{1\linewidth}
			\centering
			\includegraphics[scale=0.8]{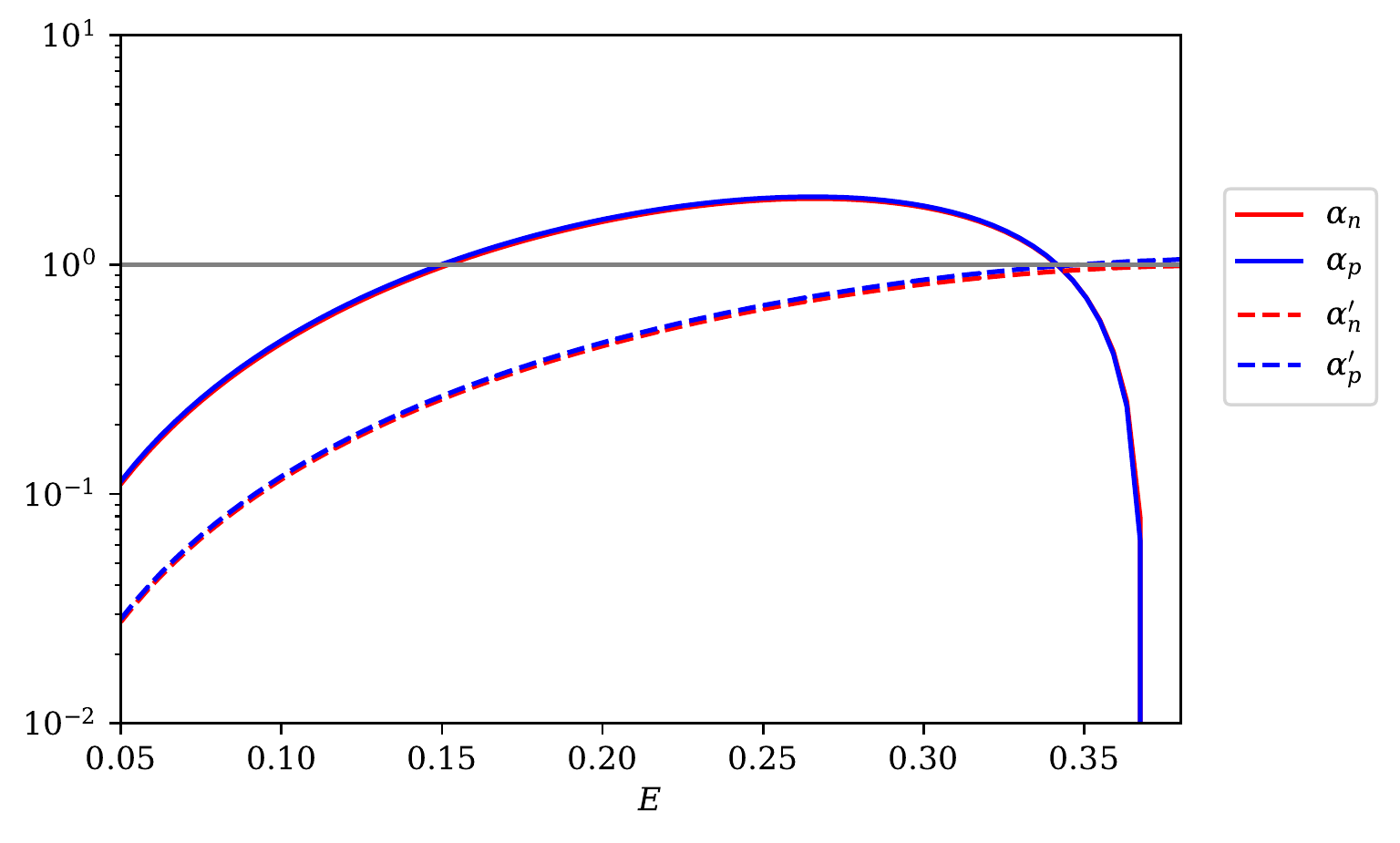}
		\end{minipage}%
	}%
	\centering
	\caption{The phase transition strength as a function of $E$ at the nucleation and percolation temperature for the quartic toy model ($A=0$) with different definitions.
		Solid lines represent the values with conventional definition.
		Dashed lines denote the values with alternative definition.
	The horizontal gray line indicates that the phase transition strength parameters are equal to one.}\label{ste3n}
\end{figure}

We present the phase transition strength as a function of $E$ at the nucleation and percolation temperature for different definitions in figure~\ref{ste3n}.
Solid lines represent the values with conventional definition.
Dashed lines denote the values with alternative definition.
We find there exists a strange behavior for the strength parameters calculated by conventional definition.
As shown in figure~\ref{ste3n}, the strength parameter of the conventional definition first increases with the increasing value of $E$, then decreases and even becomes negative for large enough $E$.
According to the conventional definition eq.~\eqref{convenalpha}, we find $TdV_{\rm eff}/dT$ can become dominant for a large $E$.
For this situation, the released vacuum energy (the numerator of eq.~\eqref{convenalpha}) might be negative, this is definitely problematic.
Therefore, the conventional definition of phase transition strength based on the bag EoS is not appropriate for large enough $E$.
That indicates the reliable EoS has a large deviation from the bag EoS, and we should use the alternative definition of phase transition strength which is more appropriate for this case.
Hence, for a model that can generate a large thermal correction,\footnote{Especially, for models contains large number of extra bosons.} we should be carefully choose the definition of phase transition strength.
In this model, it is more appropriate to use the alternative definition.
It is hard to realize ultra supercooling in this model since the phase transition usually becomes weaker without the zero-temperature potential barrier.

\begin{figure}[t]
	\centering
	
	\subfigure{
		\begin{minipage}[t]{1\linewidth}
			\centering
			\includegraphics[scale=0.8]{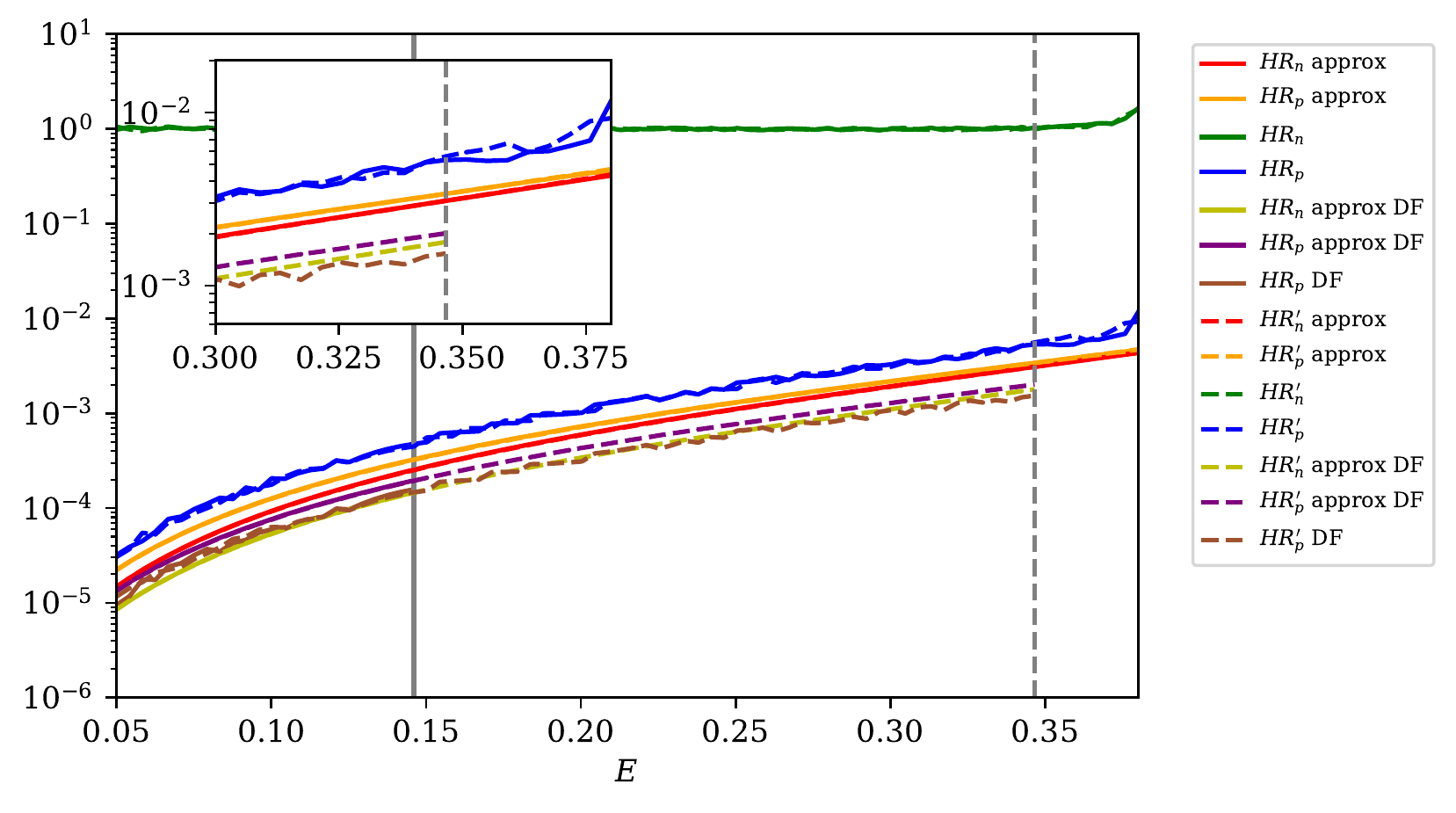}
		\end{minipage}%
	}%
	\centering
	\caption{The characteristic length scale of the quartic toy model ($A=0$). $HR_{*,ap}$ (red, orange, yellow, and purple solid lines), which is derived by the approximation, and $HR_*$ (green, blue, and brown solid lines), which is calculated by the first principle, denote the characteristic length at the nucleation and percolation temperature.
The dashed lines denote the values of characteristic length at different temperatures for different definitions of the phase transition strength.
The vertical gray solid and dashed line are the separation of the regimes of strong supercooling and ultra supercooling for both definition.}\label{hre3n}
\end{figure}

Figure~\ref{hre3n} depicts the characteristic length scale for quartic toy model ($A=0$).
$HR_{*,ap}$ (red, orange, yellow, and purple solid lines), which is derived by the approximation, and $HR_*$ (green, blue, and brown solid lines), which is calculated by eq.~\eqref{HR}, denotes the characteristic length at the nucleation and percolation temperature.
The dashed lines denote the values of characteristic length at different temperatures for different definitions of phase transition strength.
The yellow, purple and brown lines show the deflagration expansion mode ($v_b = 0.3$) gives negligible modifications to the characteristic length scale for regime of a SFOPT with $\alpha < 1$.
The vertical solid and dashed gray lines are the separation of the regions of a SFOPT with strong supercooling and ultra supercooling at the percolation temperature for both definitions.
According to figure~\ref{ste3n}, there exists a parameter region with large value of $E$ that can generate $\alpha_p < 1$ and $\alpha_n < 1$.
Since it is problematic for using the conventional definition of phase transition strength, we do not consider this regime here.
In this model, we can also find the approximated length scale fits the strict calculation very well for the deflagration mode.
However, the approximation for the detonation mode is less well compared to deflagration mode.

\begin{figure}[t]
	\centering
	
	\subfigure{
		\begin{minipage}[t]{0.5\linewidth}
			\centering
			\includegraphics[scale=0.5]{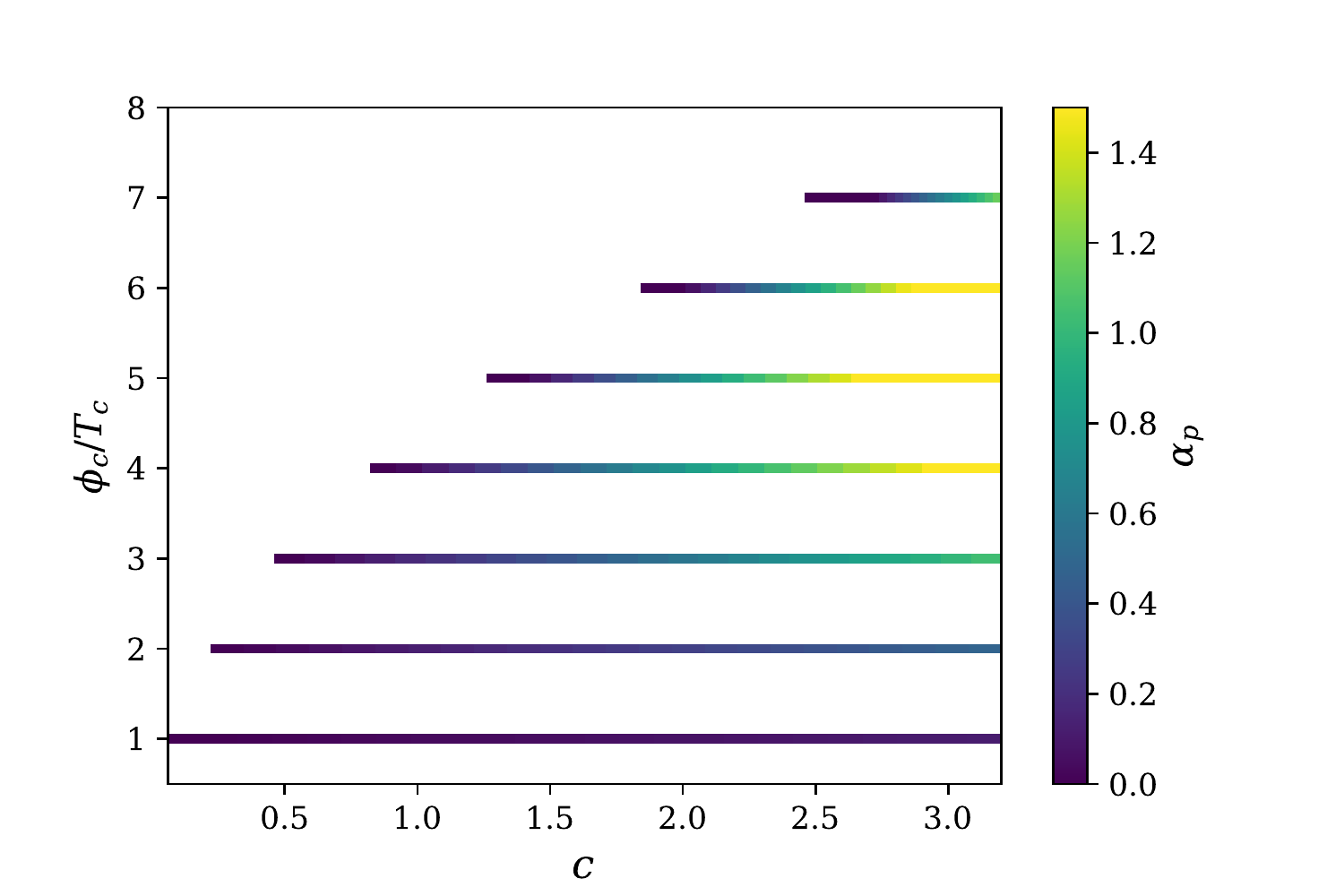}
		\end{minipage}%
	}%
	\subfigure{
		\begin{minipage}[t]{0.5\linewidth}
			\centering
			\includegraphics[scale=0.5]{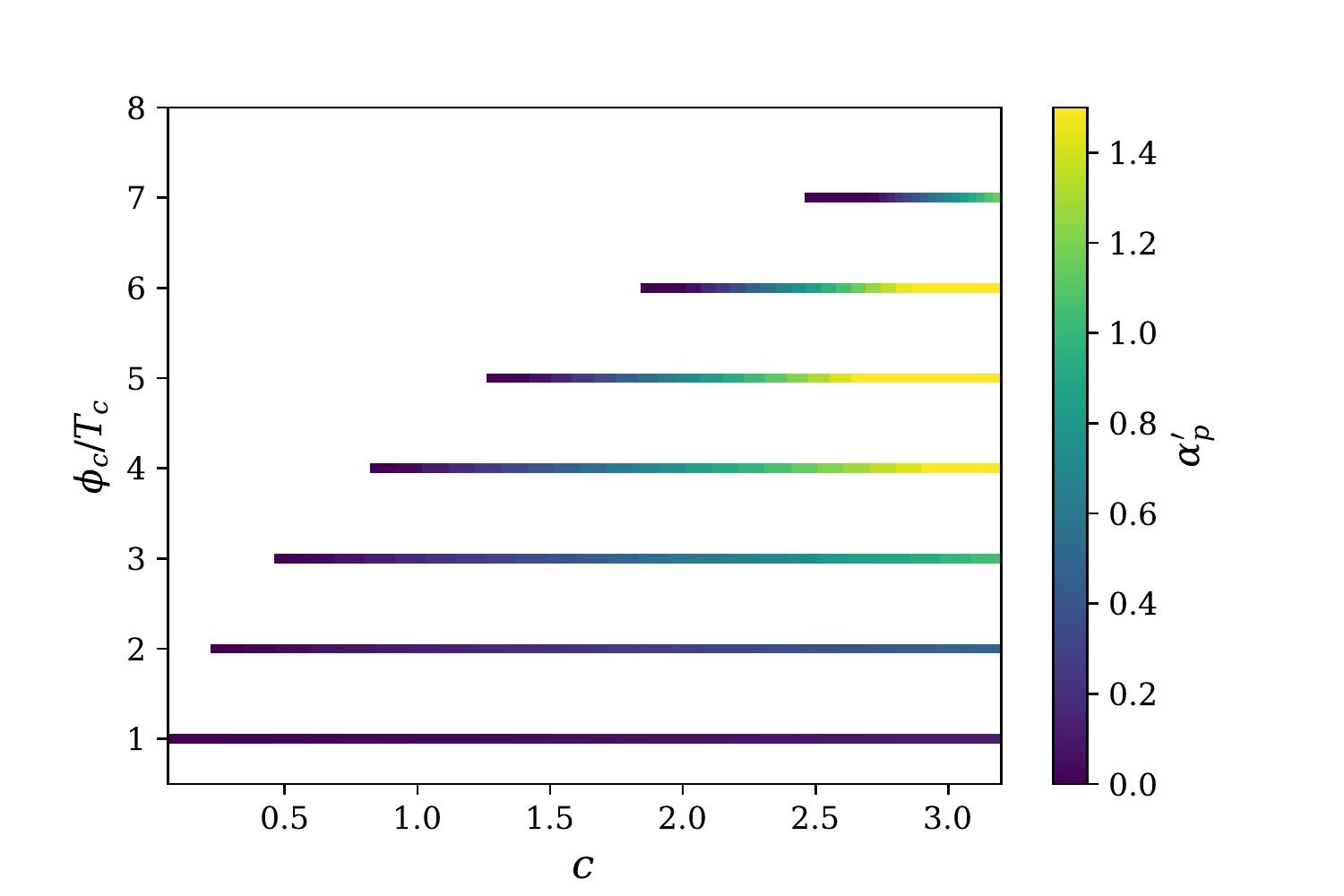}
			%\caption{fig2}
		\end{minipage}%
	}%
	\centering
	\caption{The relation between the washout parameter $\phi_c/T_c$, the thermal coupling $c$, and the phase transition phase transition strength for the quartic toy model with $A=0$.
		The multicolored lines (from bottom to top) for $E=0.05,0.1,0.15,0.2,0.25,0.3,0.35$.}\label{vte30}
\end{figure}

Figure~\ref{vte30} represents the relation between the washout parameter $\phi_c/T_c$, the thermal coupling $c$ and the phase transition strength for the quartic toy model ($A=0$).
The multicolored lines denote $E=0.05,0.1,0.15,0.2,0.25,0.3,0.35$ from bottom to top ($\lambda$ is set as 0.1).
For this model, we find even the washout parameter is large, a relatively small $c$ can not produce a ultra supercooling case.
Therefore, we can conclude that only both the washout parameter and the thermal coupling are large enough, a supercooling case can be generated.

In table~\ref{tb5}, we show the phase transition parameters of the quartic toy model ($A=0$, $c$ and $\lambda$ are set as 1.5 and 0.1 respectively) for the conventional definition of $\alpha$ for four benchmark sets.
For the first three benchmark sets, $\alpha_p$ is larger than $\alpha_n$.
For the last benchmark set, $\alpha_p$ is smaller than $\alpha_n$.
This strange behavior originates from the same reason as in the discussion of figure~\ref{ste3n}.
In this situation, the conventional definition of the phase transition strength is not appropriate.
We also show the phase transition parameters of the quartic toy model ($A=0$. $c$ and $\lambda$ are set as 1.5 and 0.1 respectively) for the alternative definition $\alpha'$ of four benchmark sets in table~\ref{tb6}.
The alternative definition is more reliable for this model.
For all the four benchmark sets, $\alpha_p^{\prime}$ is larger than $\alpha_n^{\prime}$.

\begin{table}[t]\small%
	\centering
	\begin{tabular}{|cccccccccc|}
		\hline
		& $E$  & $T_n$ [GeV] & $T_p$ [GeV] & $\alpha_n$ & $\alpha_p$ & $\tilde{\beta}_n$ & $\tilde{\beta}_p$ & $HR_p$ & $v_b$\\
		\hline
		\multirow{2}*{$BP_{11}$}& \multirow{2}*{0.134}&47.066 &46.979&0.793&0.814&14664.855&11265.542&0.000387&1\\
		&&47.066&46.965&0.793&0.817&14664.855&10816.941&0.000125&0.3\\
		$BP_{12}$&0.150&47.620&47.504&0.981&1.006&10707.093&8359.450&0.000495&1\\
		$BP_{13}$&0.209&50.121&49.867&1.625&1.662&4388.973&3630.405&0.00128&1\\
		$BP_{14}$&0.359&63.193&61.939&0.418&0.408&835.652&765.515&0.00530&1\\
		%$BP_{10}$&9.081&18.623&15.439&8.870&16.797&47.111&-10.117&0.218&1\\
		\hline
	\end{tabular}
    \caption{The phase transition parameters of the quartic toy model ($A=0$. $c$ and $\lambda$ are set as 1.5 and 0.1 respectively) for four benchmark sets with the conventional definition $\alpha$. }\label{tb5}
\end{table}

\begin{table}[t]\small%
	\centering
	\begin{tabular}{|cccccccccc|}
		\hline
		& $E$ & $T_n^\prime$ [GeV] & $T_p^\prime$ [GeV] & $\alpha_n^\prime$ & $\alpha_p^\prime$ & $\tilde{\beta}_n^\prime$ & $\tilde{\beta}_p^\prime$ & $HR_p^\prime$ & $v_b$\\
		\hline
		\multirow{2}*{$BP_{11}^\prime$}& \multirow{2}*{0.134}&47.069&46.982&0.207&0.213&14767.245&11358.500&0.000374&1\\
		&&47.069&46.967&0.207&0.214&14767.245&10904.556&0.000120&0.3\\
		$BP_{12}^\prime$&0.150&47.624&47.511&0.260&0.268&10798.378&8487.286&0.000555&1\\
		$BP_{13}^\prime$&0.209&50.133&49.884&0.476&0.492&4432.255&3674.360&0.00130&1\\
		$BP_{14}^\prime$&0.359&63.136&61.939&0.967&1.022&832.363&765.515&0.00664&1\\
		%$BP_{10}^\prime$&9.081&18.743&15.569&7.344&15.285&49.521&-7.961&0.212&1\\
		\hline
	\end{tabular}
    \caption{The phase transition parameters of the quartic toy model ($A=0$. $c$ and $\lambda$ are set as 1.5 and 0.1 respectively) of four benchmark sets with the alternative definition $\alpha'$.}\label{tb6}
\end{table}

\begin{figure}[t]
	\centering
	
	\subfigure{
		\begin{minipage}[t]{0.5\linewidth}
			\centering
			\includegraphics[scale=0.55]{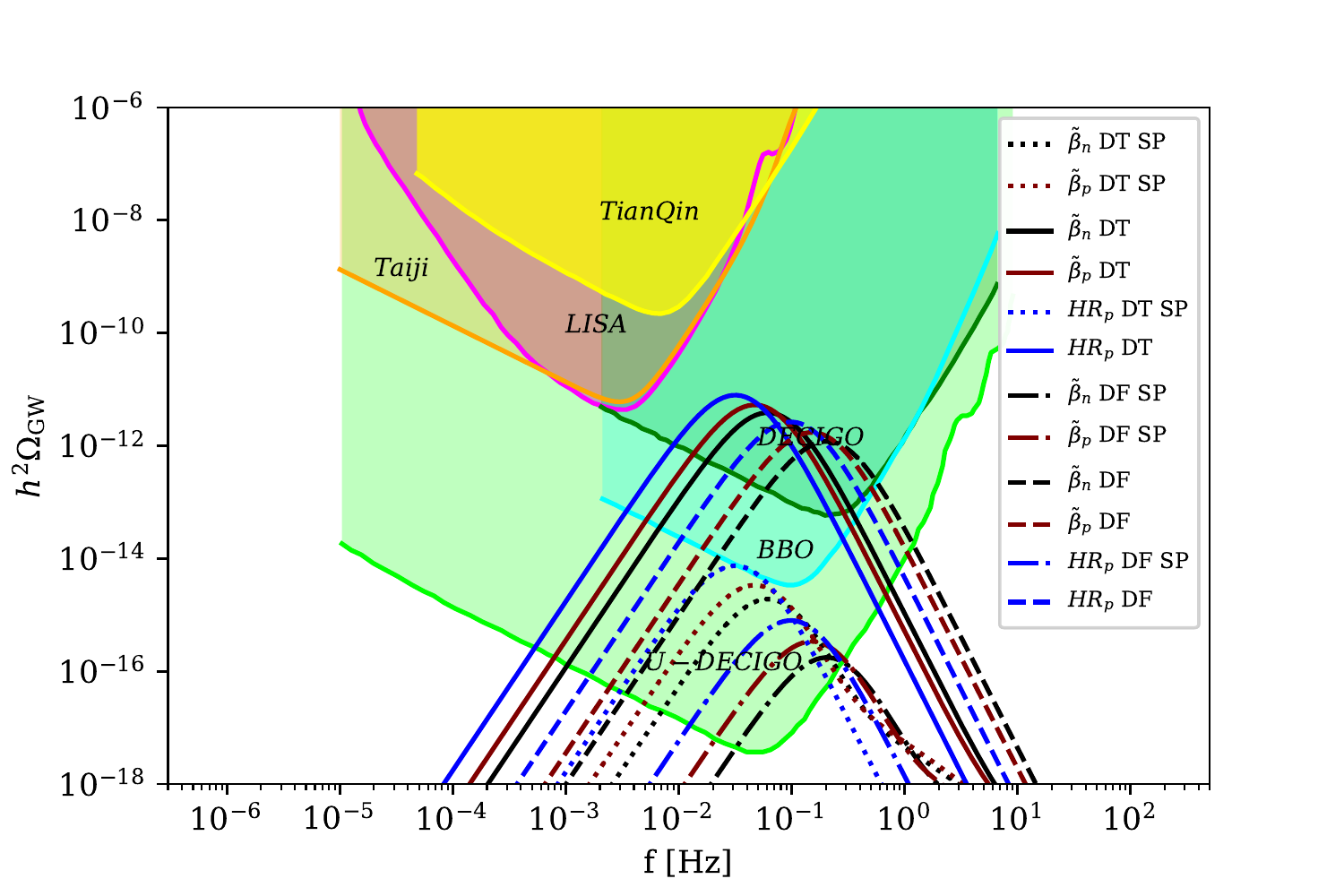}
		\end{minipage}%
	}%
	\subfigure{
		\begin{minipage}[t]{0.5\linewidth}
			\centering
			\includegraphics[scale=0.55]{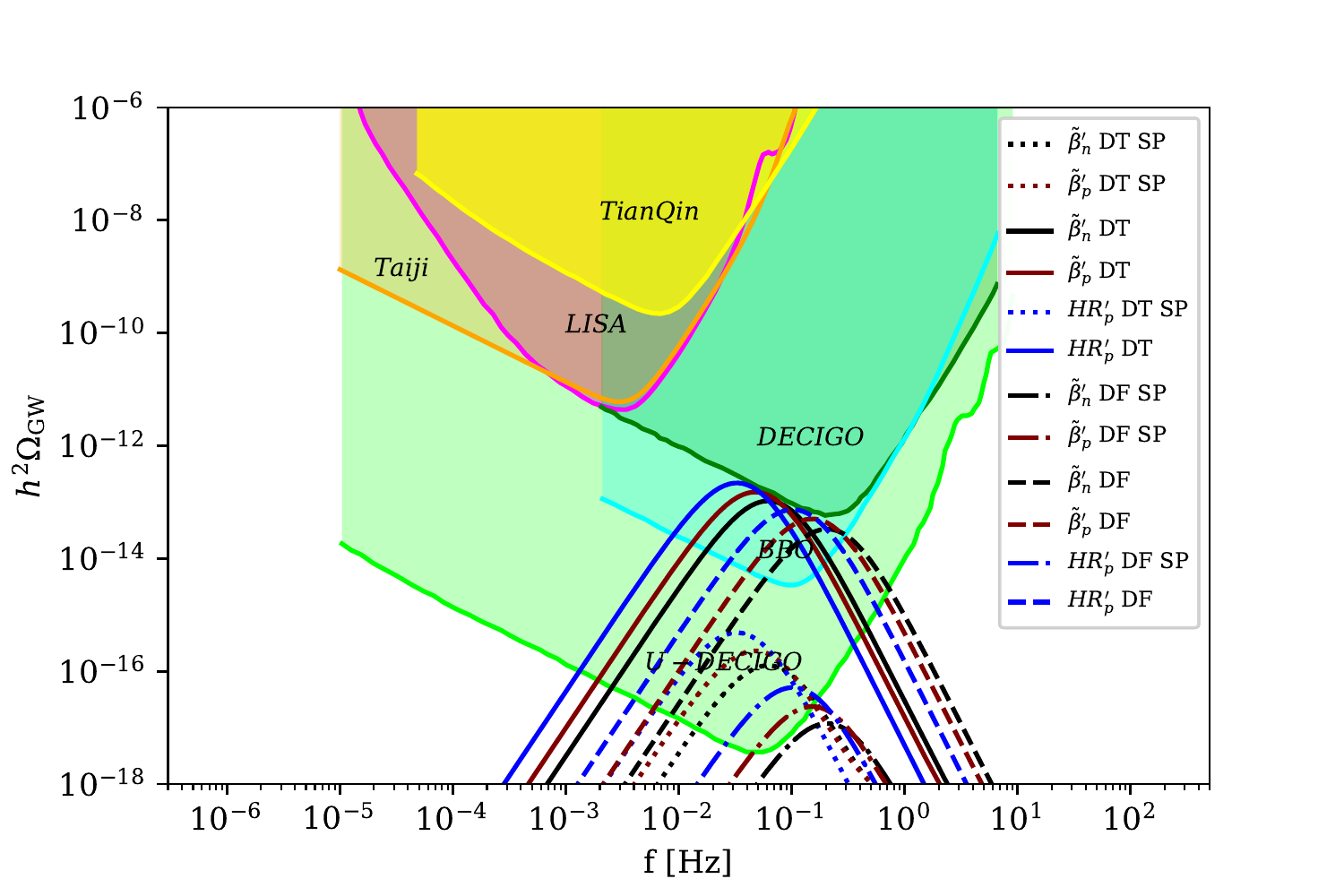}
			%\caption{fig2}
		\end{minipage}%
	}%
	\quad
	\subfigure{
		\begin{minipage}[t]{0.5\linewidth}
			\centering
			\includegraphics[scale=0.55]{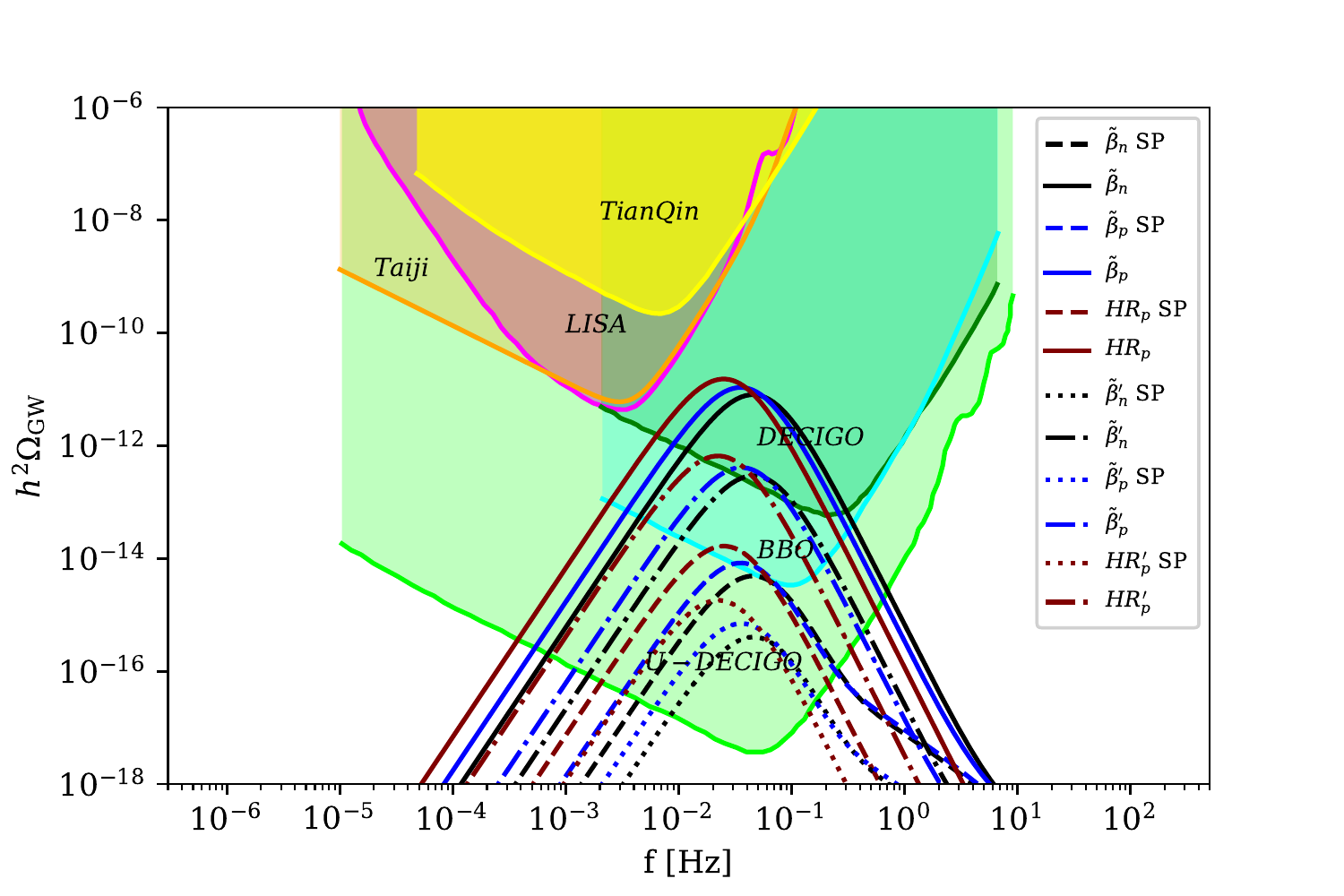}
		\end{minipage}
	}%
	\subfigure{
		\begin{minipage}[t]{0.5\linewidth}
			\centering
			\includegraphics[scale=0.55]{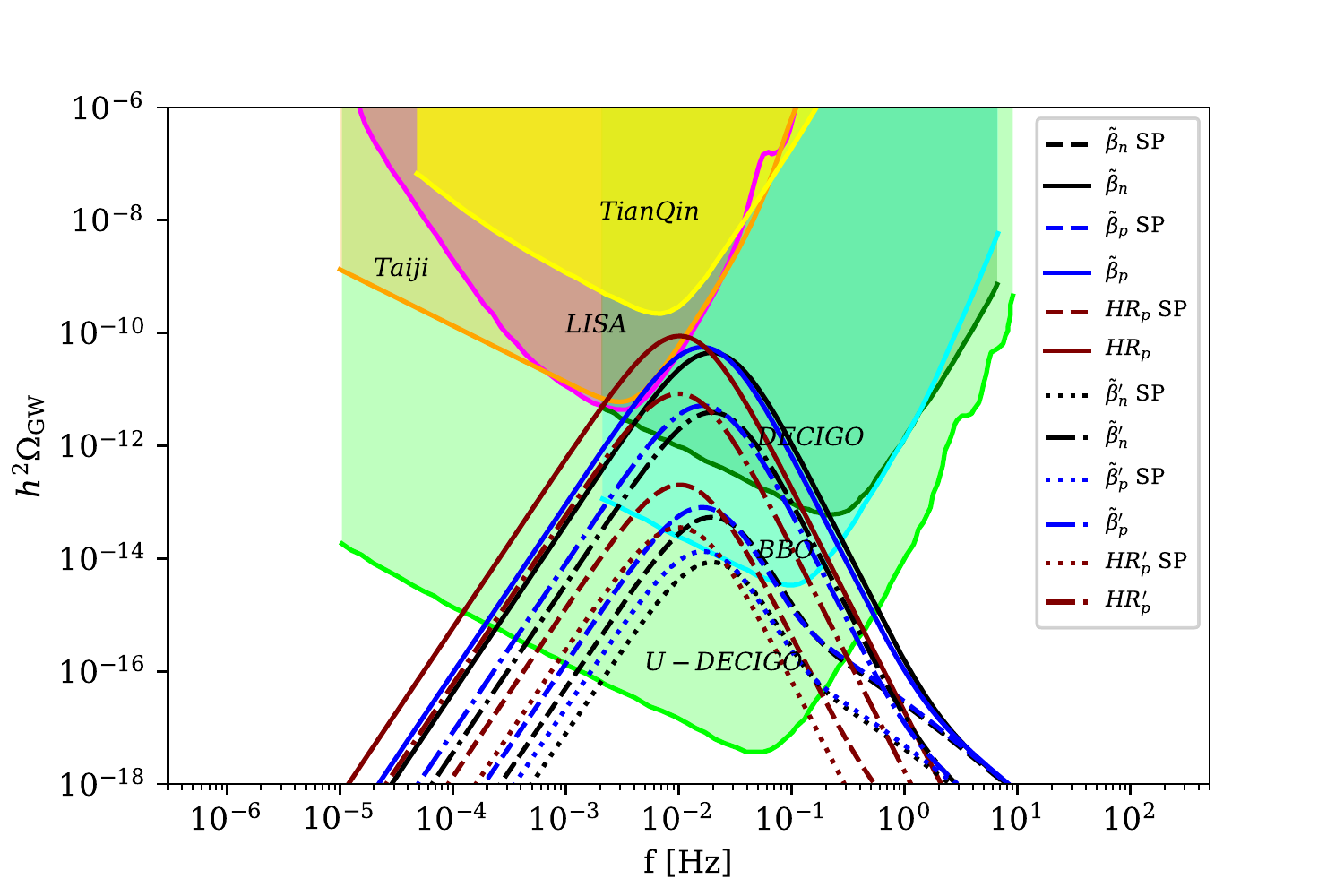}
		\end{minipage}
	}%
	\quad
	\subfigure{
		\begin{minipage}[t]{1\linewidth}
			\centering
			\includegraphics[scale=0.55]{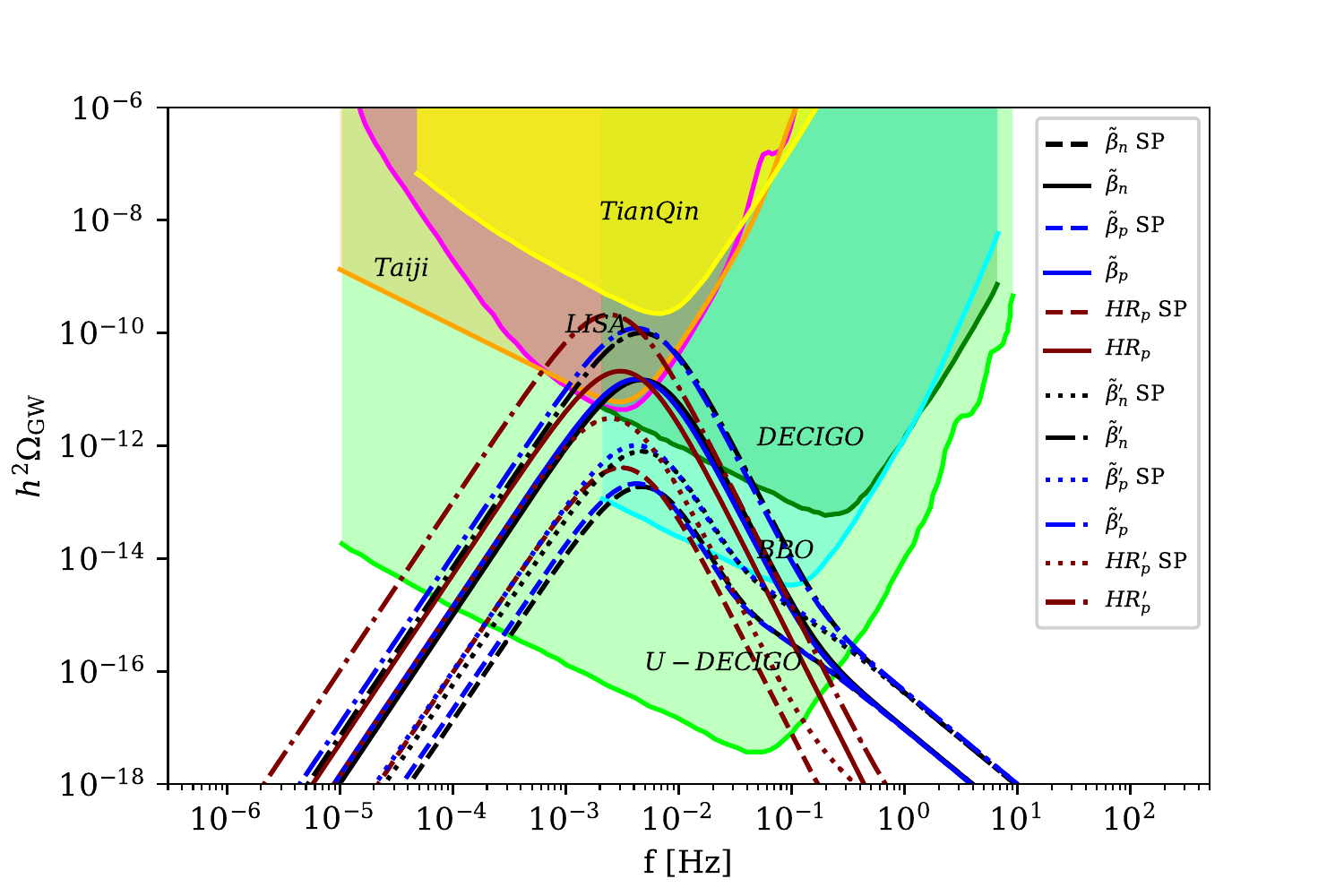}
		\end{minipage}
	}%
	\centering
	\caption{The GW spectra of the quartic toy model with $A=0$ for different benchmark sets.
	The upper left plot and upper right plot show the GW spectra of $BP_{11}$ and $BP_{12}^\prime$.
   The middle left plot denotes the GW spectra of $BP_{12}$ and $BP_{12}^\prime$.
	The middle right plot denotes the GW spectra of $BP_{13}$ and $BP_{13}^\prime$.
 The bottom  plot shows the GW spectra of $BP_{14}$ and $BP_{14}^\prime$.
		}\label{qa0gw}
\end{figure}

Figure~\ref{qa0gw} shows the GW spectra of quartic toy model with $A=0$ for different benchmark sets given in table~\ref{tb5} and table~\ref{tb6}.
The upper left plot and upper right plot show the GW spectrum of $BP_{11}$ and $BP_{11}^\prime$.
The middle left plot denotes the GW spectra of $BP_{12}$ and $BP_{12}^\prime$.
The middle right plot denotes the GW spectra of $BP_{13}$ and $BP_{13}^\prime$.
The bottom  plot shows the GW spectra of $BP_{14}$ and $BP_{14}^\prime$.
As shown in figure~\ref{ste3n}, there exist a range of $E$ that the value of $\alpha'$ is larger than the value of $\alpha$ at the nucleation and percolation temperature.
The $BP_{14}$ and $BP_{14}'$ shows that phenomenon, and the GWs spectra for these two benchmark sets give a behavior that are different from the GW spectra of other benchmark set.
Usually, for the same model parameter $\Lambda/\sqrt{\kappa}$, $A$, and $E$, the phase transition parameter with conventional definition $\alpha$ induce a stronger GW signal.
However, $BP_{14}'$ gives a stronger signal than $BP_{15}$.
This is caused by the large deviation from the bag EoS as we discussed before.
Therefore, we should choose a proper definition of the phase transition strength to give a more reliable result.

\begin{figure}[t]
	\centering
	
	\subfigure{
		\begin{minipage}[t]{1\linewidth}
			\centering
			\includegraphics[scale=0.8]{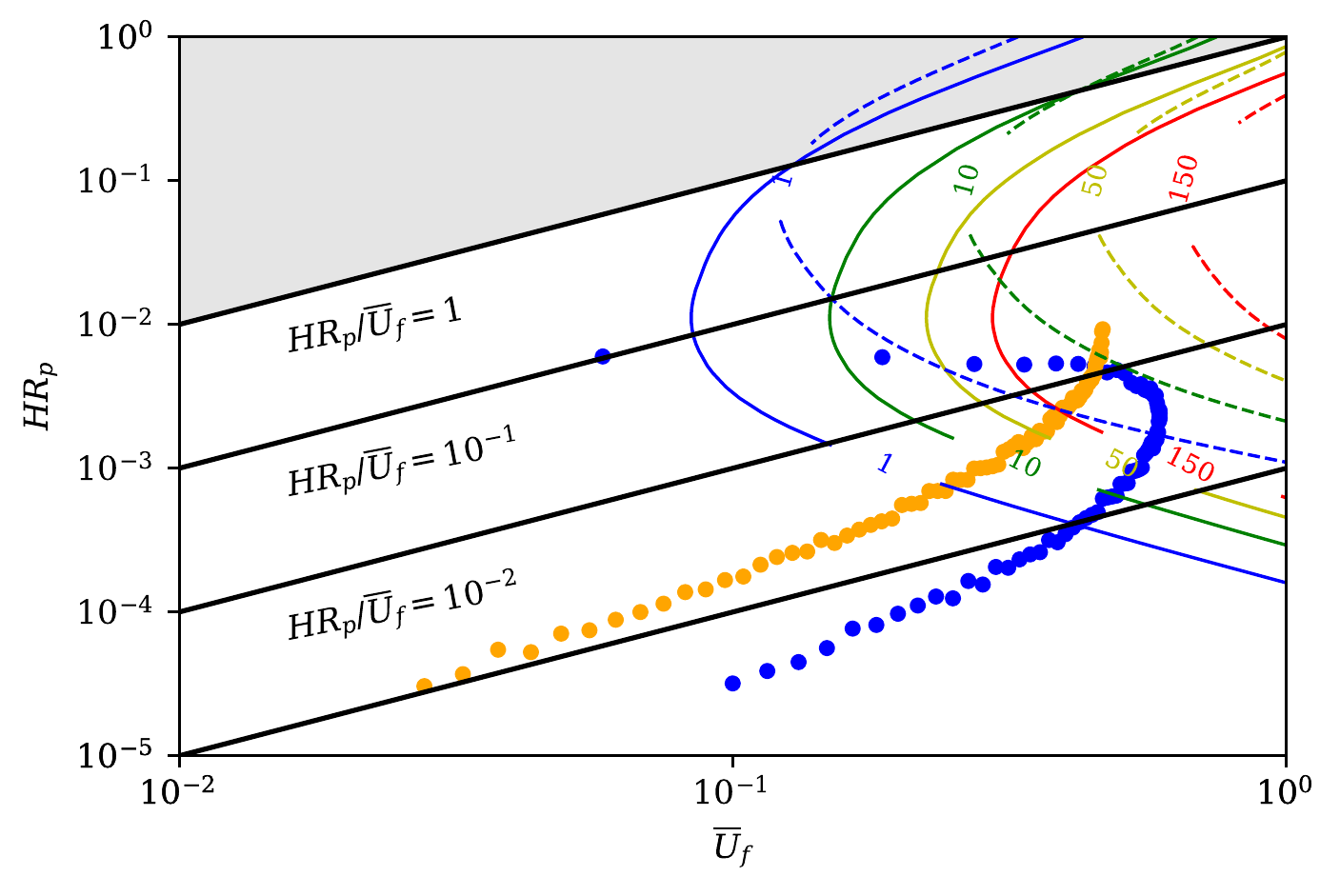}
		\end{minipage}%
	}%
	\caption{
SNR for the quartic toy model ($A=0$) suppressed (colored dashed lines) and unsuppressed (colored solid lines) power spectra with LISA configuration and  $T_p = 50 $~GeV.
The black solid lines show the magnitude of $HR_p/\overline{U}_f$.
The blue and orange dots are parameter points of the quartic toy model, with strict calculation of the length scale $HR_p$, are derived by the conventional and alternative definition of the phase transition strength, respectively.}\label{snre3}
\end{figure}

We present the SNR for the quartic toy model ($A=0$) suppressed (colored dashed lines) and unsuppressed (colored  solid lines) power spectrum with LISA configuration and  $T_p = 50 $~GeV in figure~\ref{snre3},.
The black solid lines show the magnitude of $HR_p/\overline{U}_f$.
The blue and orange dots are parameter points of the quartic toy model, with strict calculation of the length scale $HR_p$ and $\overline{U}_f$ are derived by the conventional and alternative definition of the phase transition strength, respectively.
Since the conventional definition of the phase transition strength are problematic for this model, the strange behavior of the blue dot are induced by the same reason that we discussed in the above.

\subsection{Logarithm model}
In this section, we discuss the logarithm model with the following
effective potential
\begin{equation}
V_{\rm eff}(\phi, T) \approx \frac{\mu^2 + cT^2}{2}\phi^2 + \frac{\lambda}{4}\phi^4 + \frac{\kappa}{4}\phi^4\ln\frac{\phi^2}{Q^2}\,\,,
\end{equation}
where $Q$ is the renormalization scale.
We can set the vacuum expectation value (VEV) equals the Higgs VEV $v$ at zero temperature.
This type of effective potential can come from loop corrections of singlet scalar~\cite{Espinosa:2007qk,Espinosa:2008kw}, singlet Majoron~\cite{Kondo:1991jz,Sei:1992np,Addazi:2017nmg} and Two-Higgs doublets models~\cite{Cline:1996mga,Fromme:2006cm,Hambye:2007vf}. One simple example is
the SM extended by inert singlet (no quartic terms of the singlet fields) fields as the following~\cite{Espinosa:2007qk}
\begin{equation}
\delta \mathcal{L}=-\kappa^2 \phi^2 S_i^2  \,\,.
\end{equation}
The one-loop correction at zero temperature can generate the logarithm term.

The model parameters $\mu^2$ and $\lambda$ can be expressed with the SM parameters and Higgs VEV $v$
\begin{equation}
\lambda = \lambda_{SM} - \kappa\left(\ln\frac{v^2}{Q^2} + \frac{3}{2}\right),
\end{equation}
\begin{equation}
\mu^2 = \mu_{SM} + \kappa v^2.
\end{equation}

\begin{figure}[t]
	\centering
	
	\subfigure{
		\begin{minipage}[t]{1\linewidth}
			\centering
			\includegraphics[scale=0.8]{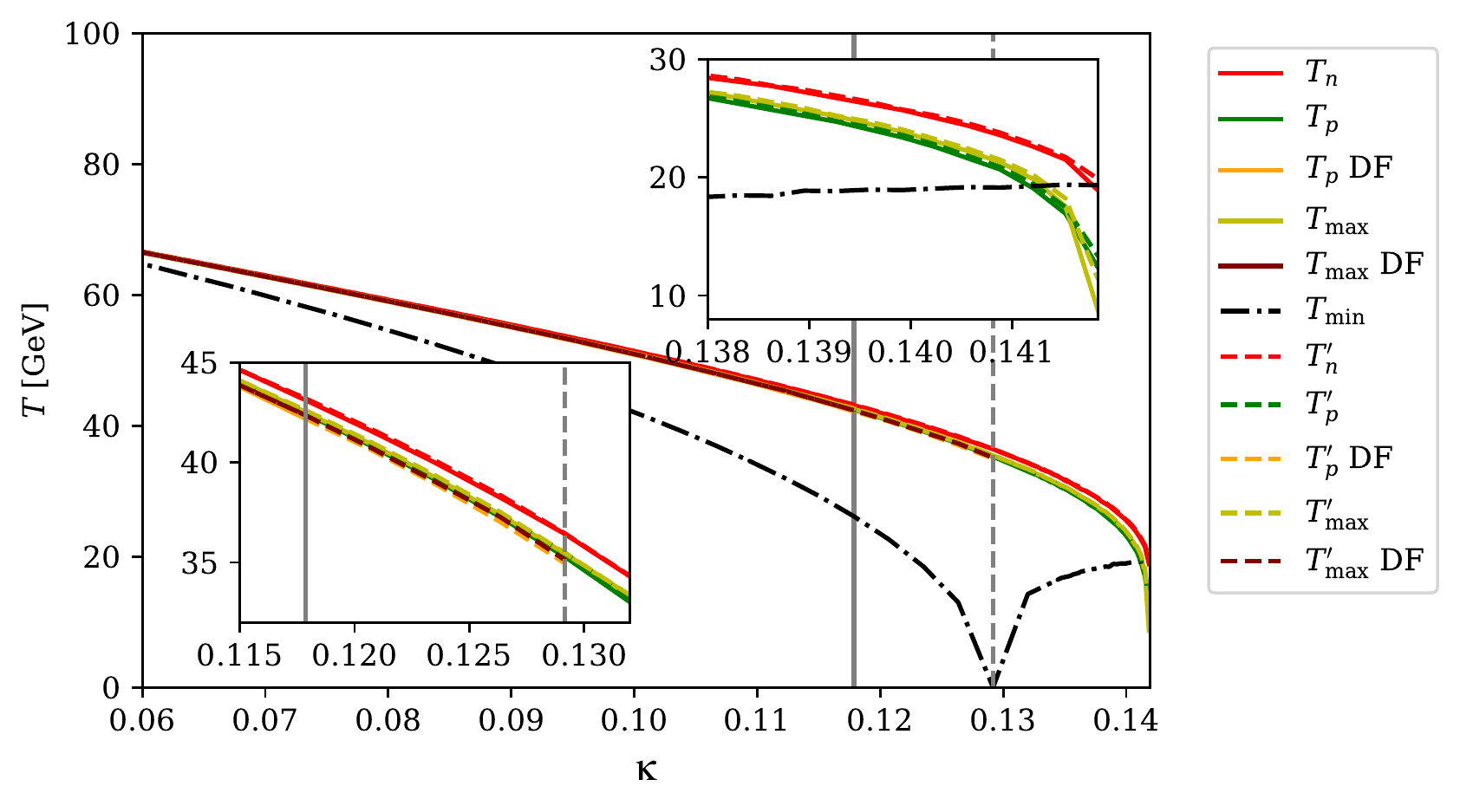}
		\end{minipage}%
	}%
	\centering
	\caption{The characteristic temperature of the logarithm model. The colored solid lines represent various characteristic temperatures for the conventional definition $\alpha$.
		The colored dashed lines denote temperatures that is derived for the alternative definition $\alpha'$.
		The black dash-dotted line denotes the minimum temperature.
		The vertical gray solid and dashed lines indicate values of $\kappa$ that can produce a SFOPT with $\alpha_p=1$ and $\alpha_p^\prime=1$, respectively.}\label{logt}
\end{figure}

Following the same approaches, we perform the numerical calculations.
In figure~\ref{logt}, we show various characteristic temperatures of the logarithm model.
We can see that the hierarchy between the nucleation temperature and percolation temperature in the same
definition increases with the increasing of the coupling $\kappa$.
The characteristic temperatures derived by different definitions of phase transition (conventional definition for solid lines, alternative definition for dashed lines) show negligible differences.
As we discussed before, the minimum temperature (the dashed-doted line) should not be affected by different definitions of phase transition strength.
This model can generate a potential barrier at zero-temperature as the quartic toy model with $A\ne0$.
Hence we should consider the completion of phase transition.
These characteristic temperature decreases with the increasing value of $\kappa$.
When $\kappa > 0.142$, the maximum temperature can not exist, and these parameter space should be excluded.

\begin{figure}[t]
	\centering
	
	\subfigure{
		\begin{minipage}[t]{1\linewidth}
			\centering
			\includegraphics[scale=0.8]{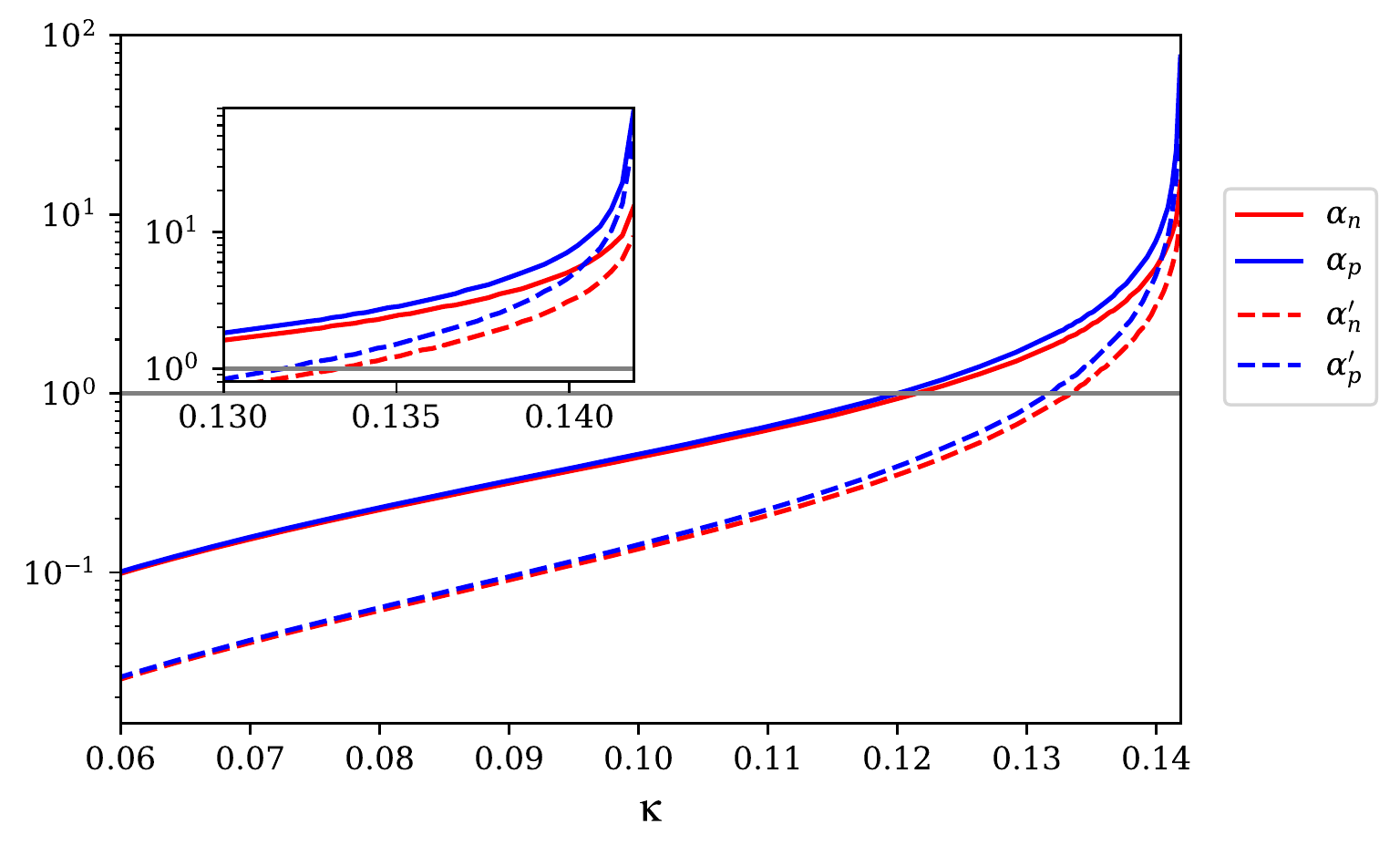}
		\end{minipage}%
	}%
	\centering
	\caption{The phase transition strength as a function of $\kappa$ at the nucleation and percolation temperature for the logarithm model with different definitions.
		Solid lines represent the values with conventional definition.
		Dashed lines denote the values with alternative definition.
	The horizontal gray line indicates that the phase transition strength parameters are equal to one.}\label{loga}
\end{figure}

In figure~\ref{loga}, we show
the phase transition strength as a function of $\kappa$ at the nucleation and percolation temperature for different definitions.
Solid lines represent the values with conventional definition.
Dashed lines denote the values with alternative definition.
For conventional definition, we can see that for $\kappa$ larger than $0.12$, the ultra supercooling occurs with the phase transition strength larger than 1.
However, the ultra supercooling can be generated with $\kappa > 0.132$ for the alternative definition.
And for the ultra supercooling, $\alpha_p$ and $\alpha_p'$ is obviously larger than $\alpha_n$ and $\alpha_n'$.
The conventional phase transition strength is stronger than the alternative one.
It is more accurate to use the percolation temperature to obtain more accurate and
larger phase transition strength.

\begin{figure}[t]
	\centering
	
	\subfigure{
		\begin{minipage}[t]{1\linewidth}
			\centering
			\includegraphics[scale=0.8]{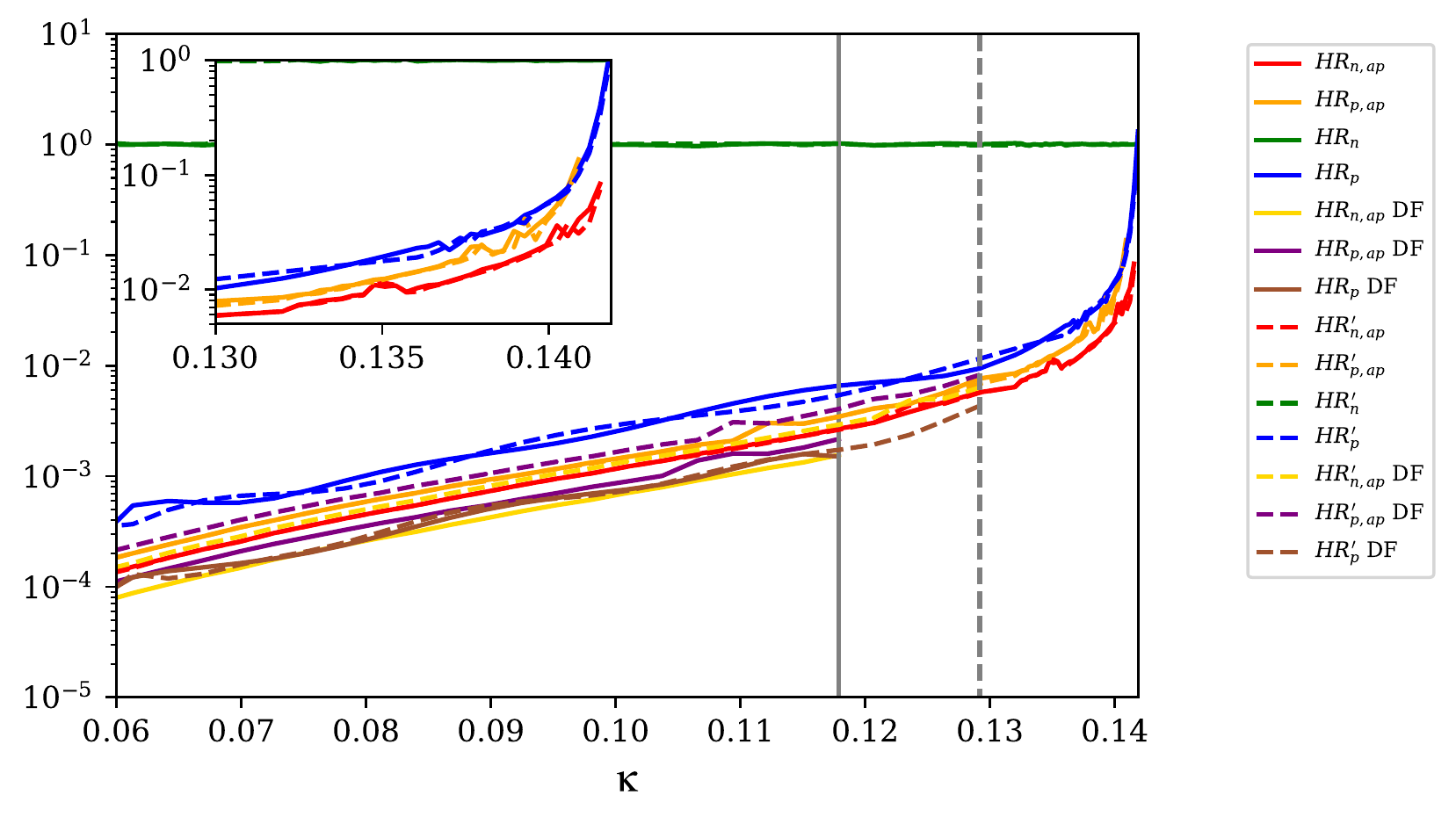}
		\end{minipage}%
	}%
	\centering
	\caption{
The characteristic length scale of the logarithm model. $HR_{*,ap}$ (red, orange, yellow, and purple solid lines), which is derived by the approximation, and $HR_*$ (green, blue, and brown solid lines), which is calculated by the first principle, denote the characteristic length scale at the nucleation and percolation temperature.
The dashed lines denote the values of characteristic length at different temperatures for different definitions of the phase transition strength.
The vertical gray solid and dashed lines indicate values of $\kappa$ that can produce a SFOPT with $\alpha_p=1$ and $\alpha_p^\prime=1$, respectively.}\label{hrlog}
\end{figure}

We present the characteristic length scale of logarithm model in figure~\ref{hrlog}. $HR_{*,ap}$ (red, orange, yellow, and purple solid lines), which is derived by the approximation, and $HR_*$ (green, blue, and brown solid lines), which is calculated by the first principle, denote the characteristic length at nucleation and percolation temperature.
The dashed lines denote the values of characteristic length at different temperatures for the alternative definition $\alpha'$.
The yellow, purple and brown lines show the deflagration expansion mode ($v_b = 0.3$) give negligible modifications to the characteristic length scale for regime of a SFOPT with $\alpha_p < 1$ and $\alpha_p' < 1$.
The vertical gray solid and dashed lines indicate values of $\kappa$ that can produce a SFOPT with $\alpha_p=1$ and $\alpha_p^\prime=1$, respectively.
Similar to the previous models, the approximation of characteristic length fits the strict calculation well for the deflagration mode.
For the detonation mode, these approximation is not as good as deflagration mode.
And we can also find the modification of these length scales that are induced by different definitions of phase transition strength is extremely small, and these results are similar to the previous models.

\begin{figure}[t]
	\centering
	
	\subfigure{
		\begin{minipage}[t]{0.5\linewidth}
			\centering
			\includegraphics[scale=0.5]{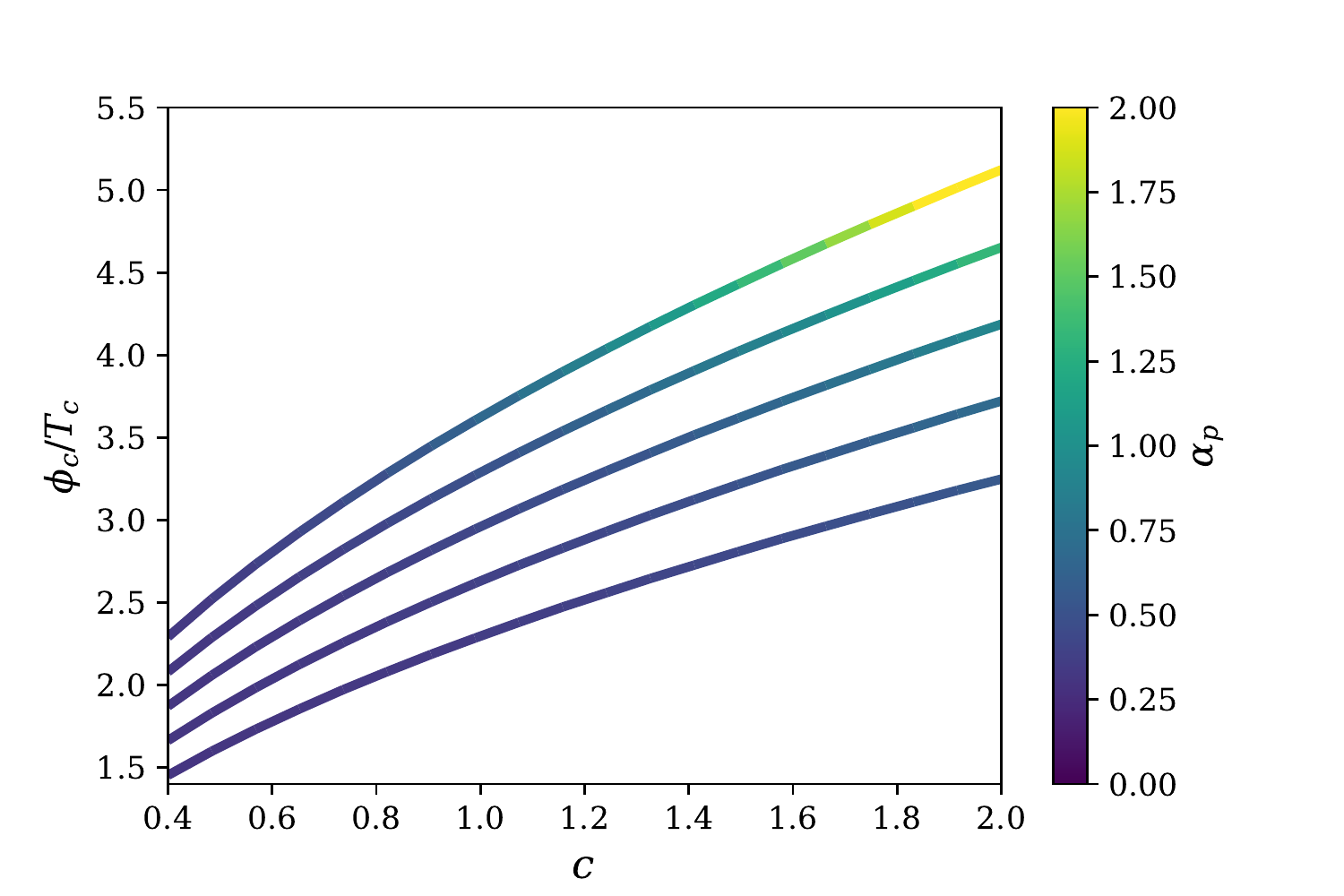}
		\end{minipage}%
	}%
	\subfigure{
		\begin{minipage}[t]{0.5\linewidth}
			\centering
			\includegraphics[scale=0.5]{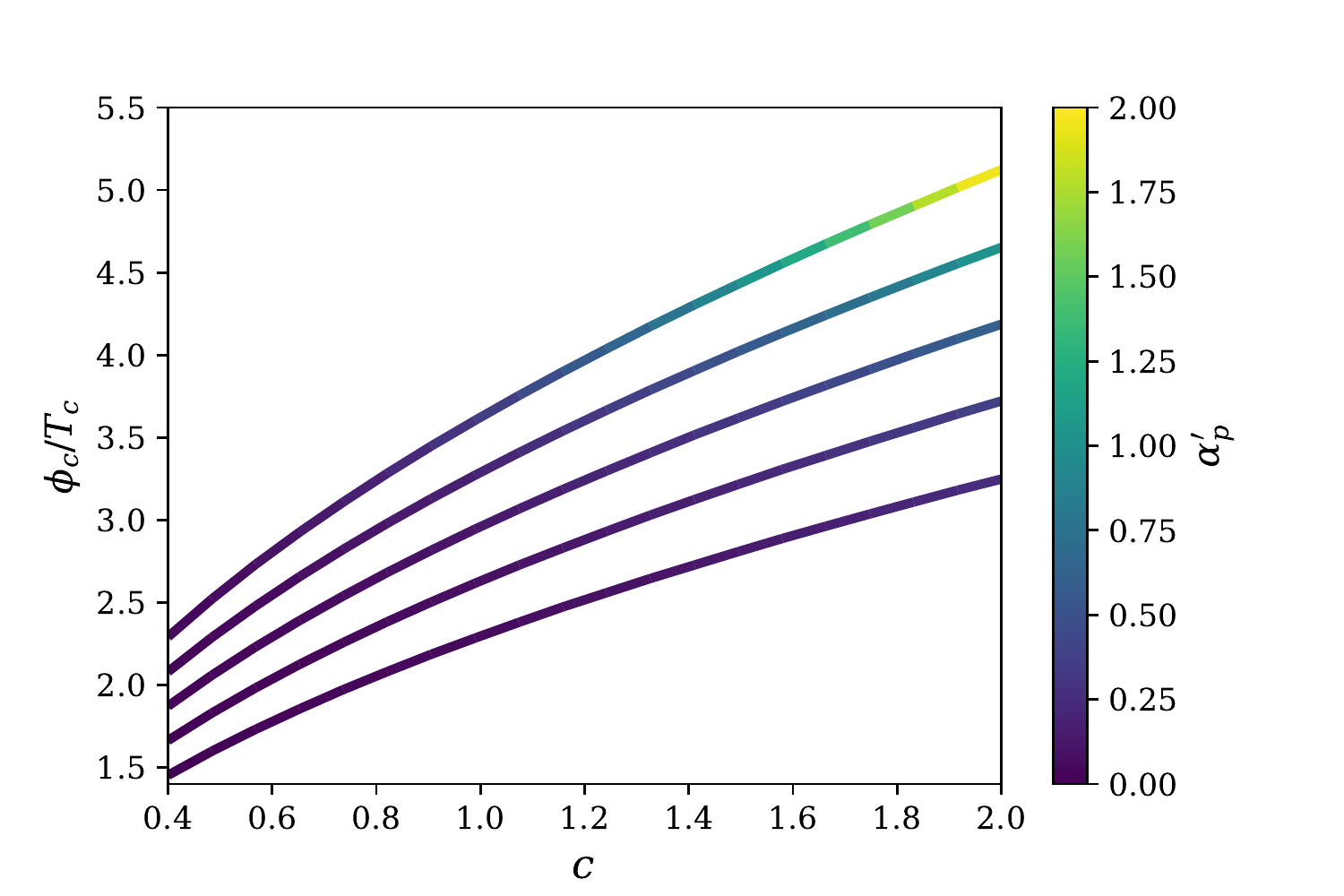}
			%\caption{fig2}
		\end{minipage}%
	}%
	\centering
	\caption{The relation between the washout parameter $\phi_c/T_c$, the thermal correction $c$, and the phase transition strength in the logarithm model. The multicolored lines (from bottom to top) are for $\kappa=0.8,0.9,1,1.1,1.2$, respectively.}\label{vtlog}
\end{figure}

In figure~\ref{vtlog}, we present the
relation between the washout parameter $\phi_c/T_c$, the thermal correction $c$, and the phase transition strength in the logarithm model.
The multicolored lines represent (from bottom to top) $\kappa=0.8,0.9,1,1.1,1.2$, respectively.
We can see that strong supercooling and ultra supercooling favor larger thermal correction $c$ just as previous model shown.

In table~\ref{tb7}, we list the  phase transition parameters of the logarithm model ($c = 1$) for five benchmark sets with the conventional definition of $\alpha$.
All the benchmark sets in the table are strong supercooling and ultra supercooling.
For the strong supercooling case, you can see obvious hierarchy  between $\alpha_p$ and $\alpha_n$.
For the ultra supercooling, the hierarchy is even larger.
For example, for the benchmark set $BP_{19}$,  $\alpha_p=16.797$ is almost two times of $\alpha_n=8.870$.
Thus, if we do not use the correct temperature and the phase transition parameter, the signal would be
much weaker than the actual one.
In table~\ref{tb8}, we list the phase transition parameters of the logarithm model ($c = 1$) of five benchmark sets with the alternative definition of $\alpha'$.
We can find similar properties as in the previous table.
However, for the alternative definition of the phase transition strength, they are weaker than the conventional definition ones.

\begin{table}[t]\small%
	\centering
	\begin{tabular}{|cccccccccc|}
		\hline
		& $\kappa$ & $T_n$ [GeV] & $T_p$ [GeV] & $\alpha_n$ & $\alpha_p$ & $\tilde{\beta}_n$ & $\tilde{\beta}_p$ & $HR_p$ & $v_b$\\
		\hline
		\multirow{2}*{$BP_{15}$}& \multirow{2}*{0.107}&48.531 &48.012&0.552&0.579&1844.170&1523.875&0.00276&1\\
		&&48.531&47.897&0.552&0.585&1844.170&1220.340&0.000779&0.3\\
		$BP_{16}$&0.121&41.590&40.839&0.965&1.034&963.705&715.219&0.00730&1\\
		$BP_{17}$&0.126&38.243&37.271&1.285&1.413&625.203&514.509&0.00758&1\\
		$BP_{18}$&0.132&34.295&33.013&1.855&2.128&457.086&343.722&0.0112&1\\
		$BP_{19}$&0.1415&21.503&16.928&9.398&22.798&35.049&-33.214&0.393&1\\
		\hline
	\end{tabular}
    \caption{The phase transition parameters of the logarithm model ($c = 1$) for five benchmark sets with the conventional definition $\alpha$.}\label{tb7}
\end{table}

\begin{table}[t]\small%
	\centering
	\begin{tabular}{|cccccccccc|}
		\hline
		& $\kappa$ & $T_n^\prime$ [GeV] & $T_p^\prime$ [GeV] & $\alpha_n^\prime$ & $\alpha_p^\prime$ & $\tilde{\beta}_n^\prime$ & $\tilde{\beta}_p^\prime$ & $HR_p^\prime$ & $v_b$\\
		\hline
		\multirow{2}*{$BP_{15}^\prime$}& \multirow{2}*{0.107}&48.588&48.070&0.178&0.191&1885.980&1553.305&0.00394&1\\
		&&48.588&47.955&0.178&0.194&1885.980&1531.797&0.00105&0.3\\
		$BP_{16}^\prime$&0.121&41.684&40.839&0.366&0.409&961.487&715.219&0.00592&1\\
		$BP_{17}^\prime$&0.126&38.364&37.393&0.536&0.607&645.778&528.301&0.0010&1\\
		$BP_{18}^\prime$&0.132&34.323&33.204&0.873&1.026&454.908&360.531&0.0156&1\\
		$BP_{19}^\prime$&0.1415&21.708&17.447&6.334&16.008&39.321&-25.396&0.333&1\\
		\hline
	\end{tabular}
    \caption{The phase transition parameters of the logarithm model ($c = 1$) for five benchmark sets with the alternative definition $\alpha'$.}\label{tb8}
\end{table}

\begin{figure}[t]
	\centering
	
	\subfigure{
		\begin{minipage}[t]{0.5\linewidth}
			\centering
			\includegraphics[scale=0.5]{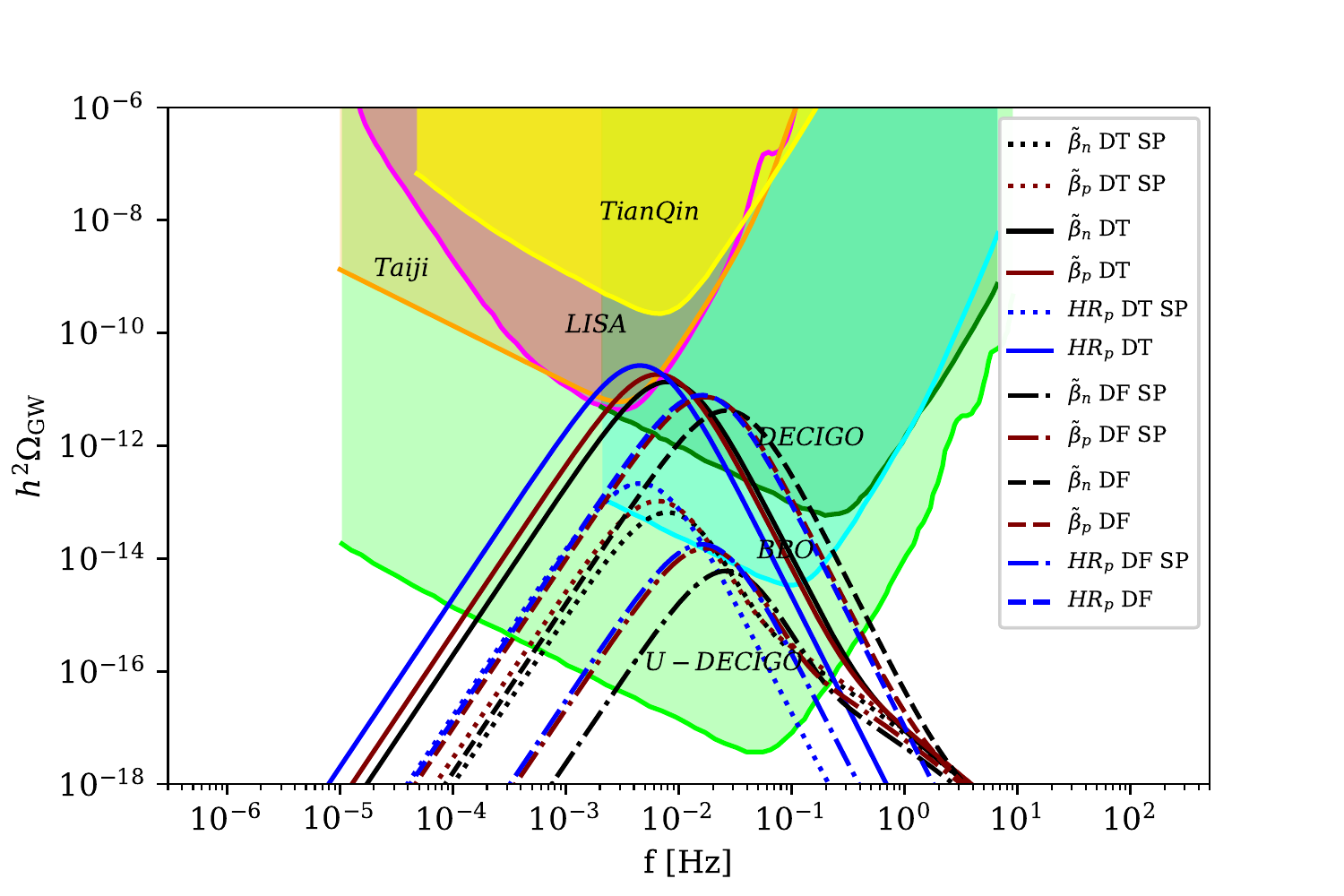}
		\end{minipage}%
	}%
	\subfigure{
		\begin{minipage}[t]{0.5\linewidth}
			\centering
			\includegraphics[scale=0.5]{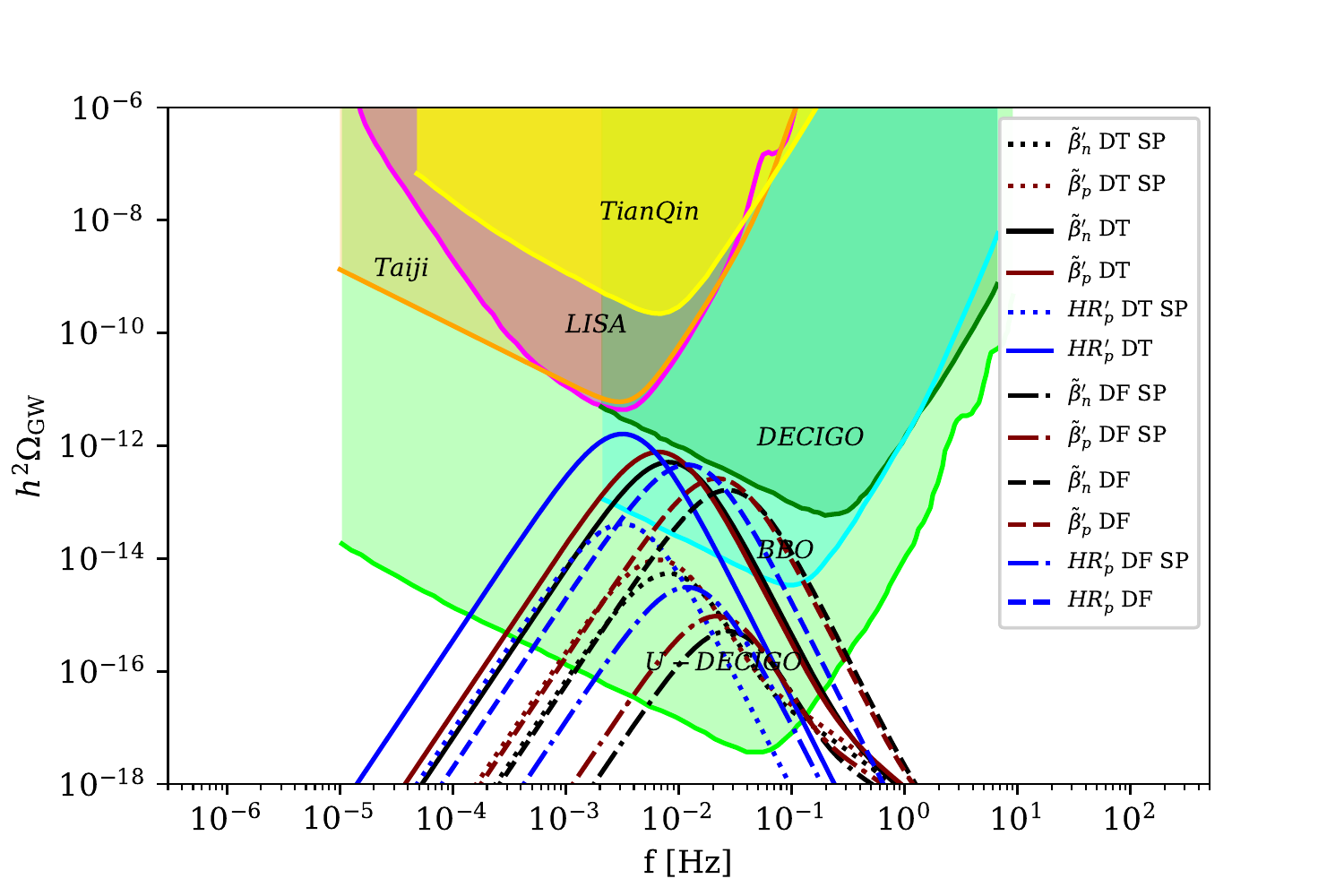}
			%\caption{fig2}
		\end{minipage}%
	}%
	\quad
	\subfigure{
		\begin{minipage}[t]{0.5\linewidth}
			\centering
			\includegraphics[scale=0.5]{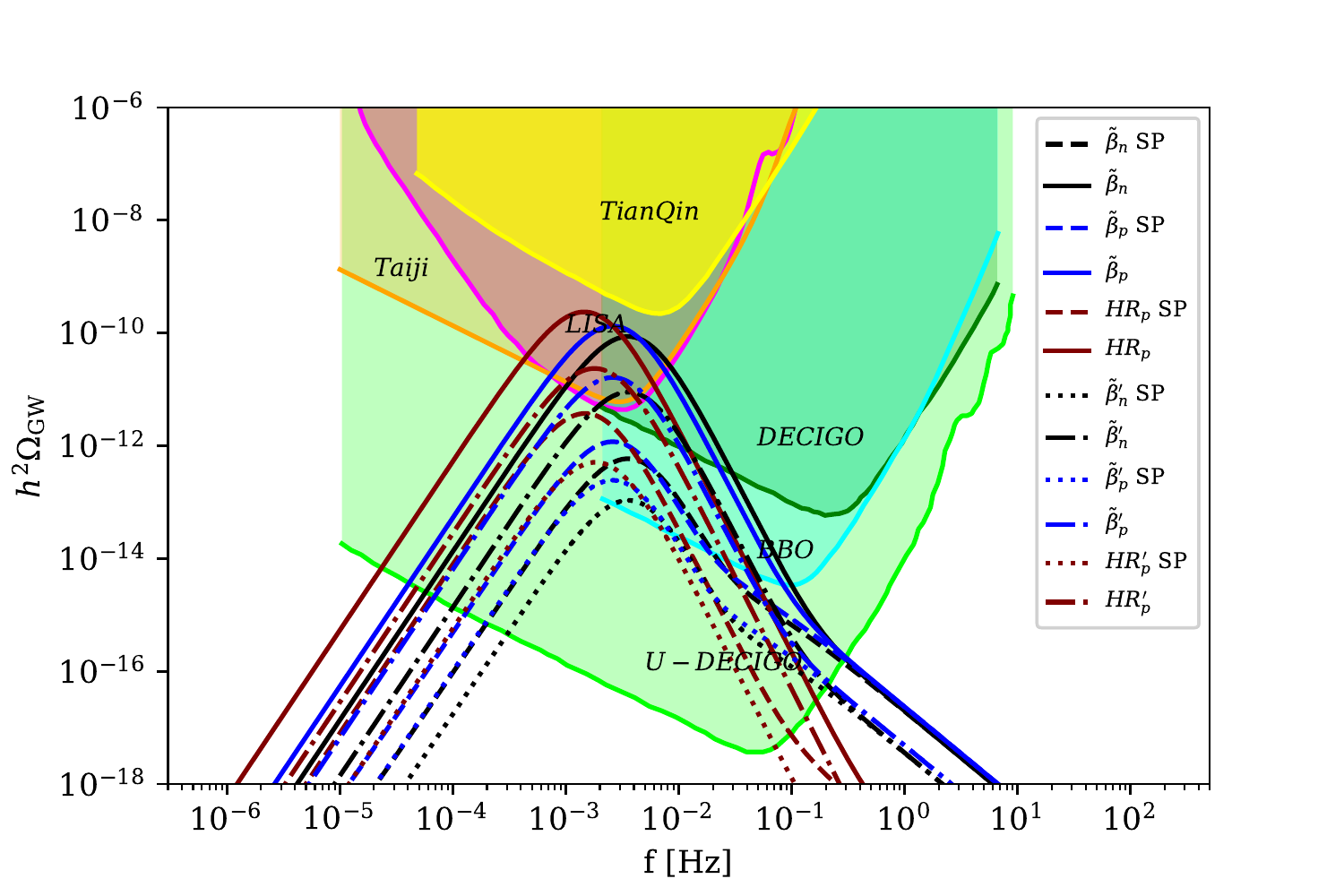}
		\end{minipage}
	}%
	\subfigure{
		\begin{minipage}[t]{0.5\linewidth}
			\centering
			\includegraphics[scale=0.5]{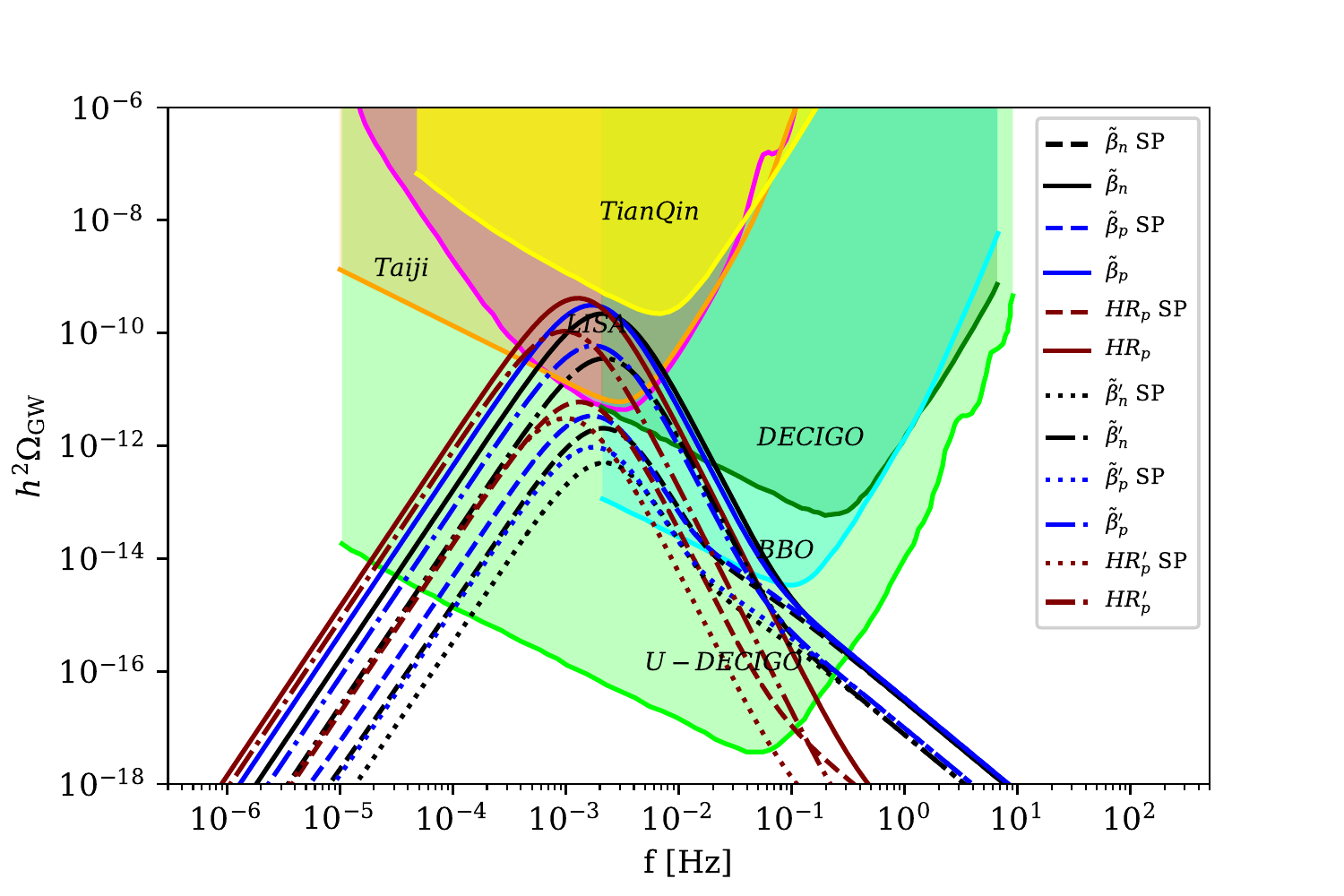}
		\end{minipage}
	}%
	\quad
	\subfigure{
		\begin{minipage}[t]{0.5\linewidth}
			\centering
			\includegraphics[scale=0.5]{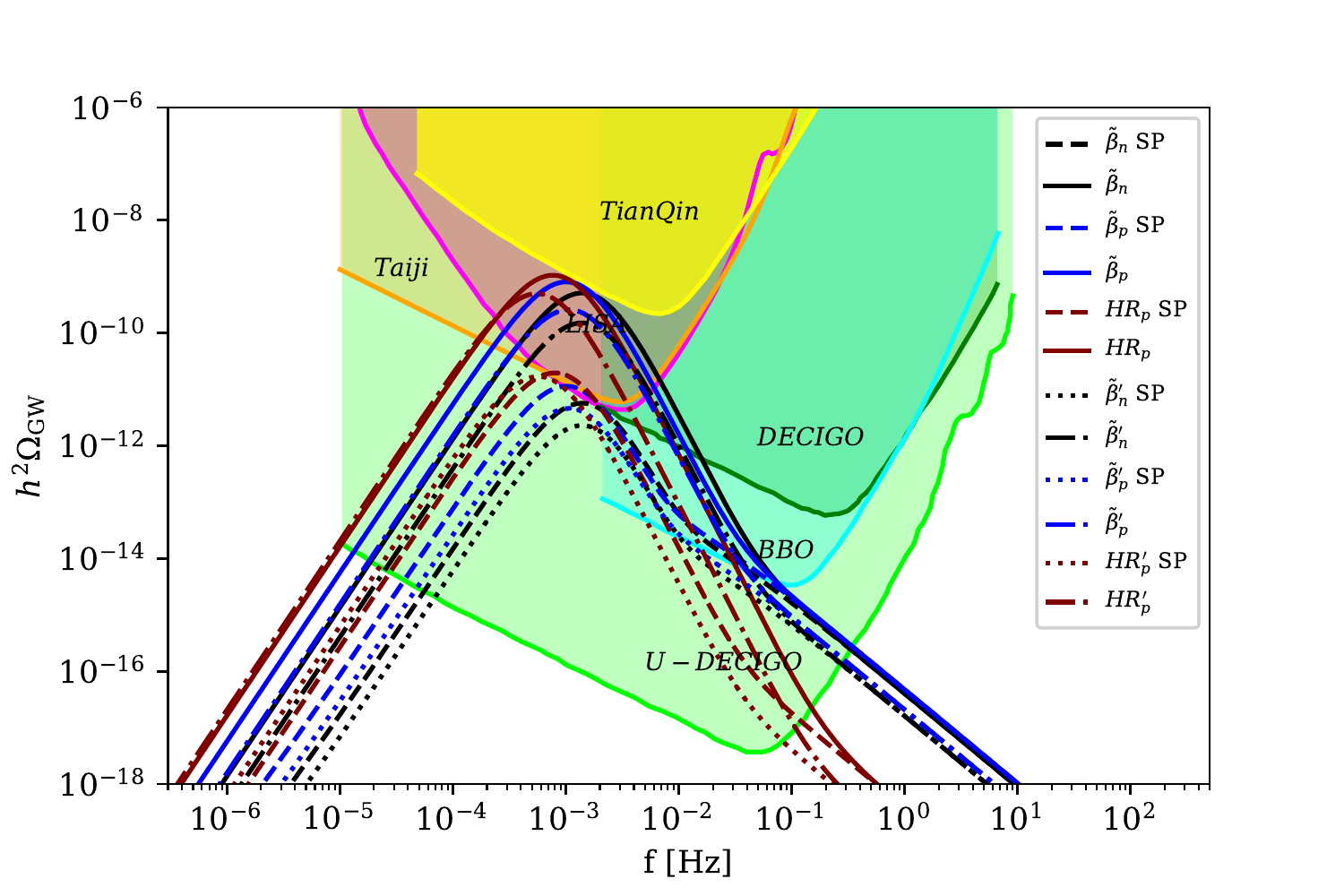}
		\end{minipage}
	}%
	\subfigure{
		\begin{minipage}[t]{0.5\linewidth}
			\centering
			\includegraphics[scale=0.5]{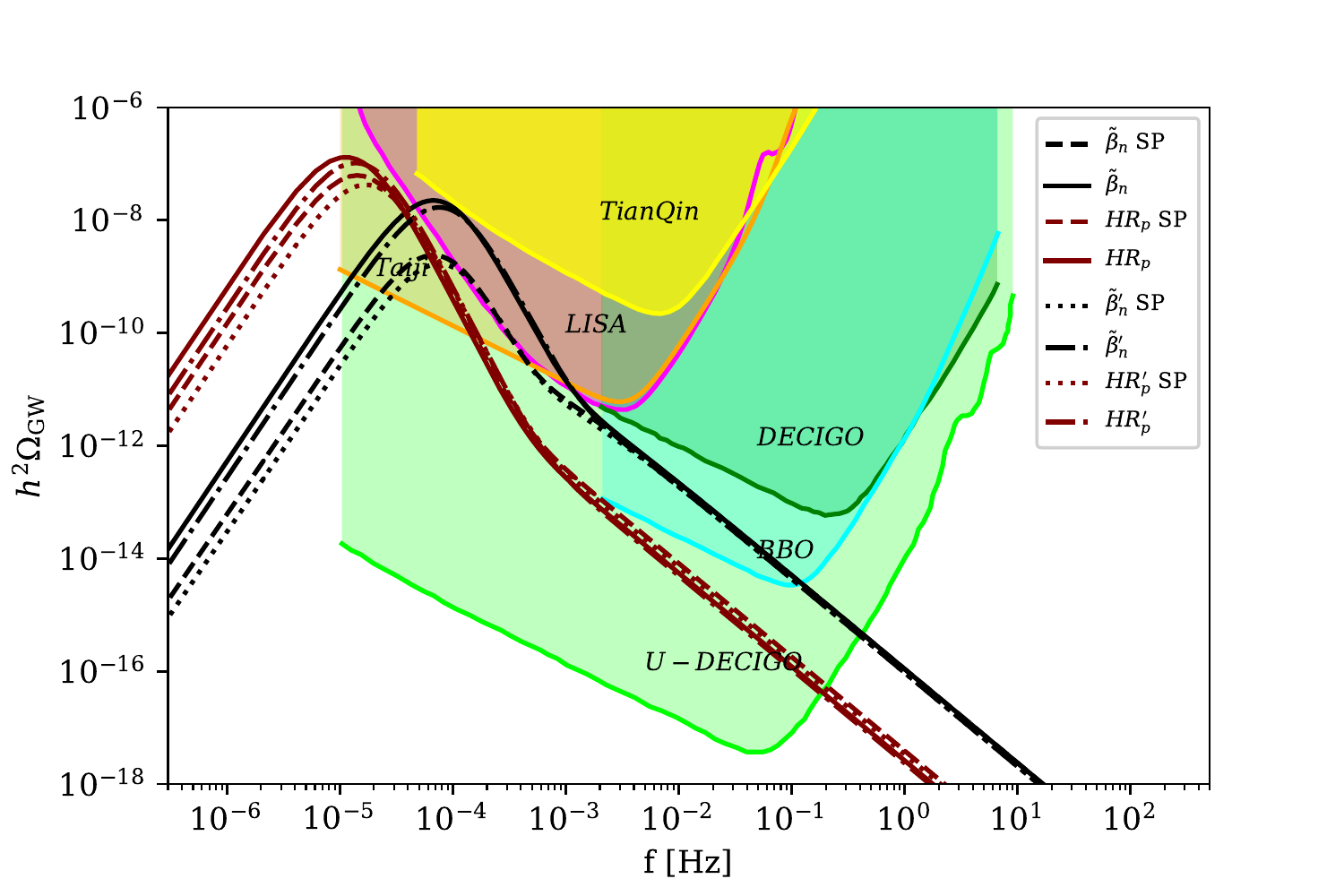}
		\end{minipage}
	}%
	\centering
	\caption{The GW spectra of the logarithm model for different benchmark sets.
The upper left plot and upper right plot show the GW spectra of $BP_{15}$ and $BP_{15}^\prime$.
The middle left plot denotes the GW spectra of $BP_{16}$ and $BP_{16}^\prime$.
The middle right plot denotes the GW spectra of $BP_{17}$ and $BP_{17}^\prime$.
The bottom left plot shows the GW spectra of $BP_{18}$ and $BP_{18}^\prime$.
The bottom right plot represents the GW spectra of $BP_{19}$ and $BP_{19}^\prime$.
	}\label{loggw}
\end{figure}

\begin{figure}[t]
	\centering
	
	\subfigure{
		\begin{minipage}[t]{1\linewidth}
			\centering
			\includegraphics[scale=0.8]{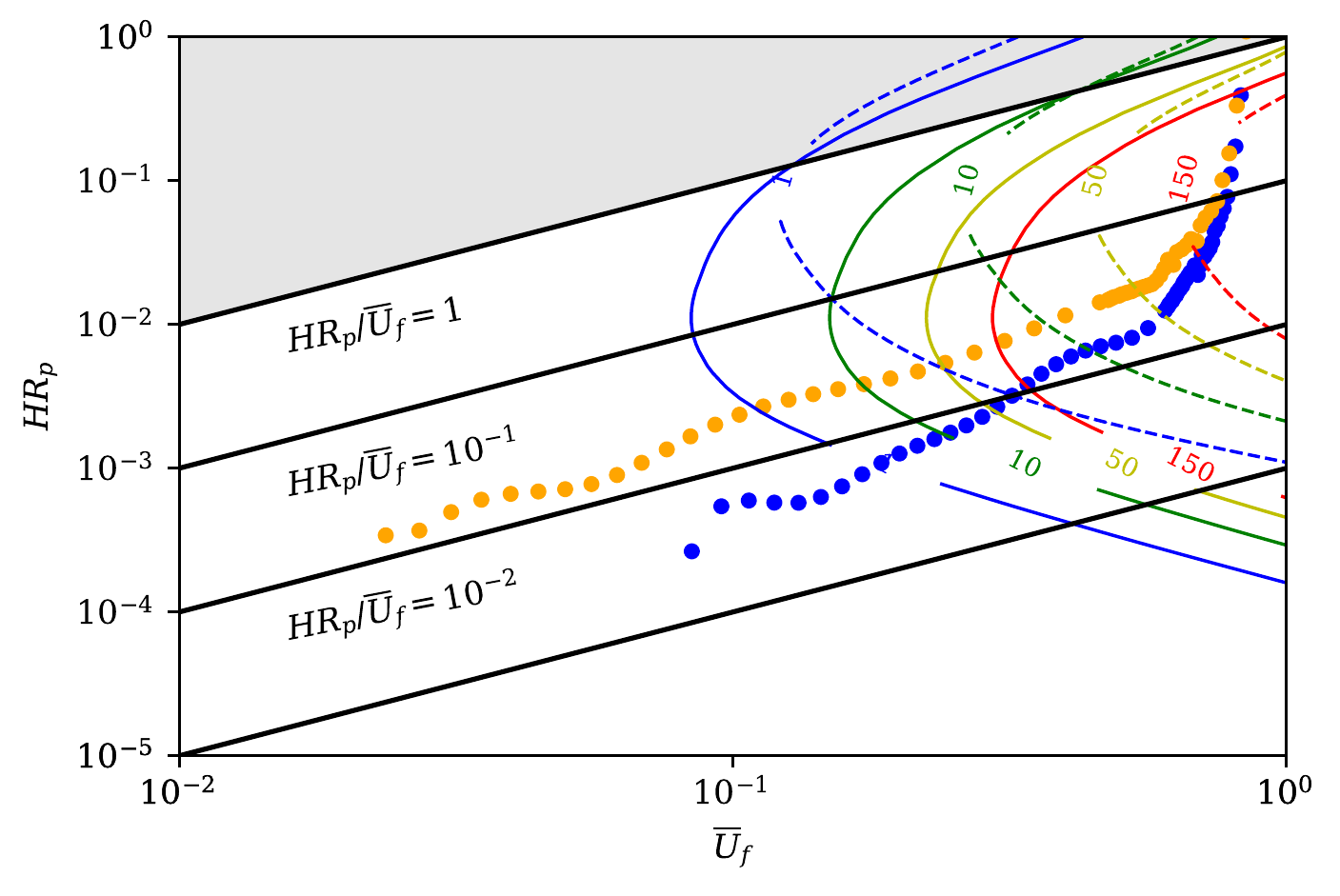}
		\end{minipage}%
	}%
	\caption{
SNR for the logarithm model suppressed (colored dashed contour) and unsuppressed (colored solid contour) power spectra with LISA configuration and  $T_p = 50 $~GeV.
The black solid lines show the magnitude of $HR_p/\overline{U}_f$.
The blue and orange dots are parameter points of the logarithm toy model with strict calculation of the length scale $HR_p$ and $\overline{U}_f$, which are derived by the conventional and alternative definition of the phase transition strength, respectively.}\label{snrlog}
\end{figure}

In figure~\ref{loggw}, we show the GW spectra of the logarithm model for different benchmark sets.
The upper left plot and upper right plot show the GW spectra of $BP_{15}$ and $BP_{15}^\prime$.
The middle left plot denotes the GW spectra of $BP_{16}$ and $BP_{16}^\prime$.
The middle right plot denotes the GW spectra of $BP_{17}$ and $BP_{17}^\prime$.
The bottom  plot shows the GW spectra of $BP_{18}$ and $BP_{18}^\prime$.
The bottom right plot represents the GW spectra of $BP_{19}$ and $BP_{19}^\prime$.
It is obvious that using the phase transition parameters calculated at the percolation temperature
produces stronger phase transition GW signals.
And the GW signal of the detonation case is stronger than the one of the deflagration case.
%For some benchmark sets, their larger GW spectra are within the sensitivity of the future GW experiments.
The GW signals with the suppressed contribution of sound wave are weaker, and we can see the suppression effect is significant for most of benchmark sets except $BP_{19}$ and $BP_{19}$.
When $\alpha$ (or $\alpha'$) becomes larger, the value of $HR_p/\overline{U}_f$ should be more close to order one.
This is direct result of the figure~\ref{snrlog}.

In figure~\ref{snrlog}, we show the
SNR for the logarithm  model suppressed (colored dashed contour) and unsuppressed (colored solid contour) power spectra with LISA configuration and  $T_p = 50 $~GeV.
The black solid lines show the magnitude of $HR_p/\overline{U}_f$.
The blue and orange dots are parameter points of the logarithm toy model with strict calculation of the length scale $HR_p$ and $\overline{U}_f$, which are derived by the conventional and alternative definition of phase transition strength, respectively.
The detectability of a given benchmark set can be significant different when the suppression effects of sound wave mechanism are considered.
And different definitions can also give a very different SNR for the same parameter set.
However, this difference should be smaller when the phase transition strength becomes larger for both definition.

\section{Discussion}
\label{Discu}

\begin{figure}[t]
	\centering
	
	\subfigure{
		\begin{minipage}[t]{0.5\linewidth}
			\centering
			\includegraphics[scale=0.5]{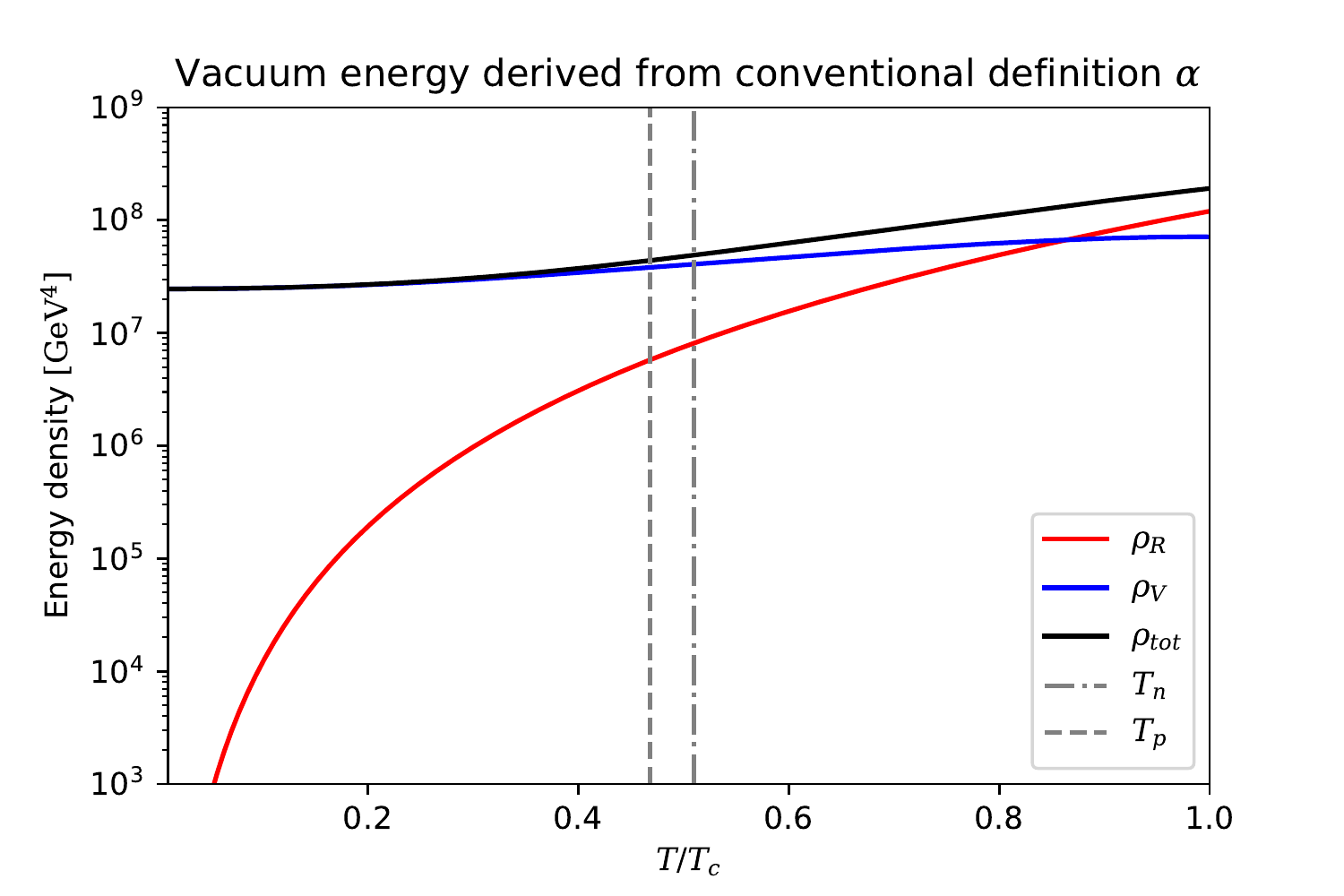}
		\end{minipage}%
	}%
	\subfigure{
		\begin{minipage}[t]{0.5\linewidth}
			\centering
			\includegraphics[scale=0.5]{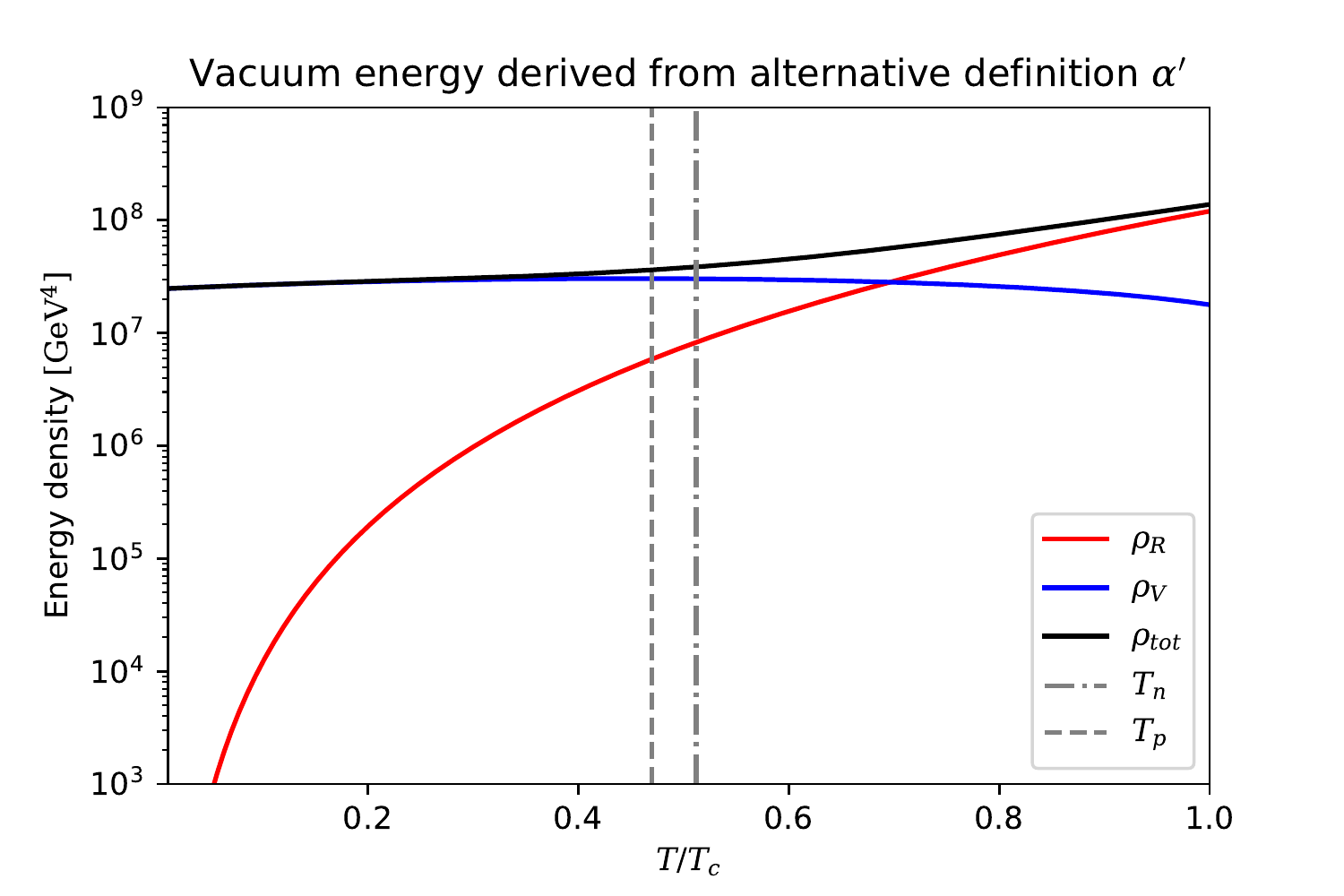}
			%\caption{fig2}
		\end{minipage}%
	}%
	\centering
	\caption{The typical evolution of different energy density for the ultra supercooling case.}\label{vandr}
\end{figure}

For the ultra supercooling case, the vacuum energy can be dominant during the phase transition process.
As shown in figure~\ref{vandr}, we find the vacuum energy density can be regarded as a constant after the vacuum energy becomes dominant for different definitions of the phase transition strength.
Therefore, the Hubble rate is an approximated constant for the vacuum energy dominant period, then the universe should experience an exponential expansion.
And different definitions of phase transition strength \footnote{Actually, different definitions give different vacuum energy density and slight modification to the radiation energy density. We take the same radiation energy density here for simplicity, and it will not give qualitative differences.} only modify the initial time of the vacuum dominant period, but not give any qualitative differences.
Here we present an extreme situation in figure~\ref{vandr} which the vacuum dominant period starts before the nucleation temperature.
The results derived from the three benchmark models show that there are two different situations for the ultra supercooling case.
For the case without zero-temperature potential barrier, this inflationary epoch should exit after the barrier disappears.
However, for the case with zero-temperature potential barrier, the `graceful exit' of inflation should be carefully considered.
And the completion of phase transition should be dealt with carefully for the ultra supercooling with a zero-temperature potential barrier.
The criterion of the completion of phase transition is presented in eq.~\eqref{EC}, and we give the viable parameter space that can obtain a $T_{\max}$ for the three benchmark models.
However, for the parameter space with $T_p<T_{\max}$, the completion of phase transition is still questionable.
And we find the parameter space which fulfills $T_n\ge T_{\min}$ coincides with the parameter space that can give a $T_{\max}$.

\begin{figure}[t]
	\centering
	
	\subfigure{
		\begin{minipage}[t]{0.5\linewidth}
			\centering
			\includegraphics[scale=0.5]{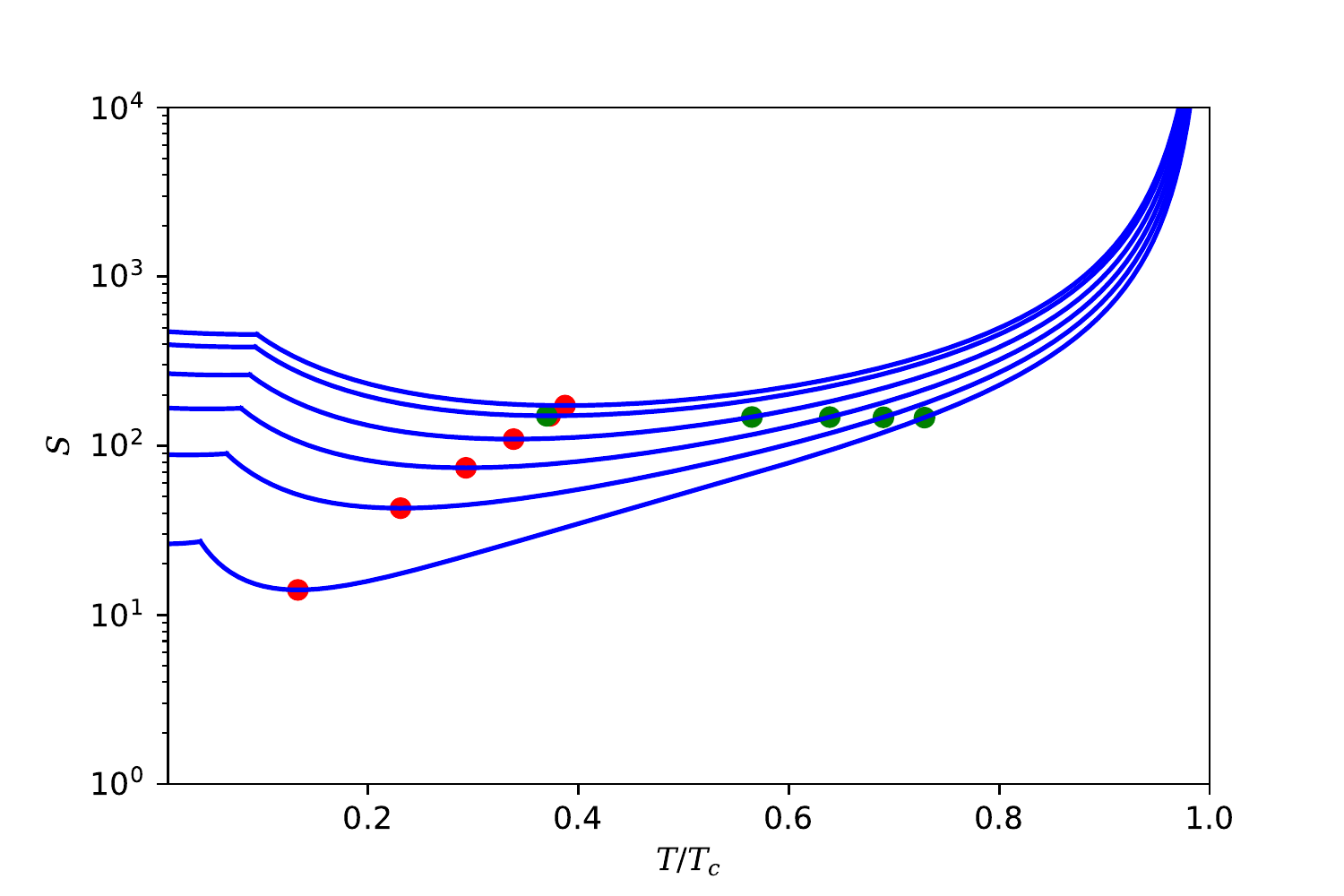}
		\end{minipage}%
	}%
	\subfigure{
		\begin{minipage}[t]{0.5\linewidth}
			\centering
			\includegraphics[scale=0.5]{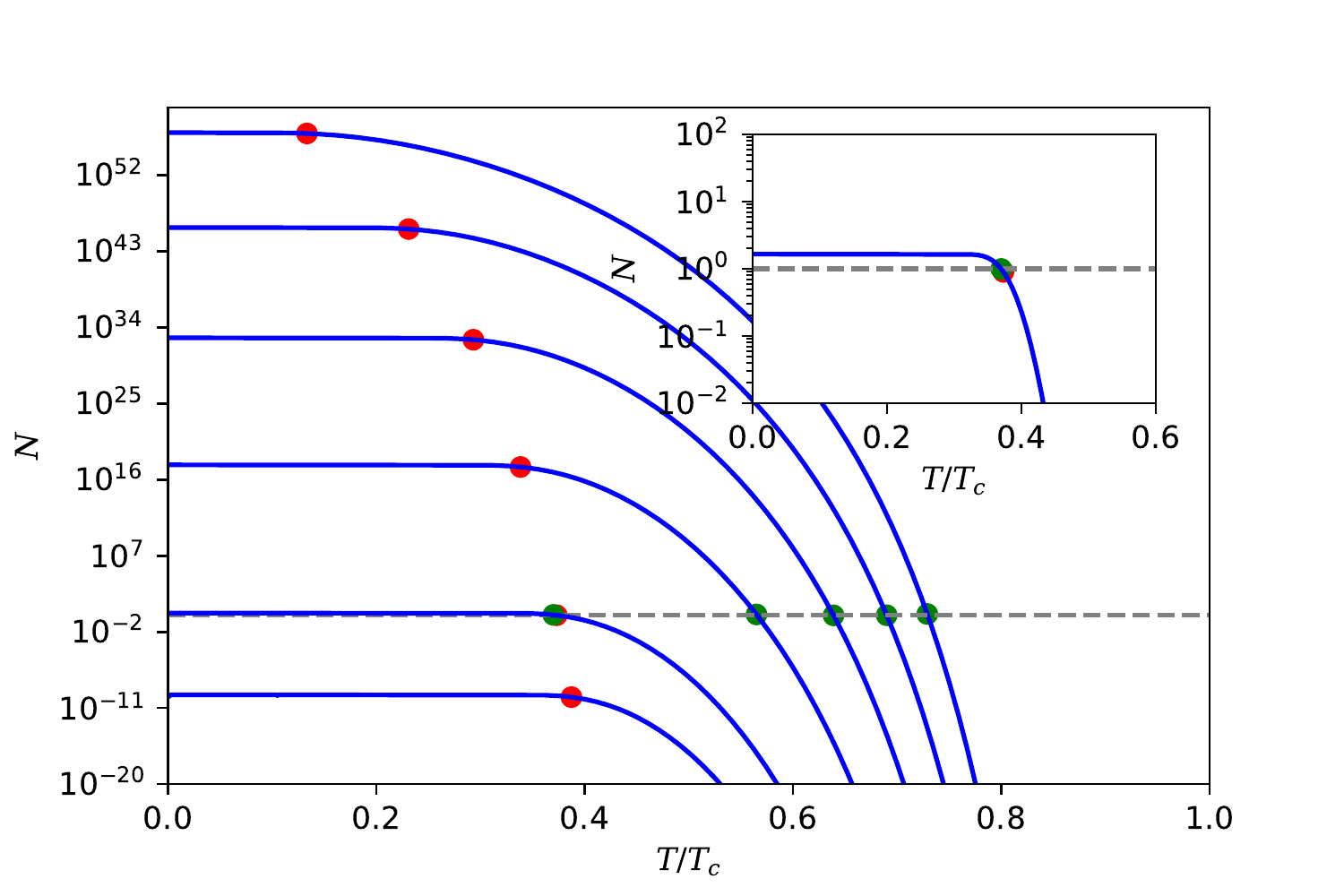}
			%\caption{fig2}
		\end{minipage}%
	}%
	\centering
	\caption{Left: the illustration of the evolution of bounce action with increasing phase transition strength (from bottom to up). Right: the evolution of the accumulated number of bubbles in one Hubble volume (for corresponding bounce action shown in left panel) with increasing phase transition strength (from up to bottom). The green and red dots indicate the nucleation and minimum temperature, respectively.}\label{NAA}
\end{figure}

For ultra supercooling case with zero-temperature potential barrier, $\Gamma$ increases with the decreasing temperature, and can obtain a maximum value at the minimum temperature $T_{\min}$, then decreases with the decreasing temperature.
However, $H^{-1}$ increases with the decreasing temperature, and becomes a constant after the vacuum energy dominates the total energy.
The typical behavior of $\Gamma/H^4$ is represented in figure~\ref{PIN}, $\Gamma/H^4$ obtains a maximum value at minimum temperature $T_{\min}$.
The integrand of eq.~\eqref{EC} should show a similar behavior of $\Gamma/H^4$ and can also get a maximum value at $T_{\min}$, but it will be suppressed by a factor of $1/T$.
Hence the integral of eq.~\eqref{EC} should approach the maximum values at the minimum temperature, as we can see from figure~\ref{NAA}.
The nucleation temperature becomes closer to the minimum temperature with the increasing of the phase transition strength as indicted in figure~\ref{NAA}.
And we can also see that the accumulated number of bubbles in one Hubble volume decrease rapidly with the increasing of phase transition strength.\footnote{The result shown in figure~\ref{NAA} should be model independent, we use the quartic model with $A\ne0$ to illustrate it here.}
When the nucleation temperature is equal to the minimum temperature or slightly larger than the minimum temperature, the accumulated number of bubbles in one Hubble volume is order one.
And for larger phase transition strength, the accumulated number of bubbles will even be less than one.
Basically, there is no bubble formed in one Hubble volume during the phase transition, hence the phase transition can not terminate.

Here, we propose an alternative criterion on the completion of phase transition for the ultra supercooling case, namely, when
\begin{equation}
T_n < T_{\min}\,\,,
\end{equation}
the completion of phase transition should be problematic.
And we can use this criterion to give a roughly estimation of the viable parameter space that a SFOPT with zero-temperature potential barrier can terminate. 
When the nucleation temperature becomes closer to the minimum temperature, the bubble number in one Hubble volume is of order one or even smaller, as shown in figure~\ref{NAA}.
Since the generation of GW is strongly related to the evolution and the number of bubbles,
for these cases, the three mechanisms of GW generation that mentioned in section~\ref{GW} may not be triggered during the phase transition process, hence the corresponding GW signals might not be generated.
Therefore, only in the parameter space that a SFOPT can terminate, the GW can indeed be generated.

\section{Conclusion}
\label{Conclu}
We have classified the SFOPT into four cases based on the properties of the phase transition strength:
\begin{enumerate}
\item Slight supercooling.  It corresponds to $\alpha_p \leq 0.1$.
In this case, $\alpha_n$ can be a good approximation to $\alpha_p$ since $\alpha_n-\alpha_p \ll 0.1$.
For slight supercooling, the GW signal is too weak and difficult to be detected by LISA.
The signal may be within the sensitivity of BBO and U-DECIGO.
\item Mild supercooling.  It corresponds to $0.1 \leq \alpha_p \leq 0.5$. For mild supercooling, its GW signal is well within the expected sensitivity of LISA, Taiji, TianQin, DECIGO, BBO and U-DECIGO. The phase transition strength at the percolation temperature is larger than one at the nucleation temperature.
\item Strong supercooling. It corresponds to $0.5 \leq \alpha_p \leq 1.0$. For strong supercooling, the phase transition GW is more stronger than the mild supercooling, and can be detected by LISA, Taiji, TianQin, DECIGO, BBO and U-DECIGO. And, the phase transition strength at the percolation temperature is obviously larger than the values at the nucleation temperature.
\item Ultra supercooling. It corresponds to  $\alpha_p \geq 1$. For ultra supercooling,  its GW signal is even stronger and more
easy to be detected by LISA, Taiji, TianQin, DECIGO, BBO and U-DECIGO.
The hierarchy of phase transition strength derived from the nucleation and the percolation temperature becomes more significant.
%The phase transition strength at percolation temperature may be much larger than one at nucleation temperature.
\end{enumerate}

We study all the four types  of supercooling. 
However, we pay more attention to the strong supercooling and ultra supercooling cases.
One hand is that they are more sensitive to the accurate characteristic temperature and length scale, and produce stronger GW signals.
On the other hand, there are several confusing points in these two cases.
Firstly, we have discussed and clarified the important aspects in calculating the phase transition dynamics and more reliable phase transition GW spectra in model independent ways. Different characteristic temperatures and length scales are clarified.
Secondly, we have studied three representative models in details.  For the dimension-six effective model, there is obvious hierarchy between the results calculated at $T_n$ and $T_p$.
They can be used as a benchmark example when we need to precisely calculate the phase transition dynamics and the corresponding GWs.
For the quartic toy model with $A\ne0$, it is also more accurate to use $T_p$ and $\alpha_p$ to predict the phase transition dynamics and GW signals.
For the quartic toy model with $A=0$, since its deviation from the bag EoS is large, it is more appropriate to use the alternative definition of phase transition strength $\alpha_p'$ and the corresponding parameters to calculated the GW signals.
However, for both cases the hierarchy between the values calculated at the nucleation temperature and percolation temperature is not as obvious as in the dimension-six effective model.
For the logarithm model, the hierarchy of the phase transition strength calculated at different characteristic temperatures can be extremely large.
And the GW signal calculated at the percolation temperature is obviously stronger than one derived at the nucleation temperature.

Based on the general systematical discussions and the concrete study of three representative models,
there are several subtle points in the calculations of phase transition dynamics and GW signals:
\begin{enumerate}
\item

We emphasize that it is more reliable to use the percolation temperature $T_p$ instead of the nucleation temperature $T_n$ to calculate the phase transition parameters and the corresponding  GWs.
Especially, when there is obvious hierarchy between $T_p$ and $T_n$, a significant hierarchy between the phase transition strength calculated at the two different temperatures can appear.
The most typical case is the ultra supercooling cosmological phase transition where $\alpha_p$ is obviously larger than $\alpha_n$ in the same definition.
And for strong supercooling and ultra supercooling,  we should use $T_p$ and $\alpha_p$ (or $\alpha_p^\prime$) to calculate the GW signals.
It originates from the fact that the GW signals mainly begin to produce when bubbles start to percolate.
The formation of bubbles just start at the nucleation temperature.
Therefore, there is no bubble collision, sound wave or turbulence, and hence no GW signals are produced at the nucleation temperature.
Thus, it is more natural to calculate all the phase transition parameters at the percolation temperature.
And in the slight supercooling case, the absolute differences between the nucleation temperature and percolation temperature and the hierarchy between $\alpha_p$ and $\alpha_n$ are small.
Due to the negligible hierarchy, the nucleation temperature can be used as simplification and approximation.
As for the mild supercooling, it is also better to calculate all the quantities at percolation temperature even the effects are not as obvious as in the strong supercooling and ultra supercooling cases.

\item The definition of phase transition strength should be carefully defined and used in the calculations of GW signal for a concrete model.
Since the conventional definition does not work well when the bag model is obviously violated in some cases, and we should use the alternative definition to obtain more reliable results.

\item  Most viable parameters of these three representative models that can induce a SFOPT can not produce a long-lasting sound wave. Hence, the GW signals should be suppressed. The conventional calculation of GW overestimates the amplitude of GW spectra.

\item When the time duration becomes large (common situation in the strong and ultra supercooling cases), using $\tilde{\beta}$ (or $\beta$) to quantify the inverse time of the phase transition is not accurate. $\tilde{\beta}$ can be very small or even negative. The conventional parameter set is ill defined.
$HR_p$ at the percolation temperature should be used instead of $\tilde{\beta}$ to calculate the GW signals.
It is better not to use $HR_n$ at the nucleation temperature.

\item For the three classes of model, we find that only
when the thermal mass related parameter $c$ and the washout parameter are large enough, an ultra supercooling case can be generated.
%more comment

\item
For the ultra supercooling case, we should carefully check whether the SFOPT can terminate.
The minimum temperature for a phase transition with the typical evolution of action as shown in the left panel of figure~\ref{fg1} can be an approximated benchmark case for the completion of phase transition.
We have found that in these three representative models, the parameter space where $T_n$ is lower than $T_{\rm min}$ is approximately equal to the parameter space where $T_p < T_{\rm max}$.
Therefore, we propose a criterion of the completion of phase transition with ultra supercooling.
That is when $T_n < T_{\rm min}$, the completion of phase transition is questionable, we should deal with it carefully.

\end{enumerate}
After considering all the above points, we can obtain more accurate prediction of the phase transition dynamics and the corresponding GW spectra for a given model with SFOPT.
Precise study of phase transition dynamics helps us to unravel the roles of phase transition in the early universe (baryogenesis, dark matter, etc.).
More precise study on the bubble wall velocity and the numerical simulation for the GW spectra of the ultra supercooling is left for future study.

\acknowledgments

We would like to thank Xiao-Jun Bi, Yi-Fu Cai and Jiang-Hao Yu for useful discussions.
XW and XMZ are supported in part by  the Ministry of Science and Technology of China (2016YFE0104700), the National Natural Science Foundation of China (Grant NO. 11653001), the CAS pilot B project (XDB23020000).
FPH is supported in part by the McDonnell Center for the Space Sciences.

\paragraph{Note added.} While this paper was under completion, we notice ref.~\cite{Ellis:2020awk} appeared on arXiv, partially overlapping with this work.

%\newpage

\end{document}